\def\preprint{0}                
\def\preprint{1}                
\def\comment#1{}
\if\preprint1
        \documentclass[usenatbib]{mn2e}
        \usepackage{times}
        \usepackage{graphicx}
        \usepackage{calc}

\else
        \documentstyle[astrop-bib,referee,times]{mn}
        \newcommand{\includegraphics}[1]{}
\fi

\def\oversim#1#2{\lower0.5pt\vbox{\baselineskip0pt \lineskip-0.5pt
     \ialign{$\mathsurround0pt #1\hfil##\hfil$\crcr#2\crcr\sim\crcr}}}

\title[A mid-infrared imaging catalogue of post-AGB stars]{A mid-infrared imaging catalogue of 
post-AGB stars\thanks{Based 
on observations made at the Very Large Telescope at Paranal 
Observatory under the  program 081D.0630 }}
\author[E. Lagadec et al.]{Eric Lagadec$^{1}$\thanks{E-mail:elagadec@eso.org},
 Tijl Verhoelst$^{2}$, Djamel M\'ekarnia$^{3}$, Olga Su\'arez$^{3,4}$, Albert A. Zijlstra$^{5}$,\newauthor
Philippe Bendjoya$^{3}$,
Ryszard Szczerba$^{6}$,
Olivier Chesneau$^{3}$,
Hans Van Winckel$^{2}$,\newauthor
Michael J. Barlow$^{7}$,
Mikako Matsuura$^{7,8}$,
Janet E. Bowey$^{7}$,
Silvia Lorenz-Martins$^{9}$,\newauthor
Tim Gledhill$^{10}$\\
$^{1}$European Southern Observatory, Karl Schwarzschildstrasse 2, Garching 85748, Germany\\
$^{2}$Instituut voor Sterrenkunde, Katholieke Universiteit Leuven, Celestijnenlaan 200D, 3001 Leuven, Belgium\\
$^{3}$Laboratoire Fizeau, OCA/UNS/CNRS UMR6525, 06304 Nice Cedex 4, France \\
$^{4}$Instituto de Astrofísica de Andalucía, CSIC, Apartado 3004, 18080 Granada, Spain \\
$^{5}$Jodrell Bank Centre for Astrophysics, School of Physics \& Astronomy, University of Manchester, Oxford Road, Manchester M13-9PL, UK\\
$^{6}$N. Copernicus Astronomical Center, Rabianska 8, 87-100 Torun, Poland\\
$^{7}$ Department of Physics and Astronomy, University College London, Gower
    Street, London, WC1E 6BT, UK\\
$^{8}$ Mullard Space Science Laboratory, University College London, Holmbury
    St. Mary, Dorking, Surrey RH5 6NT, UK\\
$^{9}$ Observatorio do Valongo, UFRJ, Ladeira do Pedro Antonio 43, 20080-090,
    Saude, Rio de Janeiro, Brazil\\
$^{10}$Science and Technology Research Institute, University of Hertfordshire,
    College Lane, Hatfield AL10 9AB, UK \\
}
\begin{document}
\def\deg{$^\circ$}
\date{}

\pagerange{\pageref{firstpage}--\pageref{lastpage}} \pubyear{2002}

\maketitle

\label{firstpage}

\begin{abstract}
   Post-AGB stars are key objects for the study of the dramatic
morphological changes of low- to intermediate-mass stars on their
evolution from the Asymptotic Giant Branch (AGB) towards the Planetary
Nebula stage. There is growing evidences that
binary interaction processes may very well have a determining role in
the shaping process of many objects, but so far direct
evidence is still weak. We aim at a systematic study of the dust distribution
around a large sample of Post-AGB stars as a probe of the symmetry
breaking in the nebulae around these systems.
   We used imaging in the mid-infrared to study the inner part of these evolved
 stars to probe direct emission from dusty structures in the core of Post-AGB stars
   in order to better understand their shaping mechanisms.
   We imaged a sample of 93 evolved stars and nebulae in the mid-infrared using 
VISIR/VLT, T-Recs/Gemini South and Michelle/Gemini North.
   We found that all the the Proto-Planetary Nebulae we resolved show a clear 
departure from spherical symmetry. 59 out of the 93 observed targets appear to 
be non resolved. The resolved targets can be divided in two categories.
The nebulae with a dense central core, that are either bipolar and multipolar. 
The nebulae with no central core have  an elliptical morphology.
The dense central torus observed likely host binary systems which triggered fast 
outflows that shaped the nebulae.
\end{abstract}


\section{Introduction}

According to our current understanding of stellar
evolution, all stars with main sequence mass in the range 
1-8 M$_{\odot}$ \, evolve via the asymptotic giant branch (AGB) phase to
the planetary nebula stage (PN). As they ascend the AGB, their mass-loss
rate increases from solar-like values ($10^{-14}$ M$_{\odot}$/yr$^{-1}$) 
up to $10^{-4}$ M$_{\odot}$/yr$^{-1}$. This mass loss  is an essential component of galactic evolution,
 as these are the main sources of s-process elements in the Universe (Gustafsson \& Ryde, 1996) and are the main productors of carbon.
AGB stars are also the main contributors to the dust phase of the Interstellar Medium, which is important for the energy balance
in Galaxies.
During the last stages of the AGB, the remains of the
convective hydrogen envelope are ejected during final violent and sporadic
mass-loss events. Dust grains and molecules, predominantly 
CO, condense in their winds, forming substantial circumstellar envelopes 
detectable in the infrared and millimetric domains.

A departure from spherically symmetric mass-loss
is observed in a substantial fraction of suspected transition objects. In
particular, multipolar structures are often associated with Proto-Planetary Nebulae (PPNe)  sources (Sahai, 2002). 
The observed morphology of these objects are projected on the sky, making it difficult to know what is the 
intrinsic morphology of PPNe and PNe. But it is estimated that
around 80\% of all  PNe show aspherical
morphologies (e.g. Manchado 1997).  Hubble Space Telescope observations of PNe,
for example, shows a large range of morphologies, including
elliptical, bipolar, multipolar  or  round nebulae (e.g. Sahai \& Trauger, 1998).  Hydrodynamical models explain many of the
observed structures from a structure-magnification mechanism, where a
fast wind from the central star of the PN 
ploughs into the earlier slow Asymptotic Giant Branch (AGB) wind (Kwok
et al. 1978), amplifying any density asymmetry already present
(Balick, Preston \& Icke 1987; Frank \& Mellema 1994): the Generalized 
Interacting Stellar Wind model or GISW. Another model has also been proposed
 by Sahai \& Trauger (1998) to explain the shaping of PNe. In their model the
 shaping of the PNe occurs at the end of the AGB phase when fast collimated jets
 are triggered and shape a bipolar nebula. If the direction of the jets changes with time,
 then multipolar nebulae can be formed. Such jets could be formed through interaction with a companion, e.g. 
in an accretion disc (for a review see Balick \& Franck, 2002).

Much of PN and PPN shaping theory relies on the presence of circumstellar material in either
a dusty torus or disc.  Our team has discovered some discs/tori in the heart of PNe (Lagadec et al.
 2006, Chesneau et al. 2006, Matsuura et al. 2006, Chesneau et al. 2007) using adaptive optics on the Very Large Telescope (VLT) and mid-infrared
 interferometry at the Very Large Telescope Interferometer (VLTI). But the role of these discs/tori for the shaping of the nebula is still unclear as
 we do not know which fraction of the total dusty mass is present in these central cores, nor the fraction of objects exhibiting such a disc/torus.

These equatorial structures are likely due to the interaction of the central star with a binary companion.
 But observation wise, there are yet no strong direct observational evidences for
this neither in the PPN phase (Hrivnak et al.,2010).
Miszalski et al.(2009, 2010)  discovered some  central binary systems in PNe, but no clear connection between the
binaries and morphological class.
Some binary post-AGB stars are known and they have very compact discs,
not resolveable with
direct imaging but only with interferrometry.
These discs are lickely Keplerian and the binary orbits revealed so
far indicate that strong interaction must have taken place on the AGB
or even RGB. The Spectral Energy Distributions (SEDs) of these objects are  specific and RV Tauri stars with dust
are mainly found in this category
(see e.g. Deroo et al.,(2006) for the interferometry; De Ruyter et al., (2006) for the SED;
Van Winckel et al., (2009) for the binarity; De Ruyter et al., (2005) and
Gielen et al. 2009  for the RV Tauri stars).

To observe the inner part of post-AGB stars, we need mid-infrared observations, as the dust optical depth
 is smaller at longer wavelengths. Mid-infrared is indeed the only wavelengths range  at which we can observe the inner
 morphology of stars from the AGB to the PPN phase.  Furthermore, the main source of radiation for these sources in the mid-infrared
 is direct emission from dust, while at shorter wavelengths this is scattered light. Mid-infrared imaging is thus  the best way to study the dusty
 structures inside these evolved stars.



Many mid-infrared imaging observations of AGB and PPNe  have been made
 in the past. But the only  mid-infrared imaging survey has been made with 3-m class telescopes (Meixner et al. 1999) and present a 
lack of angular resolution for the morphological study of the observed objects, a selection
 bias as they observed known bipolar nebulae and consists of
 only 17 resolved sources. Some work has been done using
 8-m class telescopes, but always focusing on particular individual bright well-known objects (e.g. Miyata et al. 2004).

We observed 86 evolved stars (2 observed twice, using different modes) using VISIR at the VLT, 
5 using T-Recs (2 also observed with VISIR) on Gemini South and 5 using Michelle (1 also observed with VISIR)  on Gemini North.
Taking into account that some objects were observed twice with different instrument, our total list of targets includes 93 objects.
Here we present this mid-infrared N-band imaging survey of a large number
of post-AGB stars. We aim at a systematic survey to probe the inner
dusty regions of post-AGB stars.

\section{Target selection}
The large sample of observations presented here comes from five  distinct  observing runs
 (3 VISIR/VLT runs:380.D-0630 (Normal mode), 081.D-0130 (Burst mode), 081.D-0616 (Normal mode);
 1 Michelle/Gemini North: GN-2005B-Q-16 and 1 T-Recs/Gemini South run: GS-2005A-Q-34)) and the targets selection were
done in a slightly different way for the different programs.
 The targets of the VISIR normal mode  programs  were selected from the previous mid-infrared
 catalogue by Meixner et al. (1999) and the millimetre observations compiled by Bujarrabal
 et al. (2001). We removed the AGB stars and young PNe and observed all the stars observable 
with the Very Large Telescope (VLT). 

 Most of the stars observed in burst mode were selected from the Torun post-AGB stars catalogue
 (Szczerba et al. 2007),
that list 326 known post-AGB stars. We selected all the post-AGB stars observable in July from the ESO
 Cerro Paranal observatory with an IRAS 12 $\mu$m flux larger than 10 Jy (the burst mode
works only for bright stars). These post-AGB stars includes PPNe, R CrB stars and RV Tauri stars.
R CrB are hydrogen deficient post-AGB stars with known obscuration events (Clayton, 1996).
RV Tauri are pulsating post-AGB stars, located in the high luminosity end of the Population II Cepheid
instability strip (Wallerstein, 2002). These RV Tauri stars are likely to harbour compact dusty discs (Van Winckel et al., 2003).

In addition to that, we observed the brightest Water Fountains, 
and AGB stars observable during this period.
Water fountains are oxygen-rich PPNe characterized by the presence of blue and red-shifted OH and H$_2$O masers (Likkel \& Morris, 1988).
The stars observed with Gemini were selected from their infrared spectral properties.
 We selected star with double chemistry (characterized by the presence of PAHs and crystalline silicates in their infrared spectra), the unidentified 21-micron features or hints of
the presence of an equatorial dusty disc/torus.

 We observed 93 targets: 52 PPNe, 10 Water Fountains, 11 AGB stars, 8 RV Tauri stars,
 4 PNe, 4 Massive Evolved Stars,  2 R CrB stars, 1 Be  and 1 HII regions.

\section{Observations and data reduction}
\subsection{VISIR/VLT observations}
Most of the observations presented here were obtained with the mid-infrared instrument VISIR on the VLT (Lagage et al., 2004).
All the stars were observed with 3 filters: PAH1 (8.59$\mu$m, half band width 0.42$\mu$m ),
 SiC (11.85$\mu$m, 2.34$\mu$m) and NeII (12.81$\mu$m, 0.21$\mu$m).
The PAH1 and NeII filters were chosen for their good sensitivities and their location at the blue
 and red edge of the mid-infrared N band. They also avoid the large opacities
of the SiC or silicates features and  provide information on the dust continuum for both 
oxygen and carbon-rich stars. The broad SiC filter was chosen to provide a general 
N band view. Images obtained with the SiC filter generally have a higher signal to noise 
ratio due to the better sensitivity of this broad band filter.
As all our selected targets are bright, we used the minimum integration time of 30s for
 all the filters in all our observations.
 
We used the imager in normal and burst mode, using a pixel scale of 0.075 arcsec and a field of view of 
19.2 $\times$ 19.2 arcsec. We used the standard chopping/nodding technique to remove the  noise from the sky.
 With the burst mode, all the single chopping and nodding images are recorded, allowing the reconstruction
 of quality-enhanced images using shift and add techniques and lucky imaging.

We used the standard chopping/nodding technique to remove the
background, with a perpendicular chop throw, a chopping frequency of 0.25 Hz and an amplitude of 8 arcsec. We shifted and
added the images using a maximum of correlation algorithm, after removing the bad images, selected as the one for which the
measured flux was smaller than the mean flux of all the images
minus one sigma. The great observing conditions during our run
(0.43 mm of water in the atmosphere) allowed us to obtain great
quality diffraction-limited images. 

The normal mode data were reduced using the VISIR ESO pipeline.
The pipeline first detect the bad pixels, and clean them using an interpolation with neighbouring pixels.
Nodding images are then created by averaging the images in the two positions of the chopper.
 The nodded images are then shifted and added to form
the final combined image.

\subsection{T-Recs/Gemini South observations}
Images were obtained using T-Recs/Gemini South and filters centred at  11.3$\mu$m (PAH, $\Delta$$\lambda$=0.61$\mu$m) and 18.3$\mu$m (Qa, $\Delta$$\lambda$=1.51$\mu$m).
The  chop throw was set to 15 arcseconds.
 
The IRAF MIDIR data reduction package MIREDUCE command was used to combine the different nod images.
 The resulting registered images were averaged together using the
IRAF MISTACK routine. The pixel scale of the obtained images is 0.09 arcseconds and the  field of view of is 28.8"$\times$21.6"

\subsection{Michelle/Gemini observations}

The observations were made with the mid-infrared camera Michelle on the 8-m
 Gemini North telescope (Hawaii, USA) in queue mode in different nights spread between the 26th of 
August 2005 and the 8th of January 2006. We observed
 BD\,+30 3639, IRAS\,21282+5050, OH\,231.8 and the Red Rectangle  with
 three N-band filters (centred at 7.9 ($\Delta$$\lambda$=0.7$\mu$m) 8.8 ($\Delta$$\lambda$=0.9$\mu$m), 9.7($\Delta$$\lambda$=1.0$\mu$m) and 11.6 $\mu$m ($\Delta$$\lambda$=1.1$\mu$m)) and the Qa filter 
(centred at 18.5 $\mu$m, $\Delta$$\lambda$=1.9$\mu$m). HD 56126 was observed with the same filters and the 
N-band 7.9$\mu$m filter. The N and Q band observations were made at different 
dates due to the more stringent weather requirements at Q band. The standard 
chopping-nodding technique was used to remove the sky, with a chop throw of 15 arcsec. 
The spatial resolution measured from standard stars was typically $\sim$0.4 arcsec 
at 10$\mu$m and $\sim$0.6 arcsec at 18.5$\mu$m. The field of view was 32$\times$24'' 
and the pixel scale 0.099''.


Michelle data files contains planes consisting of the difference for each
 chopped pair for each nod-set. Using the Gemini IRAF  package, these difference
 images were combined to create a single frame.

 We thus observed in a quasi uniform way 93 evolved stars in the mid-infrared.
 The names, coordinates, observing modes used and generally accepted type
 classification
of all the stars are presented in Table\,\ref{infostars}. Table\,\ref{photostars} presents
 photometric measurements from the stars from Two Microns All Sky Survey (2MASS)  and the InfraRed Astronomical Satellite (IRAS).

%
\subsection{Flux calibration, deconvolution and artefacts}
The VISIR mid-infrared detector used for most of the observations suffers from striping and the appearance of ghosts for  bright stars. The stripes are horizontal
and repeated every 16 pixels, while the ghosts are distributed vertically every 16 columns. This produces
artefacts that could bring confusion for the image morphology classification.

For all the targets,  we observed standard stars just after or before the observations. 
These standards stars were selected to have a similar airmass to our targets.
These observations were reduced in a similar way as the science targets. They were used 
as a measurement of the point-spread function (PSF)
and for flux calibration purposes. The flux calibration was performed using standard
 aperture photometry methods, applied to the program and reference stars.
All the resolved targets were deconvolved using a Richardson-Lucy algorithm and
 $\sim$30 iterations, depending on the quality of the images.




  
 \section{Results}
\subsection{Measurements}
For each star and each filter observation, we estimated the ellipticity of the
obtained image using an ellipse fitting procedure to all the signal larger than 3 times
 the standard deviation of the distribution. This gives us an orientation of the object, as well as its dimensions
 along its major axis and the axis perpendicular to this axis. 
To estimate whether the objects were resolved or not, we fitted a Gaussian to the observed
 radial profiles for the objects (on the non-deconvolved images) and their associated PSF standard for the different filters. Given
 the seeing stability due to the exceptionally good weather condition for the VISIR burst mode run, 
it is straightforward to estimate whether an object is extended or not for objects observed
 in burst mode.
For objects observed in service mode, we estimated that the objects were extended when
 the object was 50\% more  extended than its associated PSF.
For some dubious cases, where there were hints of an extension in a given direction, we 
also checked  the literature for similar structures observed at different wavelengths.

\subsection{Observed morphologies}
We observed 93 objects, and according to our measurements, 59 are point sources.
A brief description of the properties of the resolved targets is presented in Table \ref{resolved}.
Among these extended targets, we resolved a wealth of different structures, such as resolved central cores,
dark central lanes, detached shells, S-shaped outflows.
 The asymmetrical object (IRAS 12405$-$6219) was misclassified
 as a post-AGB star and is in fact an H~{\sc II} region (Suarez et al., 2009).
One object (IRAS 18184$-$1302) appears  square-shaped object and is also a misclassified post-AGB. It  is a Be star (Tuthill \& Lloyd, 2007). 
 
If we consider only the PPNe  from our sample, we end up with a sample of 52 detected objects.
 29 of this objects are not resolved and 6 are marginally resolved. Among the 17 clearly resolved objects, we find 3
 ellipticals, 10 bipolars and 4 multipolars.

All  11 AGB stars that we observed are unresolved, as well as the 8 RV Tau stars. This seems to indicate
 that no dust shell is present around these stars or at least no
 large and bright shells.
Half of the 10 Water Fountains we observed are resolved. None of the resolved ones are spherical. We
 observed 4 bipolar Water Fountains, most of them with a dark equatorial lane, and
 one multipolar.
For the RCrB stars, one (IRAS 14316$-$3920) appears unresolved, and the other (IRAS 19132$-$3336) appears to be an unresolved
 central source with a more or less spherical dust shell around (very weak).
 
 The morphologies of the clearly resolved objects are summarized in Table \ref{resolved}.




\section{Resolved  objects}
\subsection{Proto Planetary Nebulae}
\subsubsection{IRAS 06176$-$1036}
This object, dubbed the Red Rectangle (AFGL 915, HD44179), has been well 
studied since its discovery by Cohen et al. (1975).
It certainly harbours a binary system and exhibits a 
dual dust chemistry, with the presence of PAHs and
crystalline silicates, as revealed by its ISO spectrum 
(Waters et al., 1998). CO observations reveal the presence of
a Keplerian disc in the equatorial plane (Bujarrabal et al., 2005).
Observations in the optical and the near infrared of the
 Red Rectangle reveal an X-shaped nebula, projection of a bicone (Osterbart et al. 1997,
  M\'ekarnia et al., 1998,
Tuthill et al., 2002). HST images of the nebula reveal a very
 complex morphology, with a ladder-like  structure inside the
 X-shaped nebula (Cohen et al., 2004).
The central star is not seen in these images, as it is obscured
 by a dark lane, certainly due to a dusty disc. The rung of the
 ladder show
a quasi-periodic spacing, indicating a periodic mass-loss from
 the central star.

Hora et al. (1996) observed the Red Rectangle in the mid-infrared with 
UKIRT. Their multiwavelengths images (between 8 and 20 $\mu$m)
show a bright core surrounded by a rectangular shaped nebula.
Lagadec et al. (2004) obtained TIMMI/ESO 3.6m images of the Red Rectangle
 in the N band.
The observations reveal an extended ($6" \times 8"$) rectangular-shaped
 nebula elongated along the North-East/South-West direction
and symmetrical along this direction. The central core appears unresolved.
 The X-shape seen at shorter wavelengths is clearly seen
at 8.4$\mu$m and is less clear at longer wavelengths. The Red Rectangle
 was also imaged with the SUBARU telescope (Miyata et al. 2004).
A similar morphology is revealed, and their 8.8$\mu$m image reveals a similar
 morphology as the one observed in the near-infrared (M\'ekarnia et al., 1998). 
The N band emission is dominated by UIR  emission, attributed to PAHs, while the central
bicone seen at the shortest wavelengths is predominantly due to emission from hot dust  and /or from stochastically heated 
nanoparticles (Gledhill et al., 2009).

\begin{figure*}
\begin{center}
\includegraphics[width=23cm,angle=90]{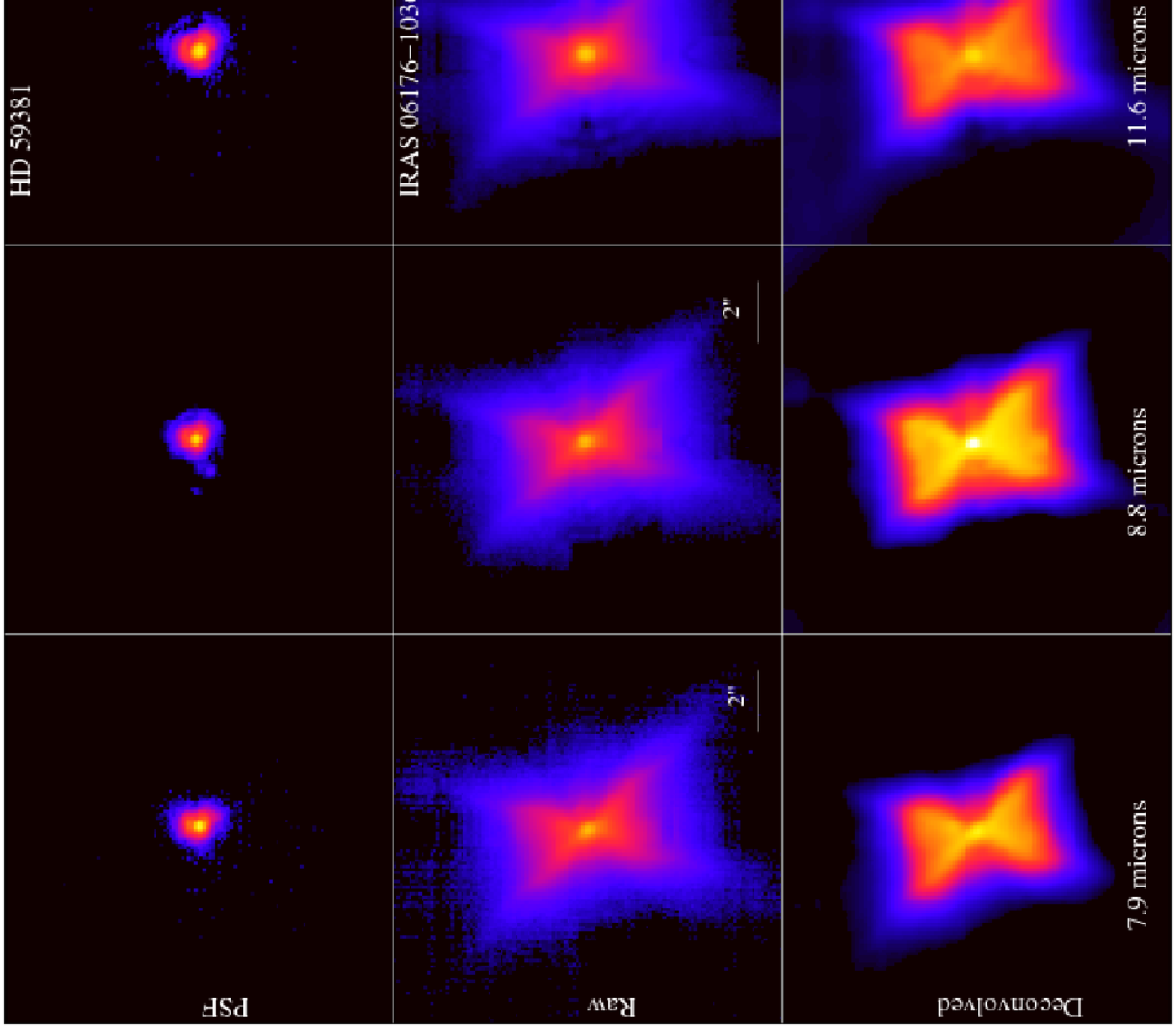}
\caption{ Michelle/Gemini North imnages of IRAS 06176 (The Red Rectangle). North is up and East left.}
\label{im_i06176}
\end{center}
\end{figure*}

Our Michelle images of the Red Rectangle at 7.9, 8.8, 11.6, 12.5 and 18.1$\mu$m 
are displayed Fig.\ref{im_i06176}. The images at  7.9, 8.8, 11.6 and 12.5$\mu$m 
are quite similar and display the well-known rectangular shape of the Red 
Rectangle, with dimensions 3.3$\times$5.9''. The X-shape is clearly seen in our deconvolved images up to 11.6 $\mu$m.
At 18.1$\mu$m, the emission is dominated by an unresolved point source.

  \subsubsection{IRAS 07134$+$1005}
HD 56126 is a well studied post-AGB star exhibiting 
the unidentified 21-micron  dust feature (Kwok et al., 1989). 
It is a carbon rich post-AGB star and its envelope has been already resolved
 in the mid-infrared with TIMMI2 on 
the ESO 3.6m telescope (Hony et al. 2003) and OSCIR 
on Gemini North (Kwok et al. 2002). They clearly detected the central star at
 10.3$\mu$m and marginally at 11.7 and 12.5$\mu$m. 
They resolved a shell-like  envelope of $\sim$5 arcsec in diameter, clearly
 elongated toward the south-west. The shell is not complete  and shows an
 opening in the direction
of its elongation. 

\begin{figure*}
\begin{center}
\includegraphics[width=18cm]{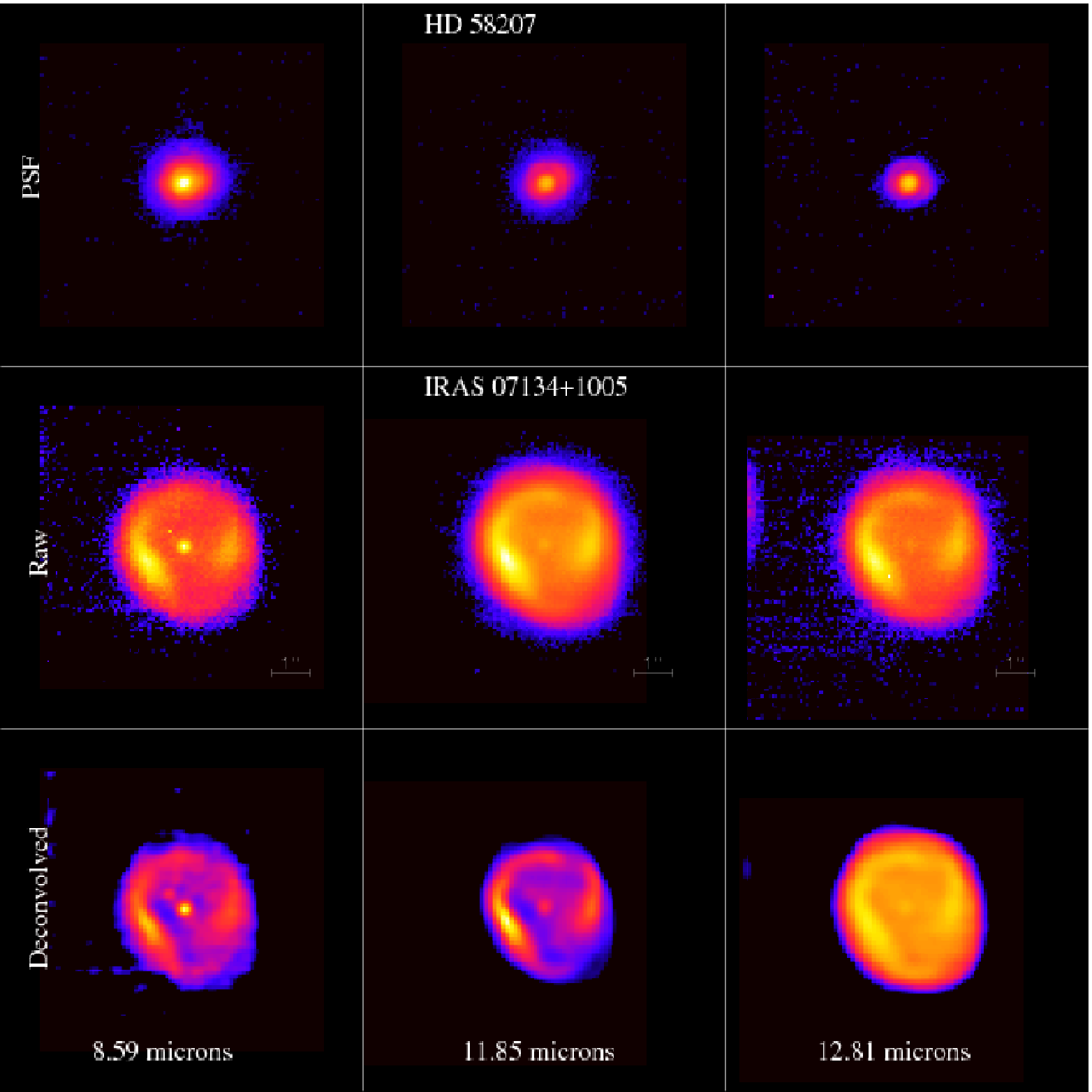}
\caption{Visir images of IRAS 07134 (HD 56126).}
\label{im_07134}
\end{center}
\end{figure*}

\begin{figure*}
\begin{center}
\includegraphics[width=23cm,angle=90]{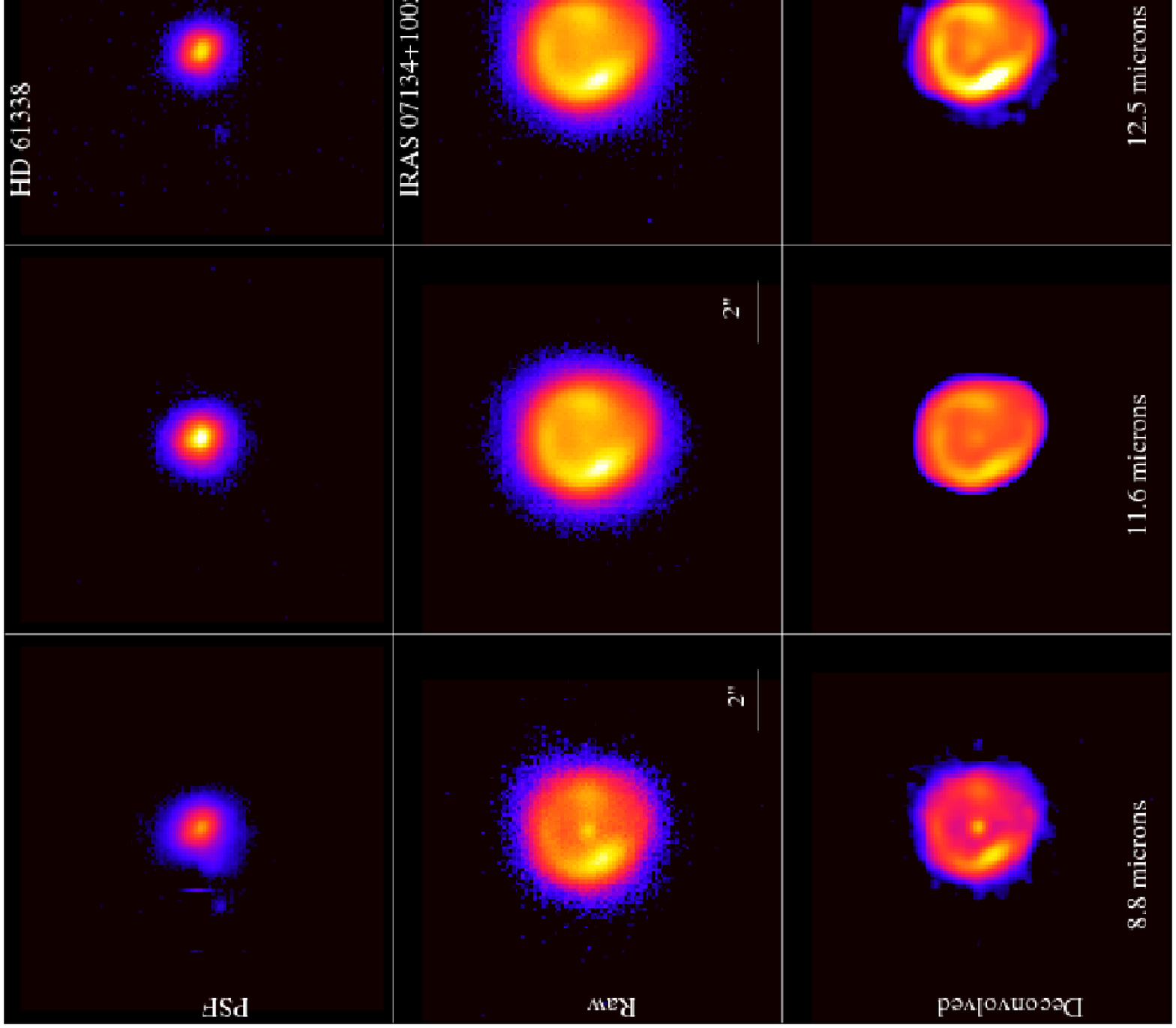}
\caption{ Michelle/Gemini North imnages of IRAS 07134 (HD 56126).}
\label{im_07134_2}
\end{center}
\end{figure*}

The VISIR images of IRAS 07134 at 8.59, 11.85 and 12.81$\mu$m are displayed Fig. \ref{im_07134}.
Our Michelle/Gemini Images of HD 56126 at  8.8, 11.6, 12.5 and 18.1$\mu$m are displayed
 Fig.\ref{im_07134_2}.
The images have a similar morphology and dimensions of 4.4 $\times$4.8''
 at all the observed wavelengths. The envelope has a roughly elliptical
 shell with a PA $\sim$25\deg, that is wider at the North. The brightest part
 of the nebula is a U-shaped structure located in the inner part of this
 elliptic envelope. This U-shaped structure shows a decrease in emission in the
 North-West. The central star is clearly detected from 8.6 to  11.9$\mu$m
 weakly at 12.8$\mu$m and not at longer wavelengths.  Its relative brightness 
 clearly decreases with increasing wavelength.  Our images are very similar to those obtained previously,
 but are certainly sharper. This is clear in our deconvolved images that show the presence
 of filamentary structures and
holes that are very similar in the three filters. We thus clearly see a hole in emission $\sim$1.4 arcsec south of the central star,
which could be due to a decrease in the dust density, or to the presence of  a cold dusty blob.
 
\subsubsection{IRAS 07399$-$1435}

\begin{figure*}
\begin{center}
\includegraphics[width=23cm,angle=90]{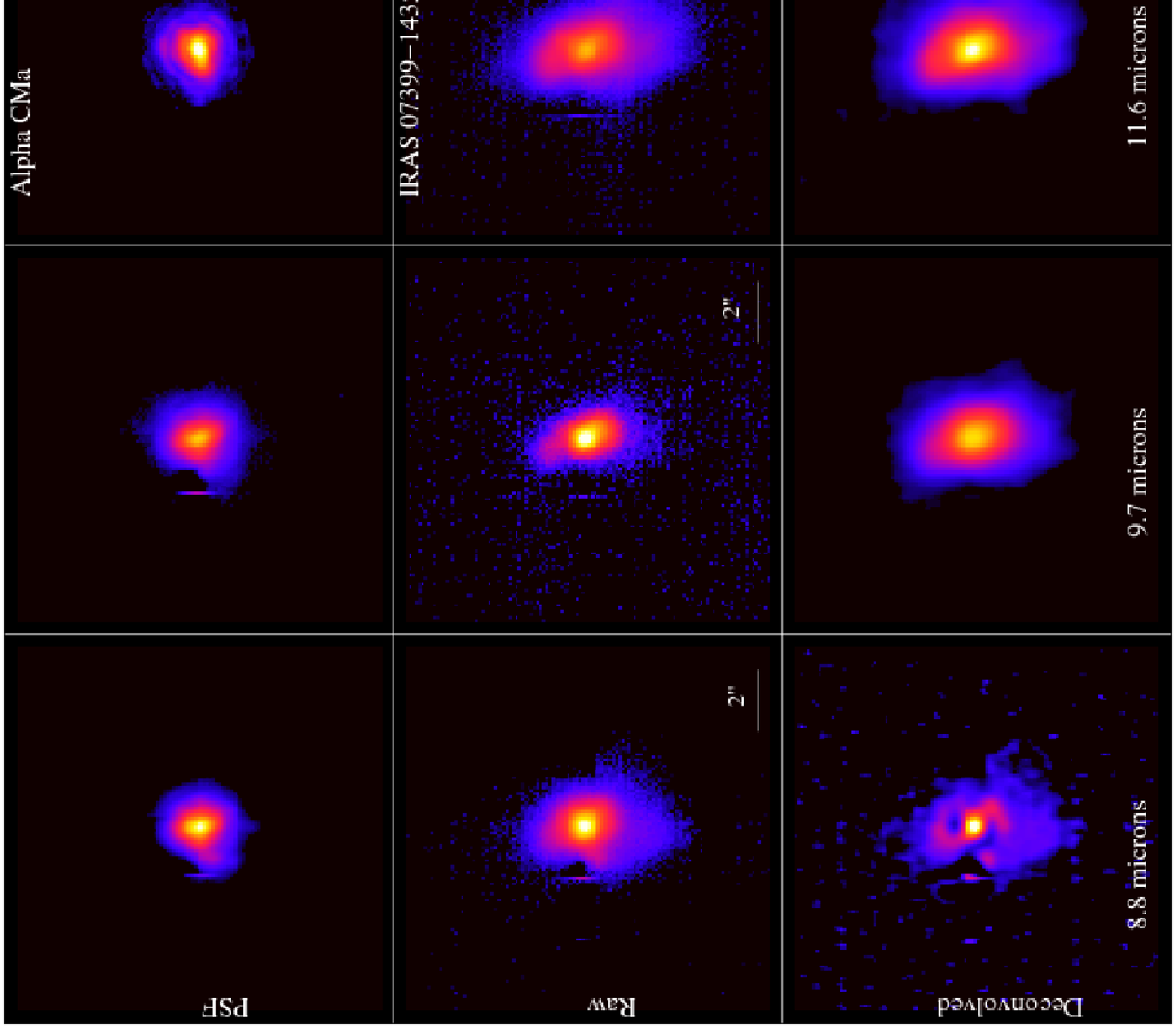}
\caption{ Michelle/Gemini North imnages of IRAS 07399 (OH 231.8+4.2).}
\label{im_07399}
\end{center}
\end{figure*}

OH\,231.8+4.2 is a remarkable bipolar nebula with a central star (QX\,Pup) 
still on the AGB with probably an A0 main sequence companion 
(Sanchez Contreras et al. 2004). Its circumstellar envelope 
has already been imaged at 11.7 and 17.9 $\mu$m using the 
Keck telescope (Jura et al. 2002), showing it to be elongated along the same direction as the bipolar outflows 
observed at shorter wavelengths with a PA $\sim$22\deg and a 
full length $>$3'', the nebula being wider toward the north-east. 
Our Michelle images of OH\,231.8 at 8.8, 9.7, 11.6 and 18.1 $\mu$m are 
displayed Fig.\ref{im_07399}. The vertical stripe at the east of the image is a detector artefact, amplified in our deconvolved 
images. Some artefacts are also seen in the 8.8$\mu$m image, seen  as "holes" north and south of the central source.
The four images are 
extended along a P.A. $\sim$ 22\deg. The dimension of the images are 
4.1$\times$6.1'', 2.6$\times$4.3'', 4.3$\times$6.7'' and 
5.4$\times$9.9''6.7 at 8.8, 9.7, 11.6 and 18.1 $\mu$m respectively.
 All the images show the presence of a bright unresolved core 
 and a diffuse halo. This halo is larger toward the south-east,
 in contradiction with observations by Jura et al. (2002). 
This is however in agreement with previous N band observations of this object
 with TIMMI on the ESO 3.6m telescope (Lagadec, Ph.D. thesis, 2005). 

The central core appears resolved at 9.7$\mu$m (its FWHM is $\sim$0.6 arcsec while its
 PSF standard has a FWHM  of $\sim$0.6 arcsec) and 18.5$\mu$m (FWHM of 0.8 and 0.6 arcsec for the object
 and its PSF respectively). These are the wavelength of silicates stretching and bending modes.

Mid-infrared spectro-interferometric observations of OH\,231.8+4.2 have resolved the central core
 (Matsuura et al. 2006). This core was over-resolved by the interferometer near
 the deep silicate absorption feature they observed around 9.7$\mu$m. 
This indicates the presence of a large silicate dust structure in the equatorial plane of OH 231.8+4.2.

  \subsubsection{IRAS 10197$-$5750}  
Roberts\,22 is a well studied PPN with dual
 dust chemistry (Sahai et al., 1999). Its envelope has been resolved in the mid-infrared with TIMMI
 observation on 
the ESO 3.6m telescope (Lagadec et al. 2005). They obtained 8.39 and 11.65$\mu$m 
images of this object. Both images show an envelope elongated in the direction North-East/South-West
 along a P.A.$\sim$45\deg. 

Images of the envelopes have also been obtained at shorter wavelengths by the Hubble Space Telescope, hereinafter HST,
 (Sahai et al. 1999, Ueta et al., 2007) or using the adaptive optics on the VLT
 (Lagadec et al. 2007).
 These images revealed the presence of an S-shaped envelope, embedded in a larger
 bipolar envelope.

\begin{figure*}
\begin{center}
\includegraphics[width=18cm]{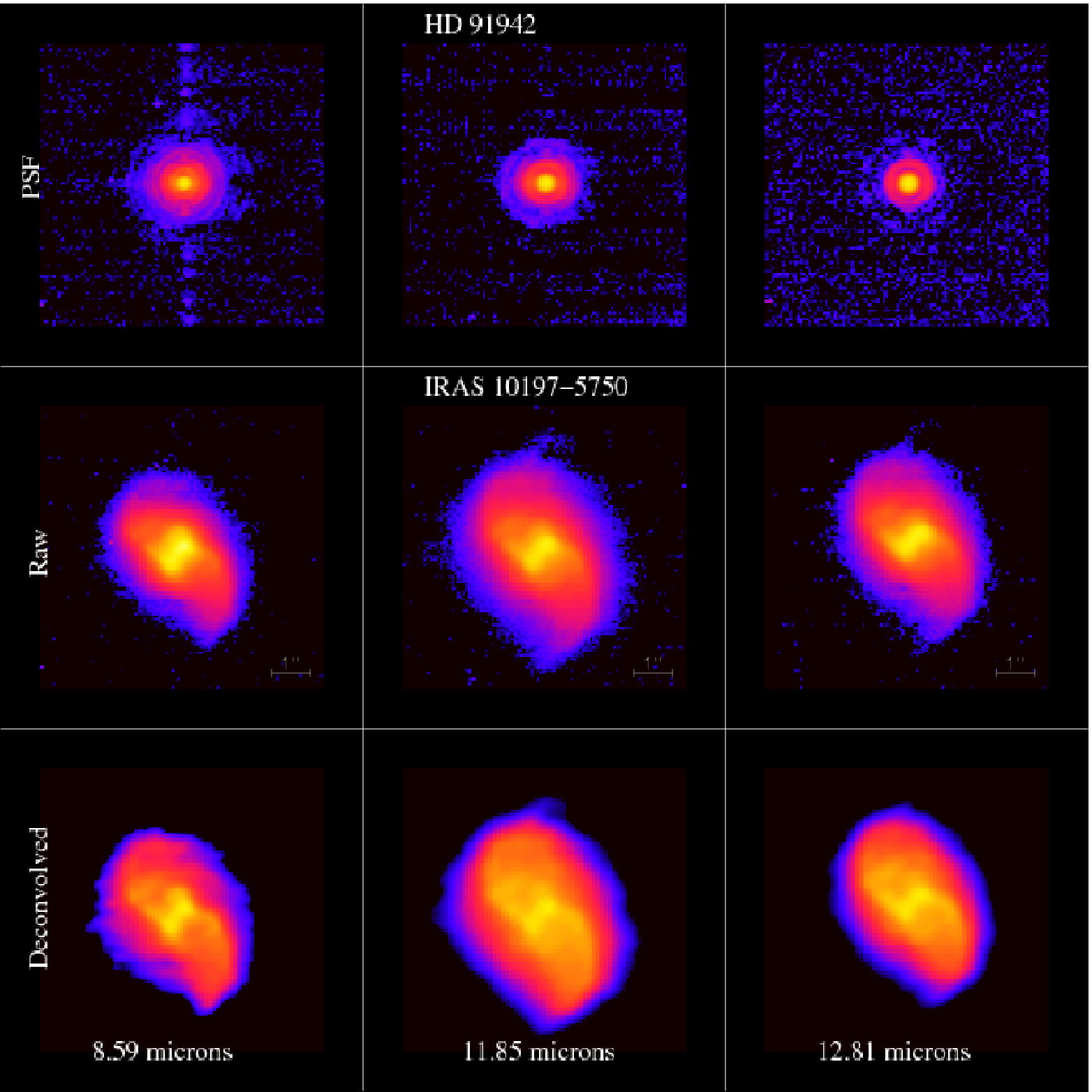}
\caption{ Visir images of IRAS IRAS 10197 (Roberts 22).}
\label{im_10197}
\end{center}
\end{figure*}

\begin{figure*}
\begin{center}
\includegraphics[width=16cm]{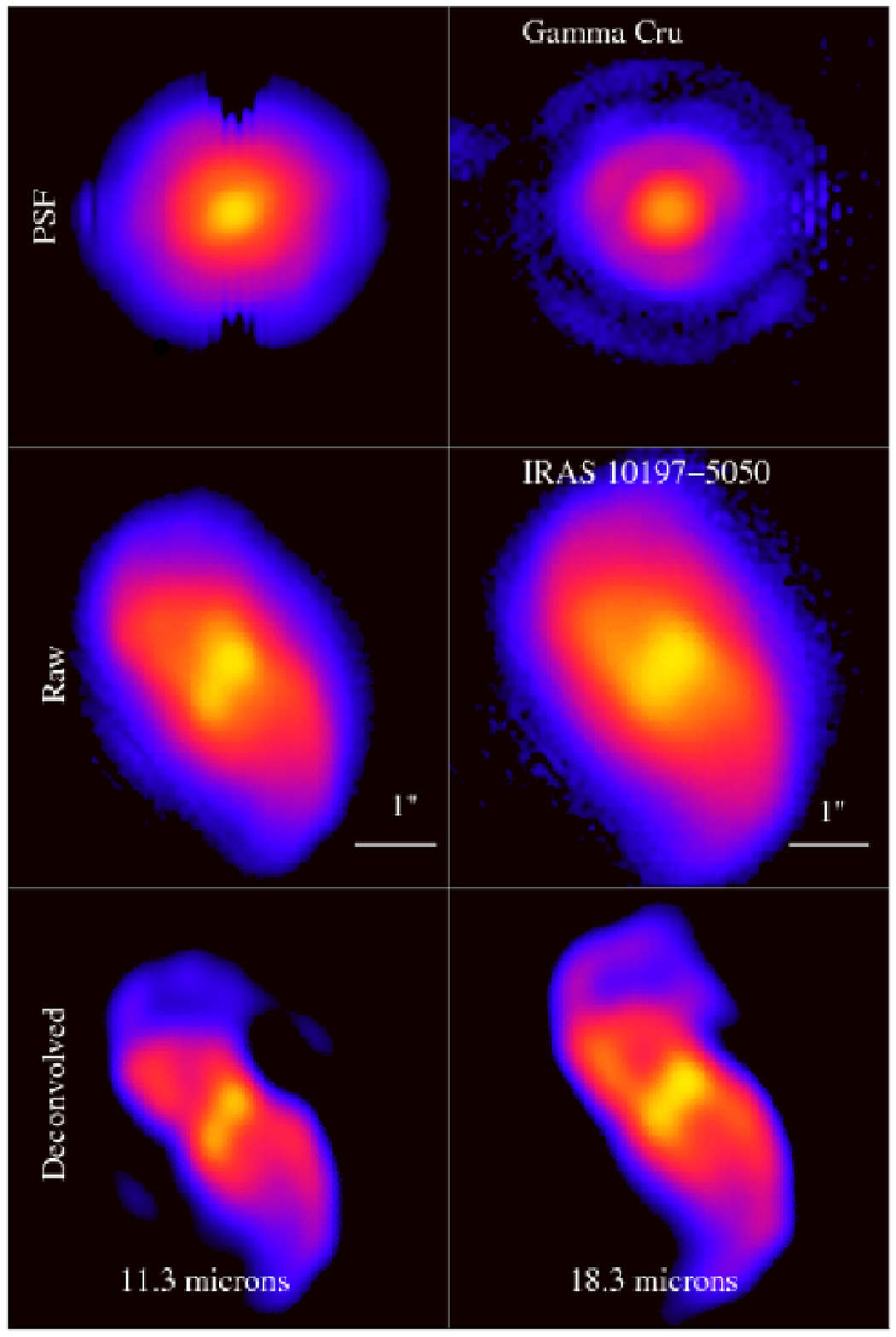}
\caption{T-Recs  images of IRAS 10197(Roberts 22).}
\label{im_10197_2}
\end{center}
\end{figure*}

Our VISIR and T-Recs images  (Fig. \ref{im_10197} and \ref{im_10197_2}) clearly show the presence of a large scale  S-shaped envelope.
 An asymmetric torus is resolved in the core of this envelope from 8 to 18 $\mu$m. The torus is
 brighter along the equator, which is more or less perpendicular to the bipolar
 nebula.
 We obtained VISIR observations of Roberts\,22 with the same filters as the present observations in November 2006.
 The orientation of the torus appears exactly the same as the one we present here. The observed S-shaped structure of
 the nebula is thus certainly not due to precession from this torus. Note that the orientation of the torus appears to be wavelength
 dependent, as indicated by the Gemini Q band image. This is a radiative transfer effect.

  \subsubsection{IRAS 15103$-$5754}   
\begin{figure*}
\begin{center}
\includegraphics[width=18cm]{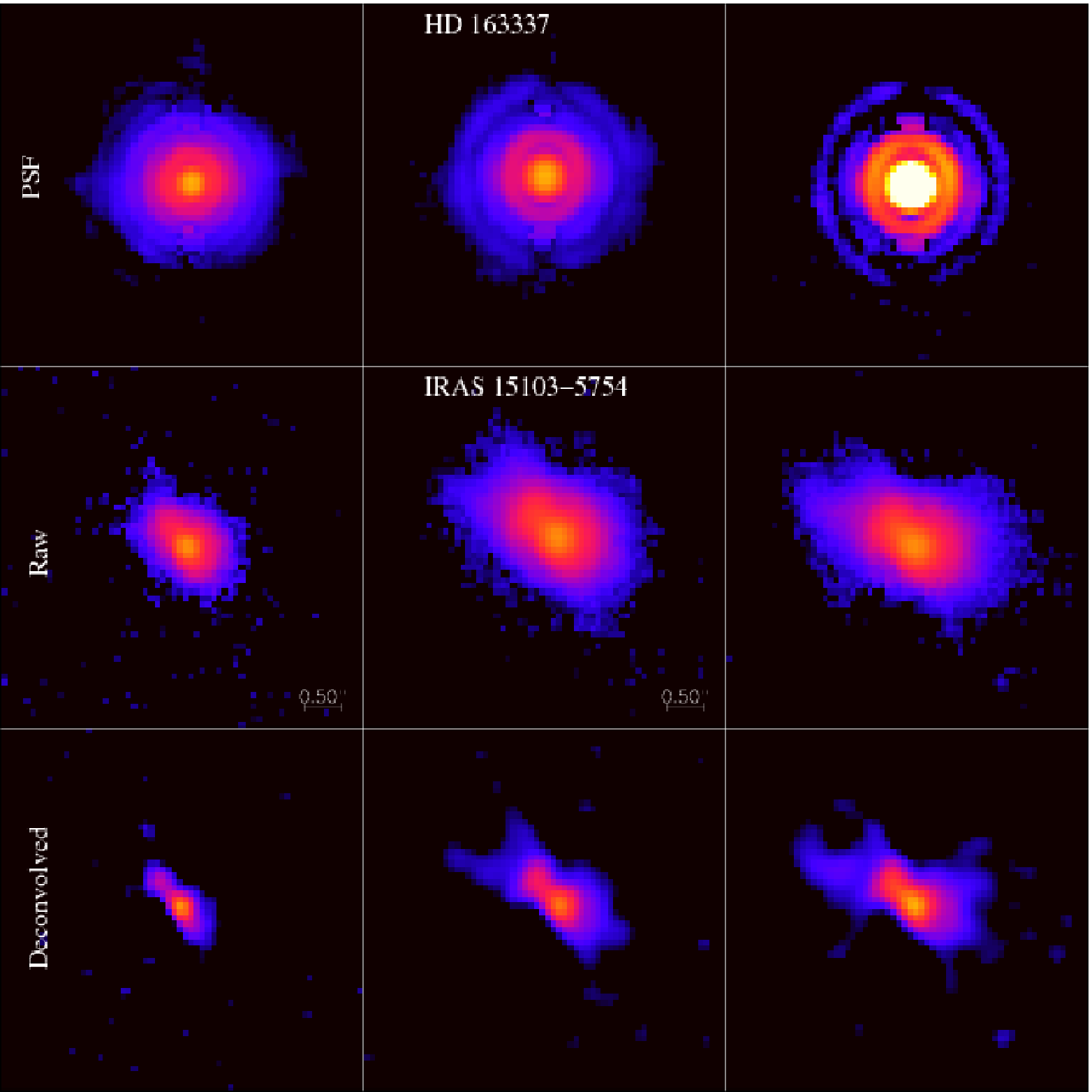}
\caption{Visir burst mode images of IRAS 15103}
\label{im_15103}
\end{center}
\end{figure*}

This is a candidate PN with  H$_2$O maser emission (Su{\'a}rez et al. 2009). The images presented Fig. \ref{im_15103} are the first ever of this object.
We clearly resolved a bipolar nebula with a narrow waist, probably due to a dense
 equatorial dusty structure. The nebula is elongated along a North-East/West
 direction
along a P.A.$\sim$35\deg. Some spurs are observed at the edges of the bipolar
 structure and are more prominent in the North-West lobe. This could be due to the presence of the
high velocity jets inferred by the water masers. 

  \subsubsection{IRAS 15445$-$5449} 
\begin{figure*}
\begin{center}
\includegraphics[width=16cm]{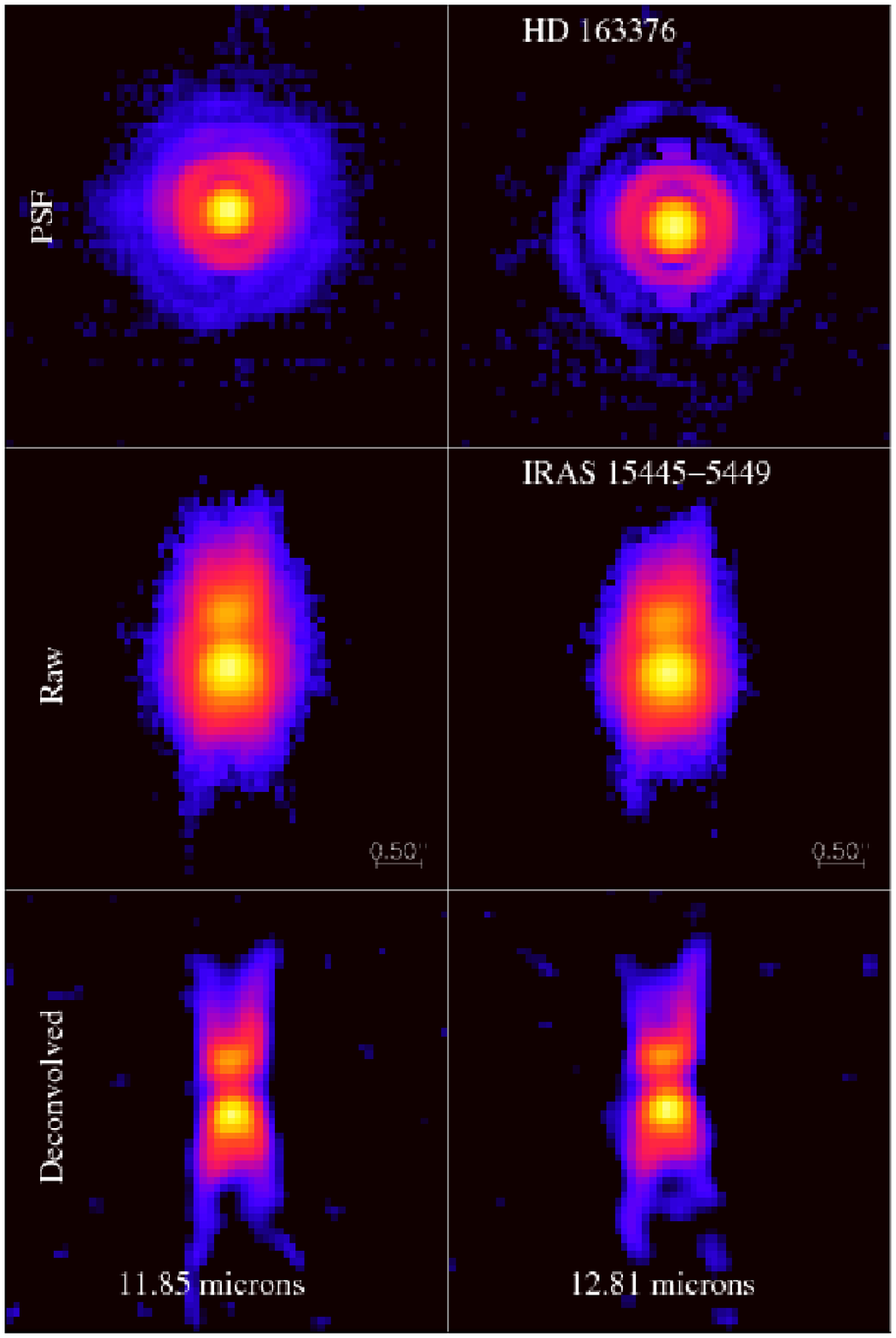}
\caption{Visir burst mode images of IRAS 15445 }
\label{im_15445}
\end{center}
\end{figure*}

  This is a water fountain nebula (Deacon et al. (2007)). We obtained the first ever images of
 this source. We resolved a compact (1.5'' $\times$ 2.3'') bipolar structure (Fig. \ref{im_15445}),
with a dark equatorial waist, indicating the presence of a dense equatorial
 structure. This equatorial lane is perpendicular to the bipolar
lobes that are elongated along a P.A.$\sim$5\deg. The edges of these lobes
 are terminated by spurs, which are probably the projection
of a biconical structure on the plane of the sky. 

 \subsubsection{IRAS 15553$-$5230}   
 This is is a poorly studied PPN. It was resolved  using optical
 HST observations (Sahai et al., 2007; Si{\'o}dmiak et al., 2008).
Their observations revealed a small (2.5 ''$\times$1.1'') bipolar
 nebula, seen nearly edge-on,  elongated along the East/West direction with a dense
equatorial waist, similar to a hourglass.
 The lobes differ in shape and size,
which is probably due to the fact that the East lobe is pointing in our
 direction. A small feature is observed in this lobe and could be a faint outflow or jet.

\begin{figure*}
\begin{center}
\includegraphics[width=16cm]{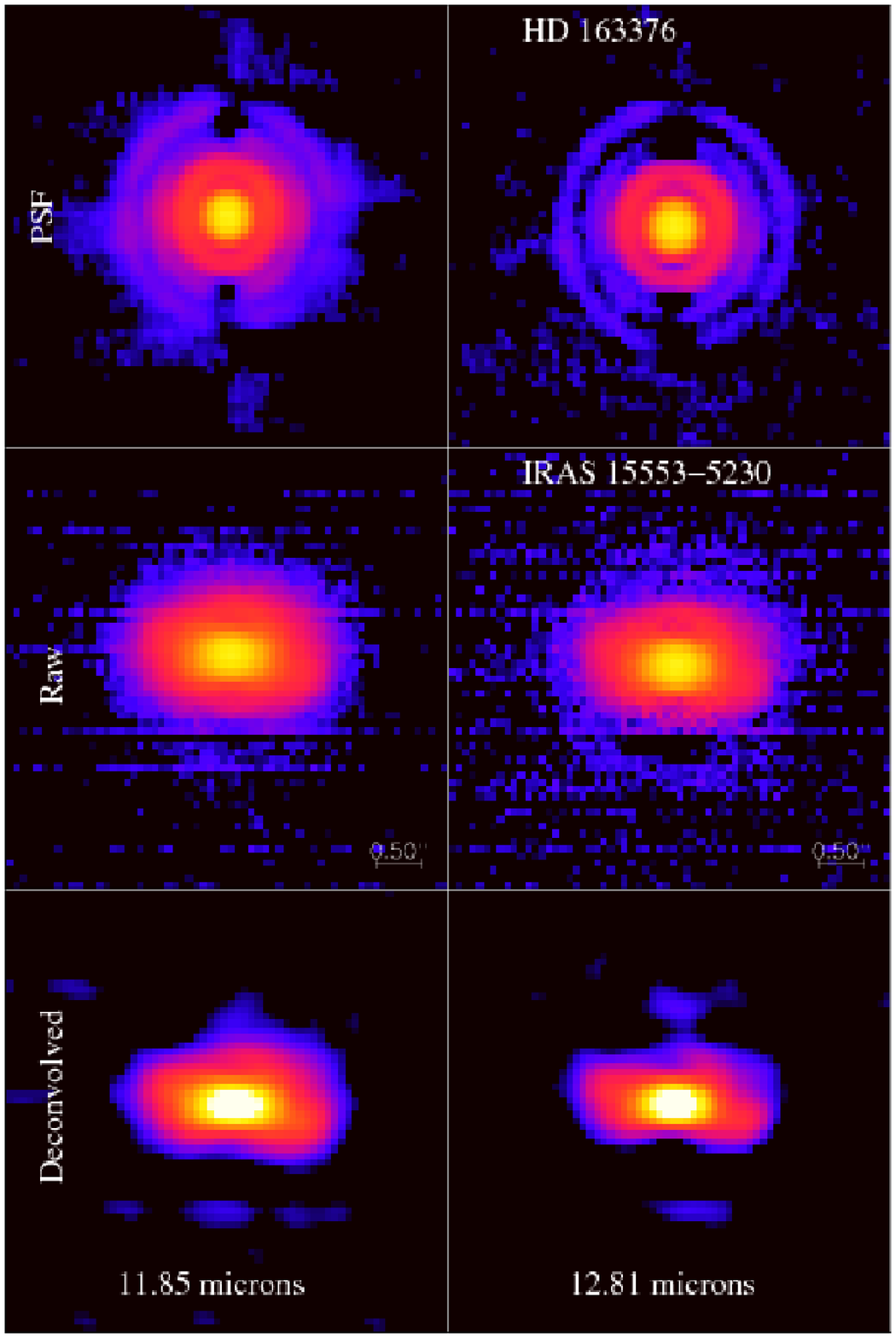}
\caption{Visir burst mode images of IRAS 15553 }
\label{im_15553}
\end{center}
\end{figure*}

Fig. \ref{im_15553} present our VISIR images of this source.
We resolved the nebula, with a point source in its core, and  it appears to be elongated along the East/West direction.

  \subsubsection{IRAS 16279$-$4757} 
\begin{figure*}
\begin{center}
\includegraphics[width=16cm]{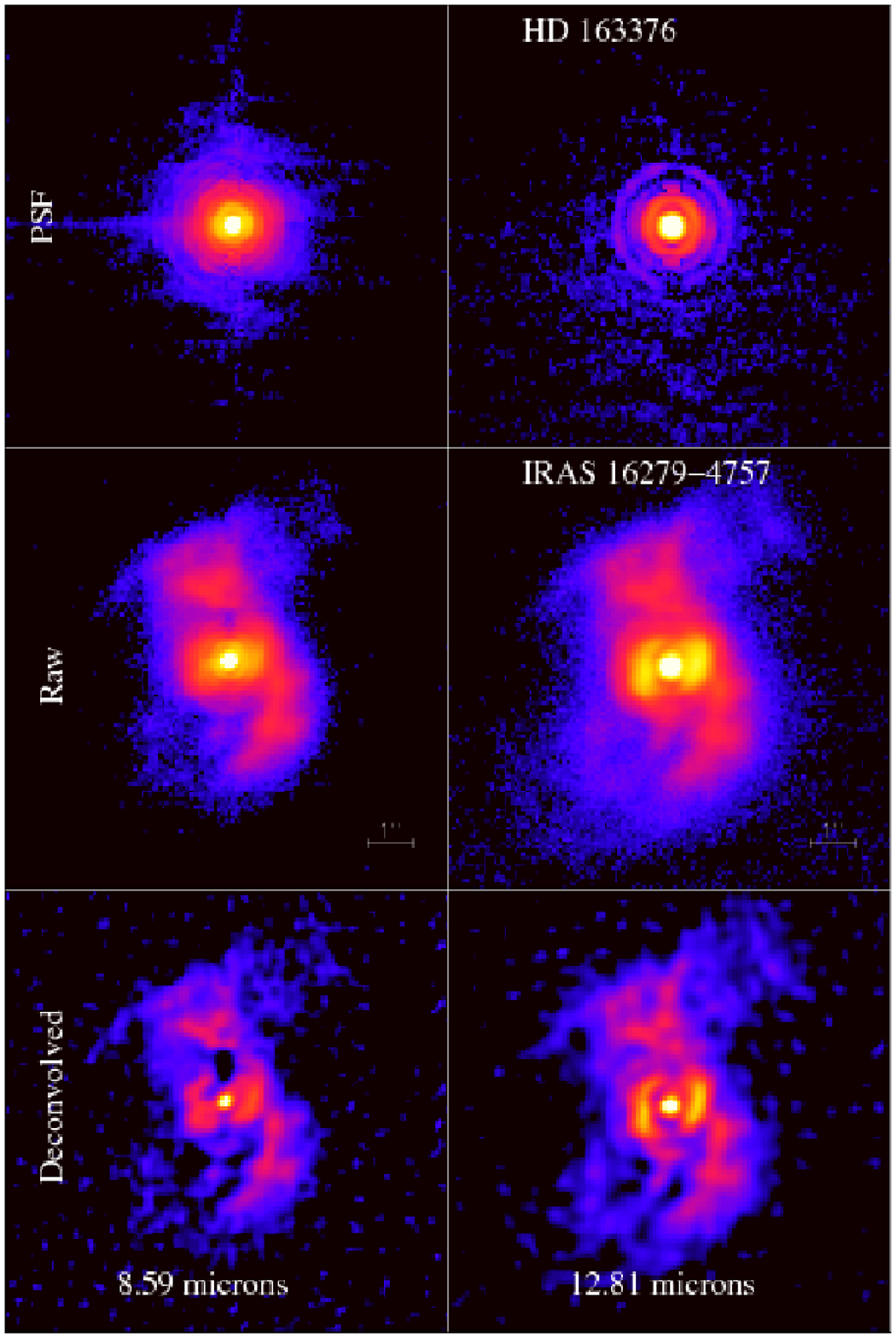}
\caption{Visir burst mode images of IRAS 16279}
\label{im_16279}
\end{center}
\end{figure*}

This object is a PPN stars with both carbon (PAHs) and oxygen-rich
 (crystalline silicates ) material in its envelope. This envelope was well studied
 by Matsuura et al. (2004)
 using TIMMI2/ESO 3.6m telescope mid-infrared images. They resolved the envelope
 and classified it as  bipolar. Our VISIR observations (Fig. \ref{im_16279}) show that the envelope 
of IRAS 16279-4757 has
 a much more complex morphology.
A large scale S-shaped structure is clearly seen in the two filters, with some
 large scale horn-like structures toward the North-West and the South-West. The
 central star is clearly
 seen in all images, surrounded by Airy rings. On the deconvolved images we can
 see that the dust is organized in filamentary structures with many holes representing underdensities
 in the dust distribution. The big hole seen on the deconvolved PAH1 just North 
of the central star is an artefact due to the detector.

  \subsubsection{IRAS 16342$-$3814} 
\begin{figure*}
\begin{center}
\includegraphics[width=16cm]{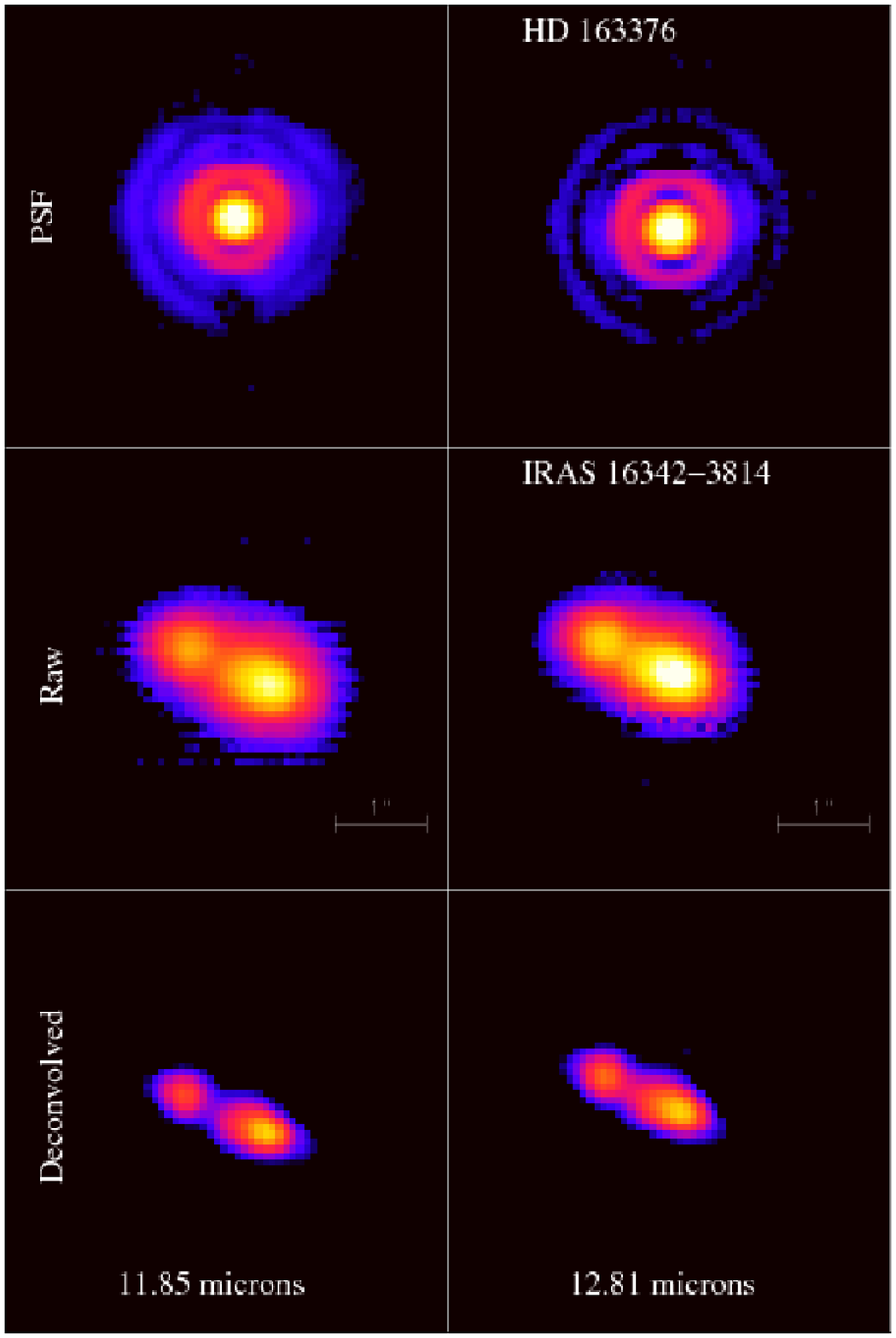}
\caption{Visir burst mode images of IRAS 16342 (The Water Fountain).}
\label{im_16342}
\end{center}
\end{figure*}

IRAS 16342$-$3814 is the prototypical object of the water fountain class (Likkel \& Morris, 1988).
It has been observed by Verhoelst et al. (2009). They resolved a bipolar nebula,
 separated by a waist dark even in the mid-infrared.
They find that this dark waist is mostly made of amorphous silicates and that its
 filling angle is fairly large and that this structure is thus not a disc.
Our two observations show the same bipolar morphologies, one of our filters being
 the same as the one used by Verhoelst et al. (2009).
But, due to the use of the burst mode on VISIR our deconvolved images (Fig. \ref{im_16342})
 are sharper
 than the ones obtained earlier.

  \subsubsection{IRAS 16594$-$4656} 
\begin{figure*}
\begin{center}
\includegraphics[width=18cm]{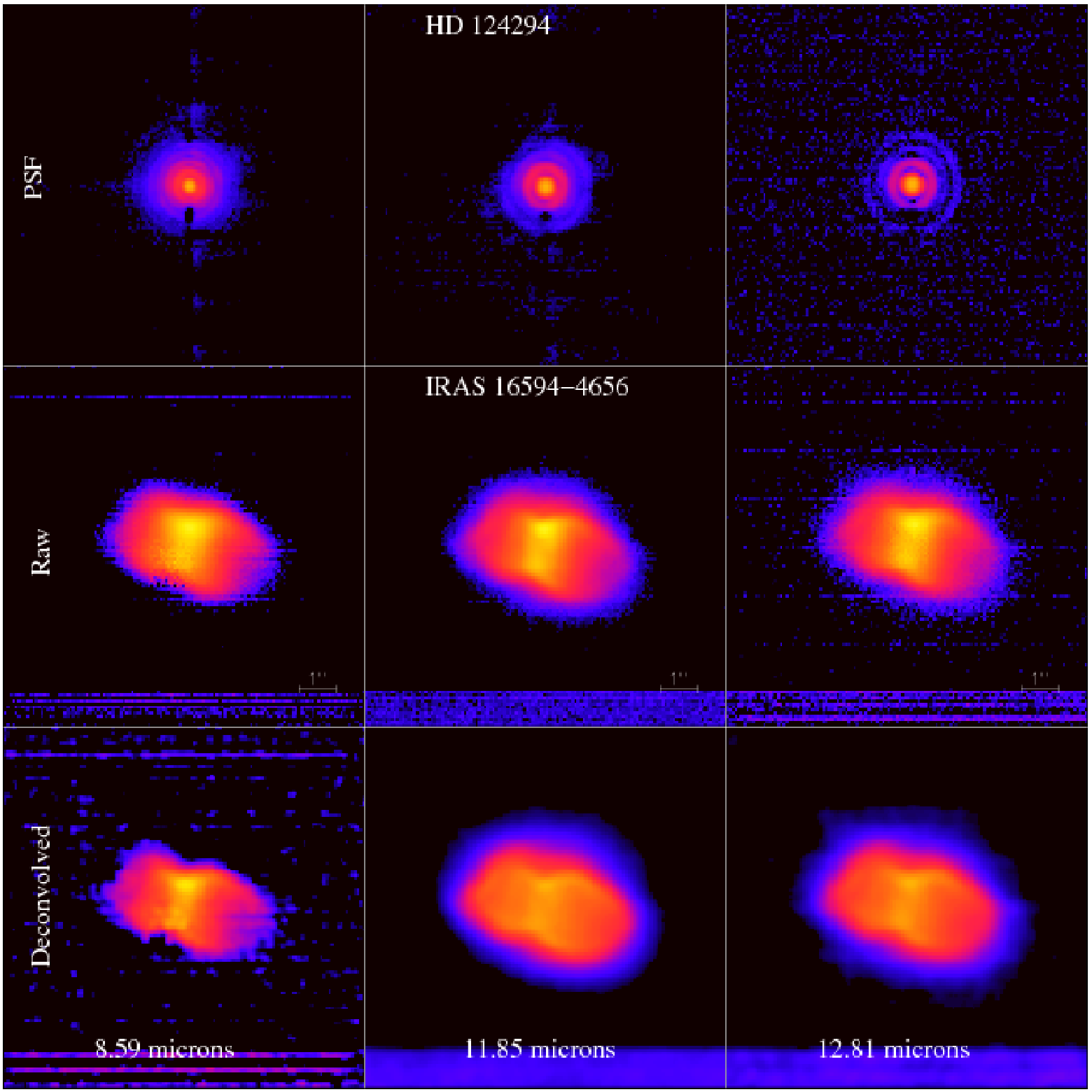}
\caption{Visir burst mode images of IRAS 16594 (The Water Lily nebula).}
\label{im_16594}
\end{center}
\end{figure*}

  The Water-Lily Nebula PPN has been observed in the mid-infrared (N and Q bands)
 with T-Recs on Gemini (Volk et al., 2006) and with TIMMI2 on the ESO 3.6m telescope (Garcia-Hernandez et al., 2006).
A bright equatorial torus, seen nearly edge-on, is clearly resolved in their
 images. No sign of the point symmetry observed in the optical
images is seen in the mid-infrared ones. 
The nebula has an overall elliptical shape and is elongated along the east-west direction (P.A.$\sim$ 80\deg).
Our mid-infrared images (Fig. \ref{im_16594}) show a similar morphology, but the larger dynamic range 
 allows us to see some more diffuse dust emission beyond the lobes.

 \subsubsection{IRAS 17106$-$3046} 
\begin{figure*}
\begin{center}
\includegraphics[width=18cm]{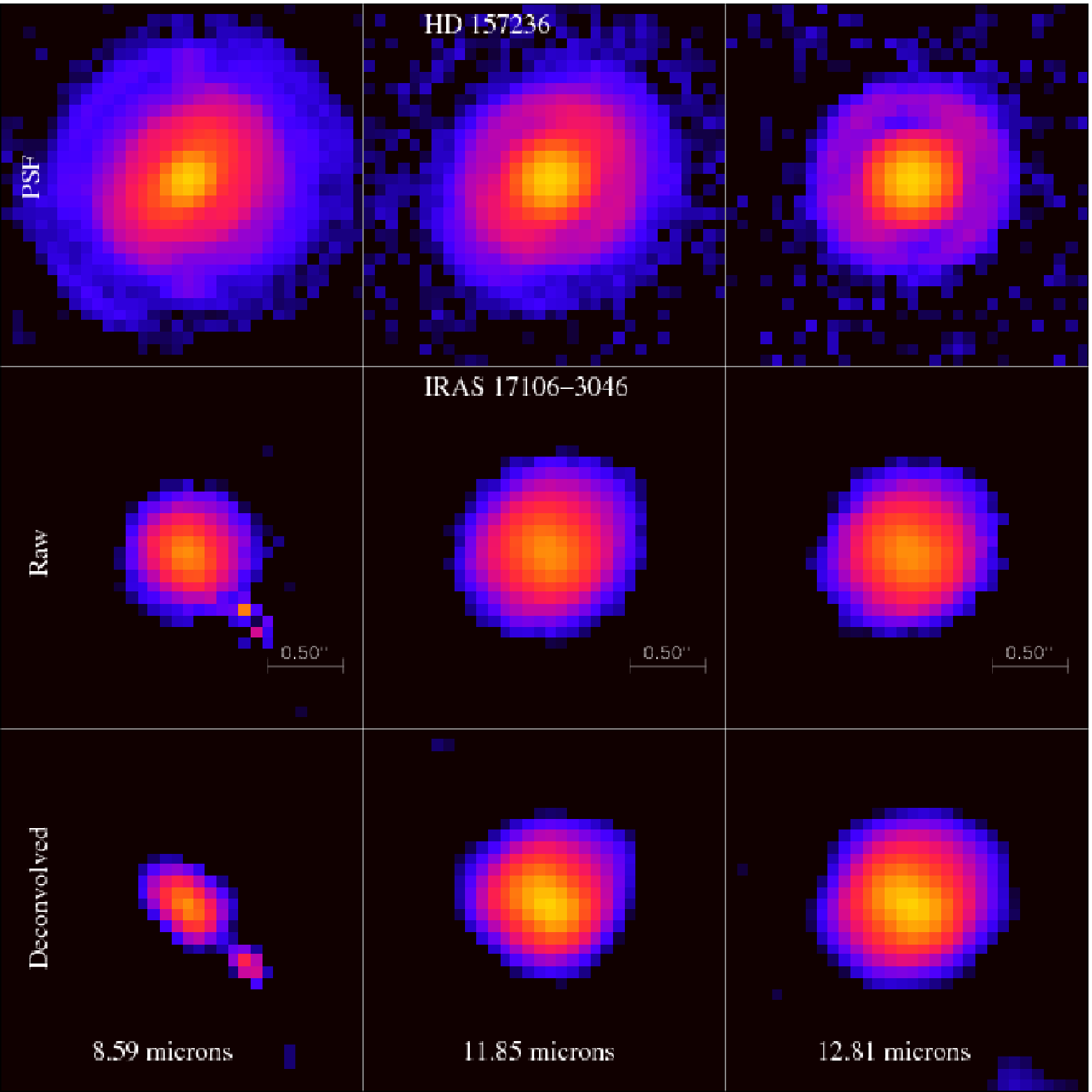}
\caption{Visir burst mode images of IRAS 17106 }
\label{im_17106}
\end{center}
\end{figure*}

Optical HST images of the PPN IRAS 17106$-$3046 (the spindle nebula) indicate the presence of a collimated
 outflow emerging from a visible disc,
 embedded in a lower
density elliptical halo (Kwok et al., 2000). The pair of lobes are collinear,
 orientated along a P.A. of 128\deg, and the disc is perpendicular to these lobes
 with a P.A.$\sim$42\deg.
Our images (Fig. \ref{im_17106}) reveal the presence of a dense central  structure along a P.A.$\sim$42\deg.

  \subsubsection{IRAS 17150$-$3224}     
The Cotton Candy nebula is a well studied bipolar PPN.
It has been observed in the optical and in the near-infrared by the HST
 (Kwok et al., 1998, Su et al., 2003).
The central star is visible in the near infrared images, but not in the
 optical where it is obscured by a dark lane.
This dark lane separates two lobes that form an extended bipolar nebula
 along a P.A. $\sim$112\deg.
Near-infrared spectroscopic observations provide evidence for the presence
 of an expanding torus in the core of IRAS 17150 (Weintraub et al., 1998).

\begin{figure*}
\begin{center}
\includegraphics[width=18cm]{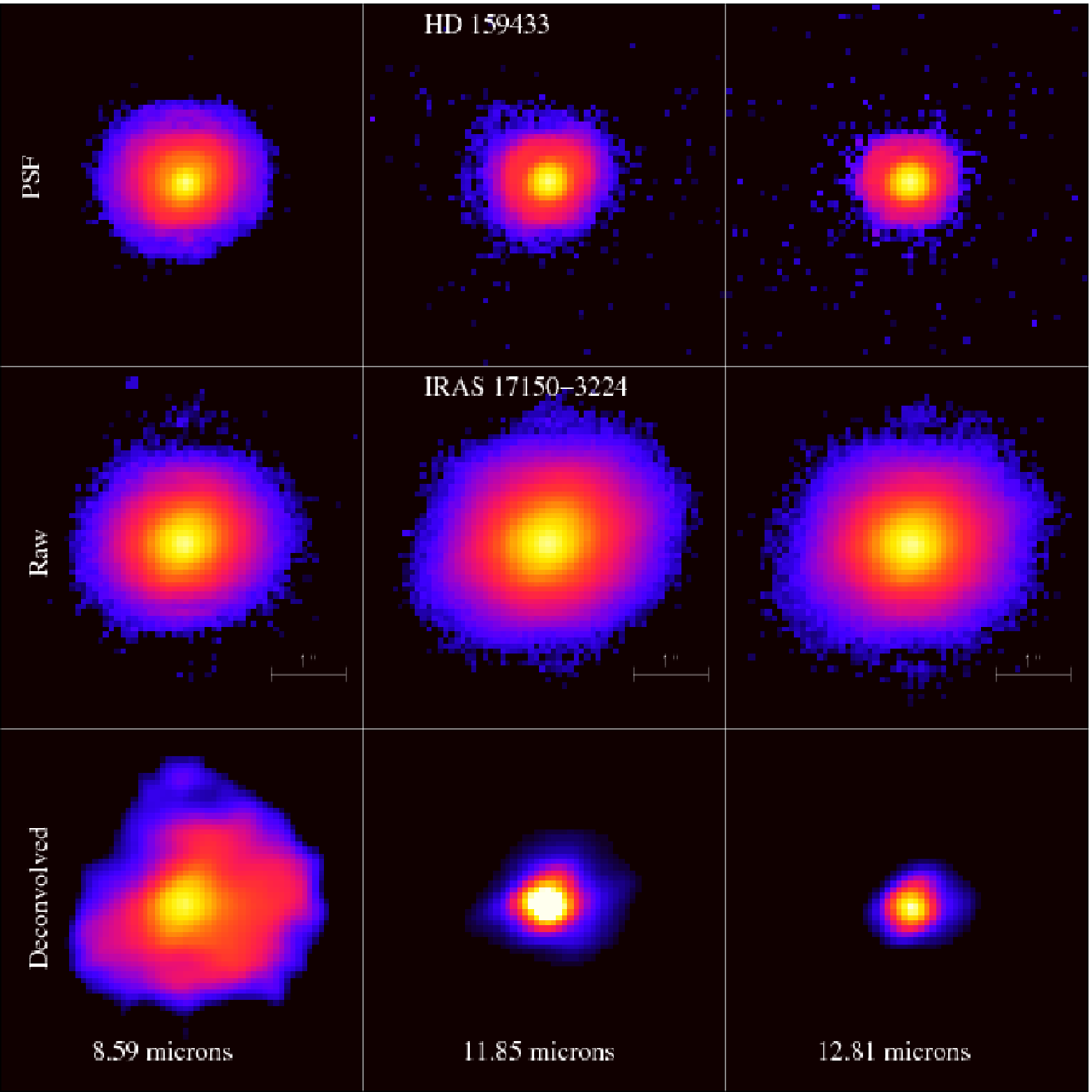}
\caption{Visir burst mode images of IRAS 17150 }
\label{im_17150}
\end{center}
\end{figure*}

\begin{figure*}
\begin{center}
\includegraphics[width=16cm]{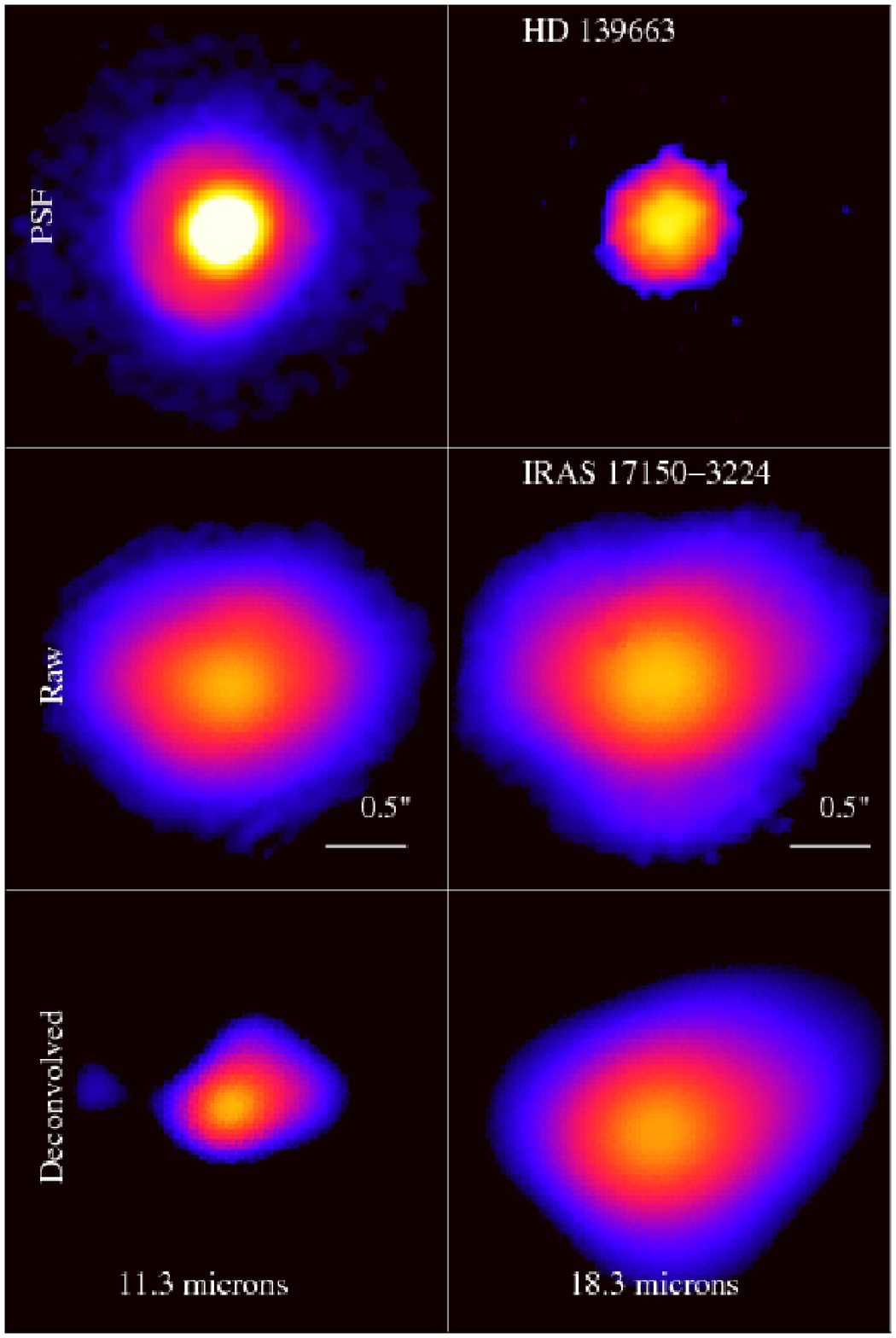}
\caption{T-Recs/Gemini images of IRAS 17150 (The Cotton Candy nebula).}
\label{im_17150_2}
\end{center}
\end{figure*}

Our VISIR and T-Recs images of IRAS 17150 are displayed Fig. \ref{im_17150} and \ref{im_17150_2}.
The central star is not seen in our images between 8 and 20 $\mu$m.
These images reveal an unresolved central peak, located at the position of the dark lane observed at shorter wavelengths.  
The nebula is elongated  along a P.A.$\sim$ 108\deg, similar to the bipolar nebula observed in the optical.
An elongation is also seen in a direction roughly perpendicular to this bipolar nebula, more clearly in the 8.59$\mu$m
deconvolved image.

  \subsubsection{IRAS 17311$-$4924}     
IRAS 17311$-$4924 is a carbon-rich PPN (Hony et al. 2002). Its ISO spectrum reveals
 the presence of PAHs, SiC and a 30-micron feature usually associated to MgS in it envelope (Hony et al. 2002).
The images we present here are the first ever obtained for this object. 
The images reveal a bipolar nebula seen edge-on with two bright lobes on both sides of the polar direction.
This is the projection of an equatorial torus, aligned with the bipolar structure.

  \subsubsection{IRAS 17441$-$2411}   
  \label{17441}
\begin{figure*}
\begin{center}
\includegraphics[width=18cm]{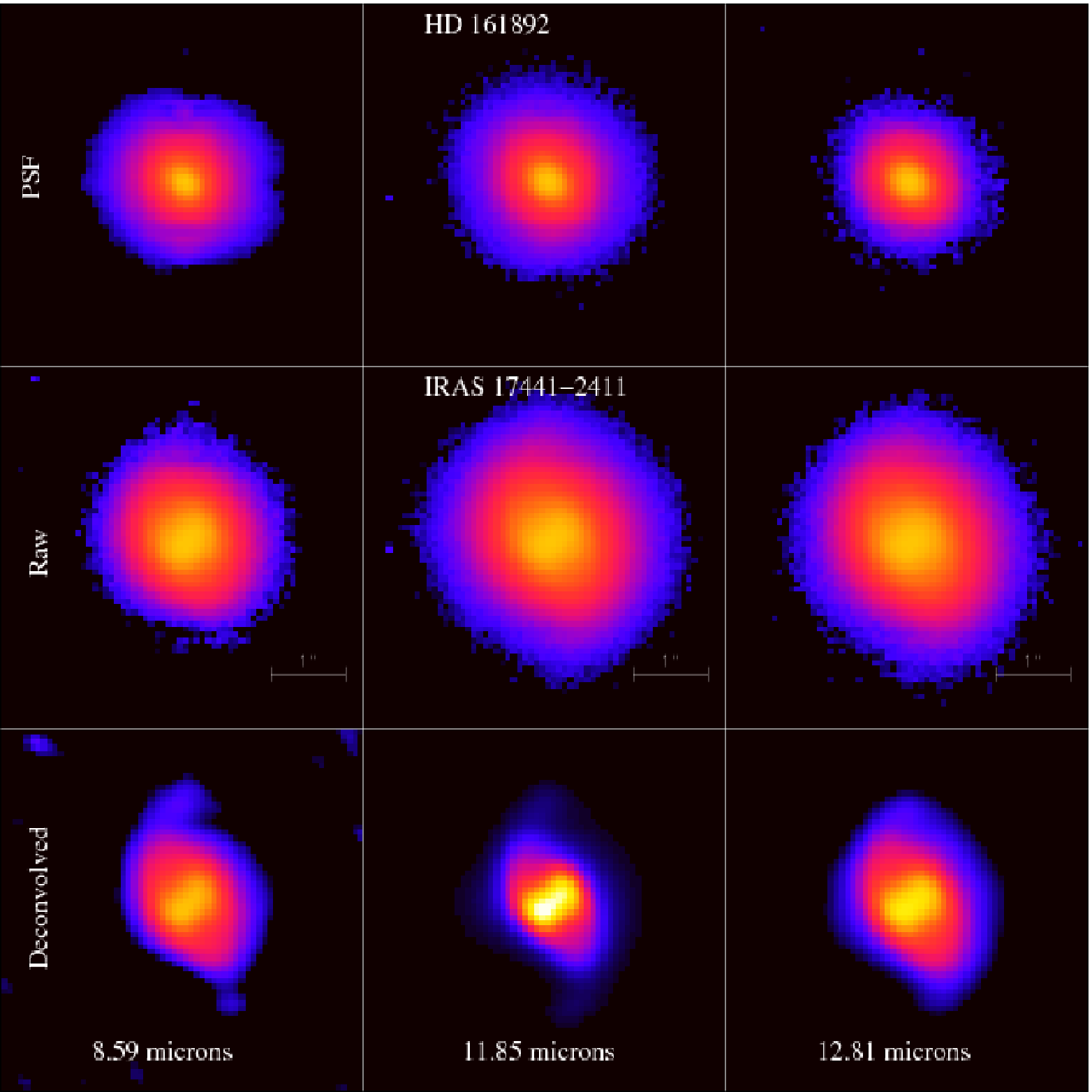}
\caption{Visir burst mode images of IRAS 17441.}
\label{im_17441}
\end{center}
\end{figure*}

The Silkworm Nebula  is a PPN with a multipolar envelope (Ueta et al., 2007). It has been
 imaged at high angular resolution with the HST (Ueta et al., 2007) in the optical
 and near-infrared,
and with Gemini in the mid-infrared (Volk et al. 2007), who resolved a torus, tilted by 23\deg with
respect to the bipolar nebula observed in the optical. They  raised the possibility that 
this torus is precessing at a rate of 1\deg.yr$^{-1}$ that could lead to the precession of the outflows.
This is supported by the possible S-shape of the nebula, which can be  seen in the near-infrared, and guessed
in the mid-infrared Gemini images. 
This S-shaped structure is clearly seen in our
 VISIR image (Fig. \ref{im_17441}), more particularly in the 11.65$\mu$m deconvolved image. Our images are also deeper and reveal
the presence of cooler dust around this structure as it in the case of Roberts\,22.

  \subsubsection{IRAS 18276$-$1431}  
   IRAS 18276$-$1431 is a star in the short transition phase between the OH/IR phase
 and the PN phase, as indicated by the the progressive disappearance of H$_2$O maser (Engels, 2002). Near
 infrared images obtained with  adaptive optics on the Keck (Sanchez-Contreras et al., 2007)
 show that the envelope of IRAS 18276 displays a clear
bipolar morphology (PA$\sim$23\deg) with two lobes separated by a dark waist.
Some strong OH masers activity  is observed in the dense equatorial region, approximately
 perpendicular to the bipolar lobes (Bains et al., 2003).
Our images show a bipolar structure  with a P.A.$\sim$9\deg, embedded in a larger
 dusty structure (more visible at 11.65$\mu$m, as this filter is more sensitive).
  \subsubsection{IRAS 18450$-$0148}
W43A is a water fountain source (Imai et al., 2002). Water maser observations of this 
source revealed the presence of a collimated and precessing jet of molecular gas.
 The H$_2$O maser
 spots are concentrated in two clusters, one receding (north-east) and one
 approaching (south-west side) (Imai et al., 2002). The two clusters are separated by $\sim$0.65 arcsec. The jets
 have a position angle of 62.7$\pm$0.5\deg. Our VISIR observations allowed
 us to resolved a compact (1.2''$\times$1.6'') bipolar dust shell in W43A, orientated along a P.A.$\sim$62\deg.
The  molecular jets have thus certainly shaped the dusty bipolar  structure we resolved.

  \subsubsection{IRAS 19016$-$2330} 
  IRAS 19016$-$2320 is a PPN (Garcia-Lario et al., 1997).
Our images of IRAS 19016$-$2330 are the first ever obtained for this object.
They show a compact structure elongated along a P.A.$\sim$25\deg,
 embedded in a structure elongated along the East/West direction.

  \subsubsection{IRAS 19374$+$2359} 
This PPN  has been observed in the optical with the HST (Ueta et al. 2000)
 and in the near-infrared with 
UKIRT, in imaging and polarimetry (Gledhill, 2005). In the optical,
 IRAS 19374$+$2359 appears bipolar, with its central star partially
 visible and limb brightened bipolar
lobes along a P.A.$\sim$6\deg. The near infrared polarimetric
 maps indicate scattering and emission from an optically thin axisymmetric
 dust shell. The nebula
appears to be brighter in the North than in the South and two brightness
 peaks can be seen  East and West of the central star.

\begin{figure*}
\begin{center}
\includegraphics[width=18cm]{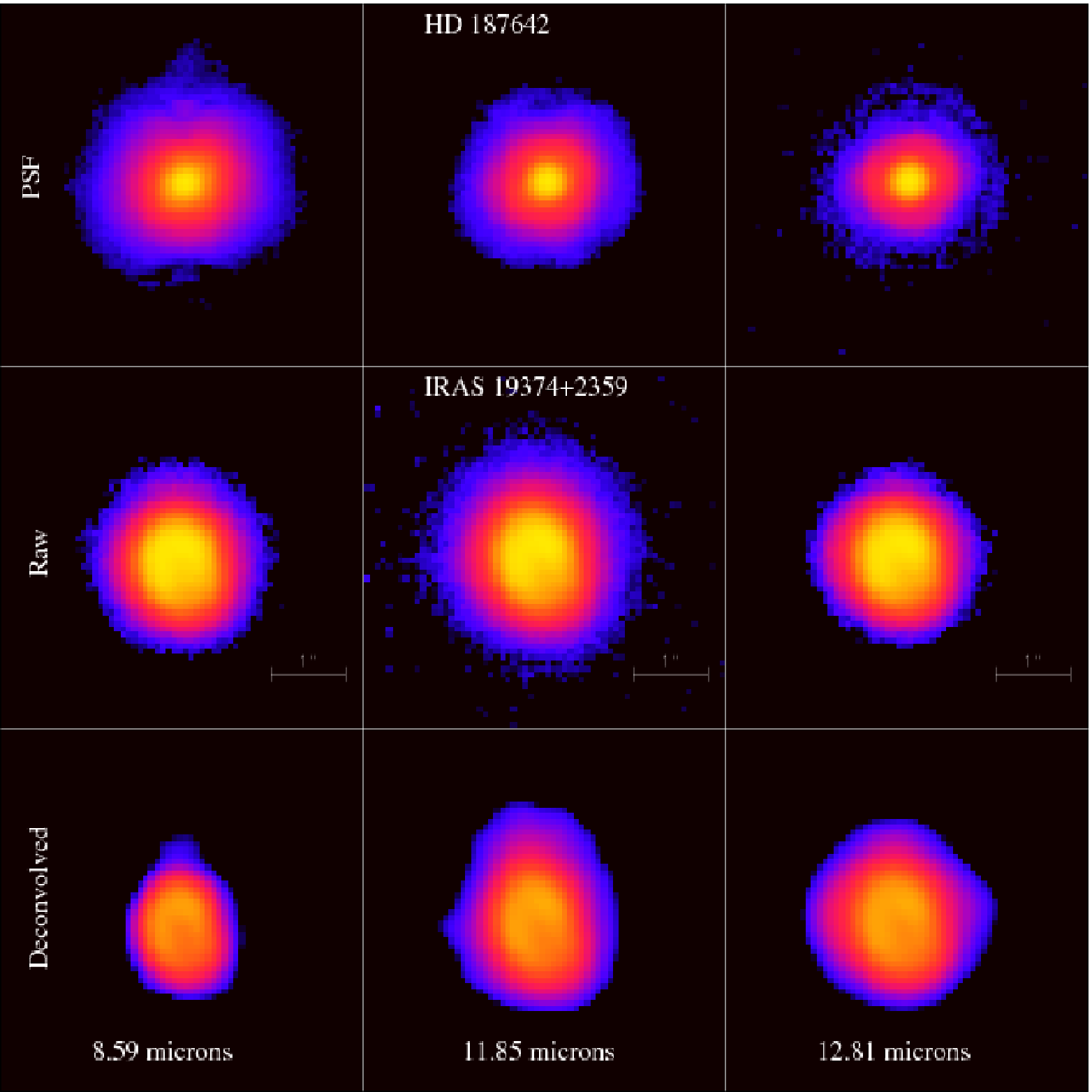}
\caption{Visir  images of IRAS 19347.}
\label{im_19347}
\end{center}
\end{figure*}

Our observations (Fig. \ref{im_19347}) show a more or less elliptical nebula along a P.A.$\sim$11\deg.
 The detached shell predicted from the polarimetric observations is clearly resolved, 
and we confirm that the nebula is brighter toward the North. An opening in the 
shell is actually observed toward the South, and this object looks very similar to IRAS 07134+1005.

  \subsubsection{IRAS 19386$+$0155}    
\begin{figure*}
\begin{center}
\includegraphics[width=18cm]{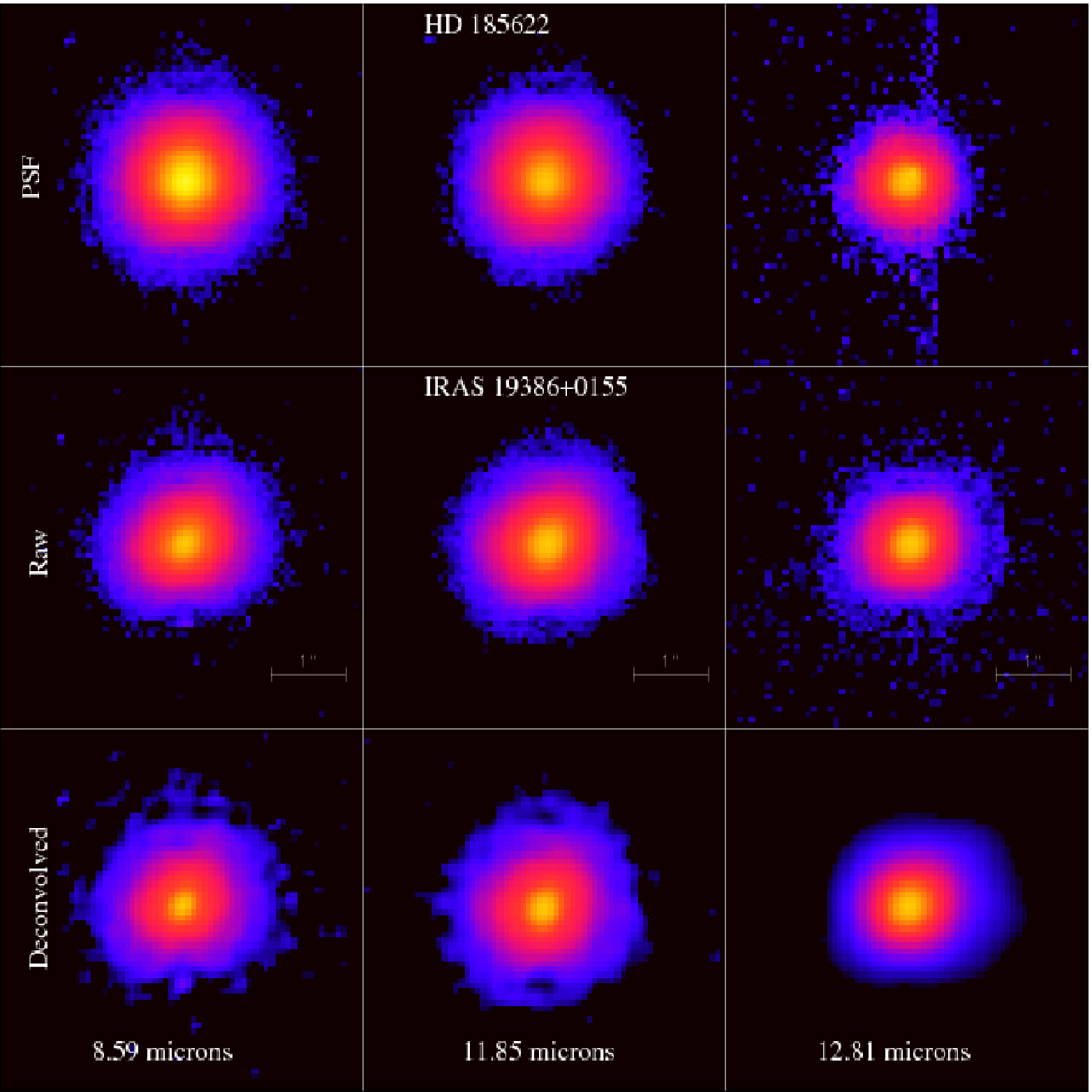}
\caption{ Visir burst mode images of IRAS 19386.}
\label{im_19386}
\end{center}
\end{figure*}

 IRAS 19386$+$0155 was observed in the near-infrared imaging 
polarimetric survey by Gledhill (2005).
Its degree of polarization is low but there is some evidence 
for scattering. The core of the polarized intensity image is 
elongated along an approximately
North/South direction, while the outer region has a 
North-East/South-West orientation. They suggest that 
this object has a bipolar morphology.
Our images (Fig. \ref{im_19386}) are the first to resolve this object in the
 mid-infrared. We can see that the nebula is elongated in
 the South-East/North-West direction, perpendicular
to the extension observed in the near-infrared.

  \subsubsection{IRAS 19454$+$2920} 
\begin{figure*}
\begin{center}
\includegraphics[width=18cm]{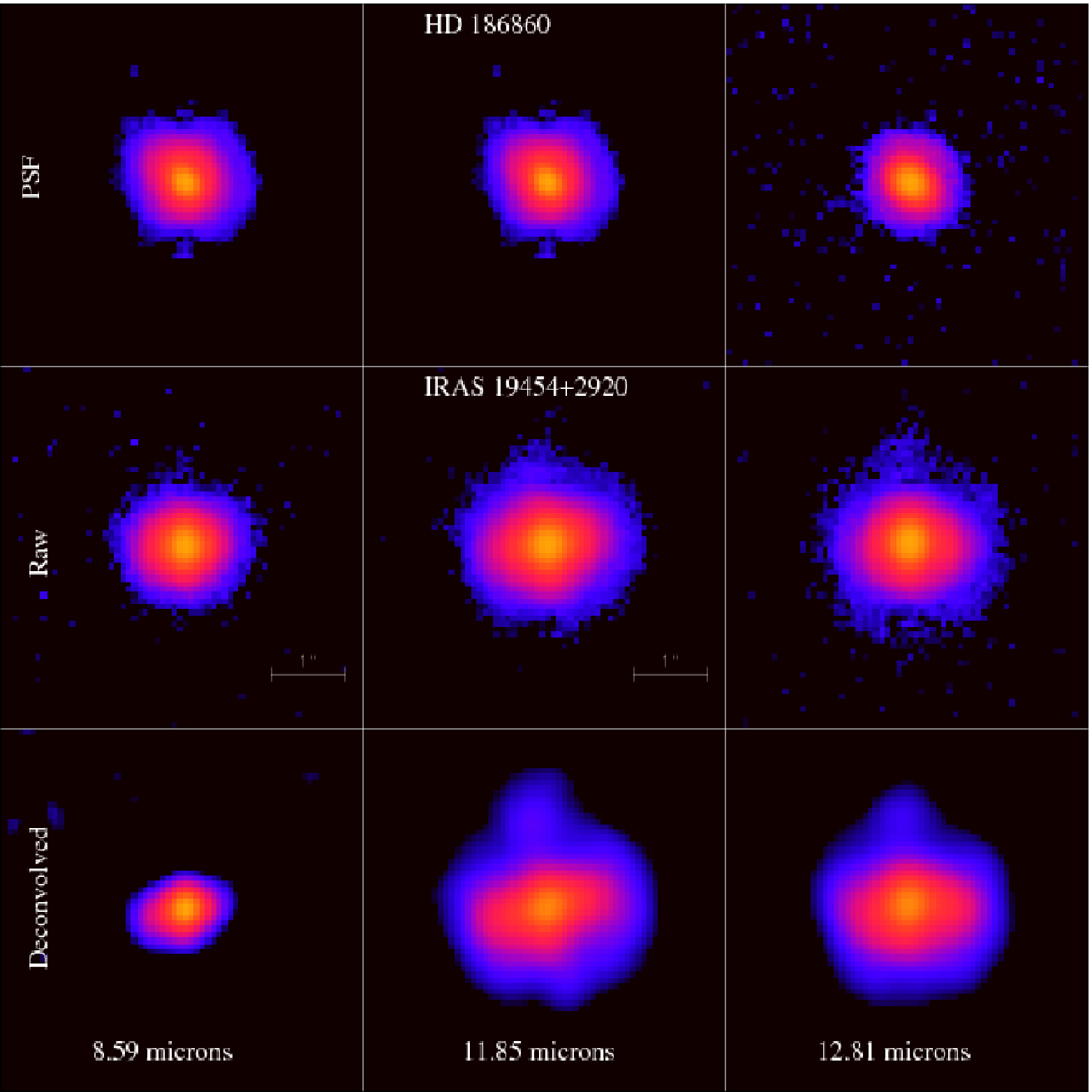}
\caption{Visir burst mode images of IRAS 10197 }
\label{im_19454}
\end{center}
\end{figure*}

This object is not very well studied.
The only study of its morphology was done in the near-infrared imaging polarimetric
survey performed by Gledhill et al. (2001). Their observations show that this object is unresolved and unpolarized.
Our VISIR images (Fig. \ref{im_19454}) reveal the presence of a bipolar nebula, elongated along a direction roughly East/West,
with a central structure elongated in a direction perpendicular to the bipolar nebula.

  \subsubsection{IRAS 19500$-$1709}     
IRAS 19500$-$1709 is a carbon-rich PPN, with 21 microns and
 30 microns features in its mid-infrared spectrum (Justtanont et al., 1996).
It has been imaged in the mid-infrared with OSCIR/Gemini
 North (Clube \& Gledhill, 2004). 
An extended circumstellar envelope is detected, and there
 are some indications that this envelope is elongated
along the North-East/South-West direction. Using radiative 
transfer modelling, Clube \& Gledhill (2004) estimated
the inner and outer radius of a detached dust shell around 
this object. They obtained an inner radius of 0.4 arcsec, not resolved with their
OSCIR observations.

\begin{figure*}
\begin{center}
\includegraphics[width=18cm]{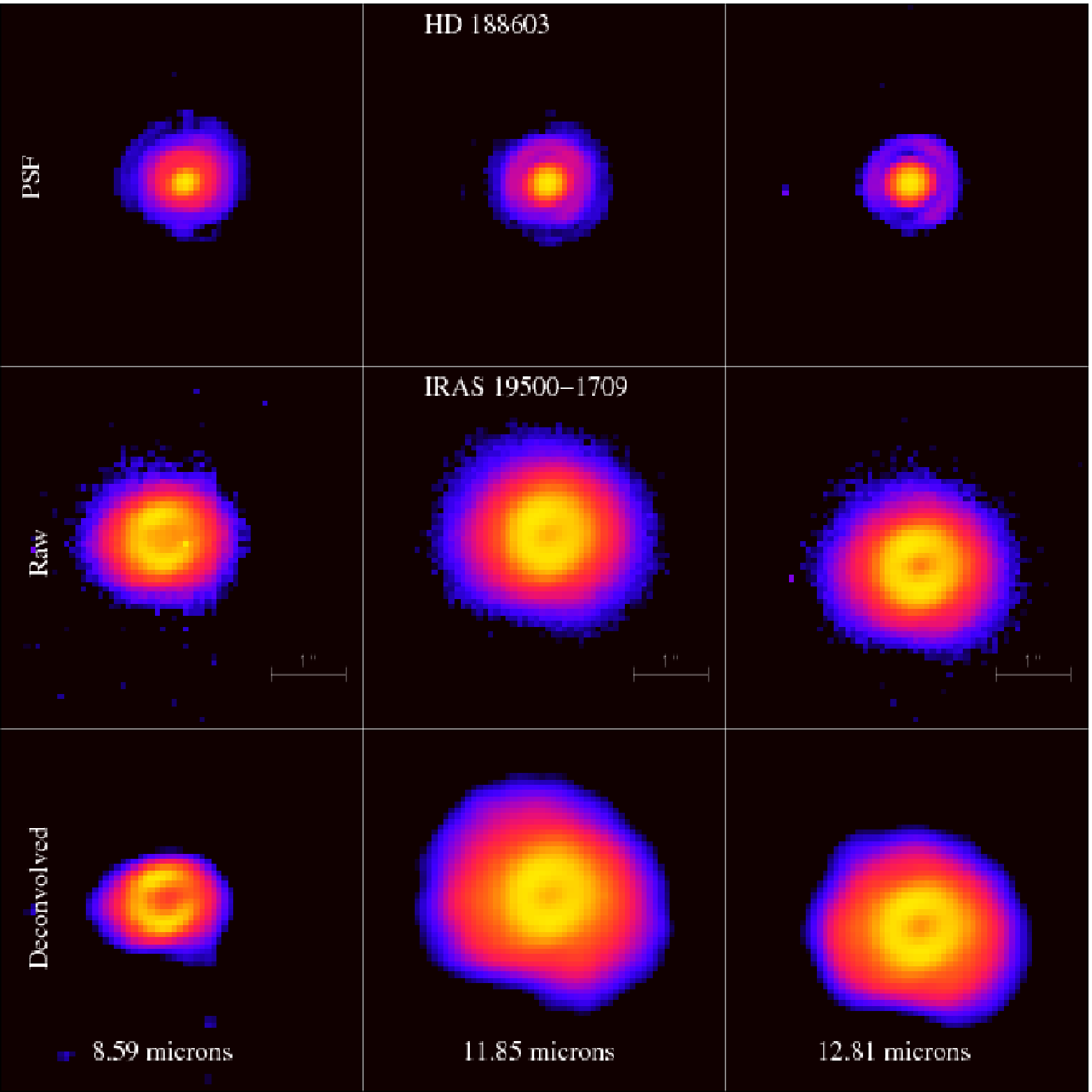}
\caption{Visir images of IRAS 19500.}
\label{im_19500}
\end{center}
\end{figure*}

Our observations (Fig. \ref{im_19500}) show that the envelope around 19500 is
 extended and elongated along two preferential direction,
 North-East/South-West as mentioned by
Clube \& Gledhill (2004) but also East/West. A detached
 shell is clearly resolved at all three wavelengths, with an inner radius of $\sim$ 0.4 arcsec.

  \subsubsection{IRAS 20043$+$2653} 
IRAS 20043$+$2653 has  been resolved in  the near-infrared
 (K band)  by Gledhill (2005) using polarimetric measurements.
The polarization map suggests that the object is bipolar. 
This is supported by the polarized flux image, which show 
an extension along a P.A.$\sim$120\deg , while the polarization
vectors are almost perpendicular to that direction. Our
 VISIR observations show that this object is slightly resolved and show an extension along a P.A.$\sim$114\deg in
all three filters, in a  similar direction as the polarisation flux image.

\subsection{Planetary Nebulae}

\subsubsection{IRAS 14562$-$5406}

Hen 2$-$113 (He 1044) is a very young PN with a [WC] central star, and displays  dual dust chemistry , with the presence
of PAHs (carbon-rich) and crystalline silicates in it envelope (Waters et al., 1998). HST observations reveal a complex morphology
for this very young PN (Sahai et al. (2000)), with the presence of a spherical halo, remnant of the AGB mass-loss.
A bipolar structure is observed along a P.A. $\sim$136\deg, with a globally elliptical morphology
with other faint lobes, the brightest along a P.A. $\sim$55\deg. A bright core is observed in this elliptical structure,
showing the presence of two rings separated by a dark lane. This object was observed by Lagadec et al. (2006) in the near and mid-infrared. 
Their observation are limited to the bright core and  show that rings are  the projection of a diabolo-like
dusty structure (Lagadec et al. 2006). The central star is visible at short wavelength, up to 5 $\mu$m, but is not detected in
the N band. 

\begin{figure*}
\begin{center}
\includegraphics[width=16cm]{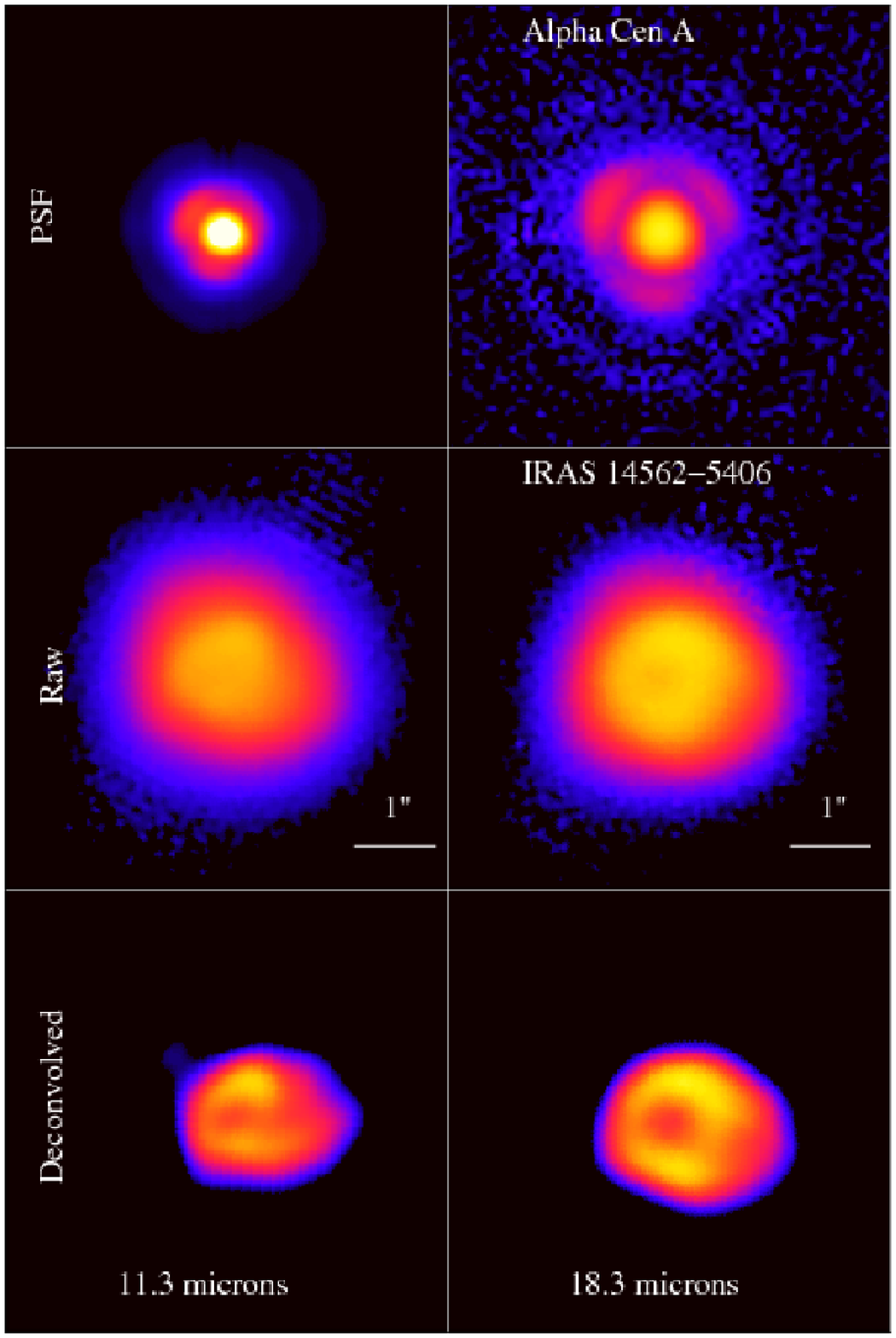}
\caption{T-Recs images of IRAS 14562 (Hen 2-113).}
\label{im_14562}
\end{center}
\end{figure*}

Our T-Recs mid-infrared  images (Fig. \ref{im_14562}) reveal a global structure similar to the one reported by Lagadec et al. (2006).
But  the better resolution of the present observations allow us to clearly resolve the brightest ring in N and Q band.
 
  \subsubsection{IRAS 16333$-$4807}   
  IRAS 16333$-$4807 is a H$_2$O PN (Suarez et al., 2009). We present here the first resolved image
 of its envelope. We resolved a compact nebula.
The envelope appears point symmetric, with an unresolved central core defining
 a very narrow equatorial waist. The holes seen
 North and South of the central star (more clearly seen in the deconvolved image)
 could be artefacts due to the detector. The point-symmetry of this source
is an indication that it has been shaped by precessing jets.

\subsubsection{IRAS 17047-5650}
CPD-56\deg8032 is a young PN with a [WC] central star and is spectroscopically 
the twin of Hen 2-113. An edge-on  disc in its core has been discovered by De Marco et al.
 (2002) through HST/STIS spectroscopy. A study combining HST imaging and
 MIDI/VLTI interferometry of CPD-56 revealed the complexity of this object (Chesneau et al., 2006).
The HST image reveals the presence of several lobes, in a shape similar
 to the Starfish nebulae described by Sahai \& Trauger (1998).
 The farthest structure is located 7 arcsec away from the central star.
A well-defined lobe is observed along a P.A. $\sim$ 53\deg.
 The mid-infrared environment of CPD-56\deg8032 is barely resolved with  single dish
 VLT images obtained during the MIDI acquisition at 8.7$\mu$m.
 It is elongated along a P.A.
of $\sim$104\deg and a dimension of $\sim$0.3 arcsec $\times$0.4 arcsec.
 This orientation corresponds to none of the observed lobes but is similar
 to that of a bow shock observed next to the central star in the HST image.
The interferometric measurements reveal the presence of a dusty disc,
 with an orientation 
along a P.A. $\sim$28\deg, corresponding to none of the structures observed.

\begin{figure*}
\begin{center}
\includegraphics[width=16cm]{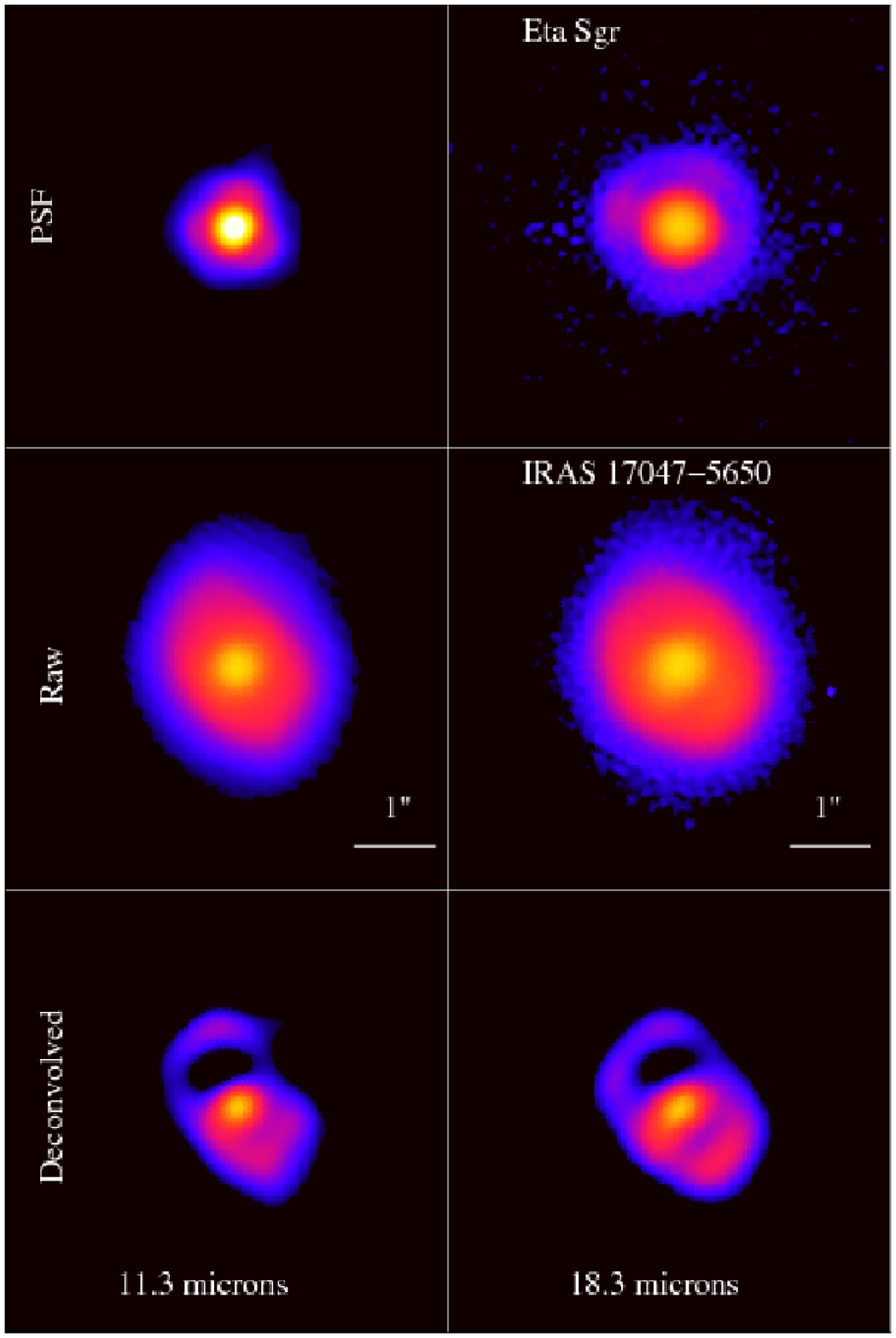}
\caption{T-Recs/Gemini  images of IRAS 17047 (CPD 56\deg8032 }
\label{im_17047}
\end{center}
\end{figure*}

Our T-Recs images (Fig. \ref{im_17047}) reveal a bipolar nebula  with a bright central structure   elongated
along a P.A. $\sim$ 123\deg and an overall bipolar structure almost perpendicular to this.
This central structure seems to indicate the presence of a dusty disc in this direction, but puzzlingly, its orientation is almost
perpendicular to the one inferred from the MIDI observations. Two bright structure are observed toward the North and South of this 
equatorial structures,a and are separated by density gaps. The shape of the Northern structure has the same orientation as the 
bow shock observed by the HST.

  \subsubsection{IRAS 17347$-$3139} 
IRAS 17347$-$3139 is a young PN with water masers. Its circumstellar envelope
 has been resolved by Sahai et al. (2007), using HST optical and near-infrared
 observations.
These reveal the presence of bipolar collimated outflows separated by a dark
 waist. The lobes are asymmetric in shape and size, the north-west lobe being 
larger and the morphologies
of these lobes seem to indicate that the jets shaping the nebula are precessing.
A faint spherical halo can be observed at shorter wavelengths, up to a radius of 2 arcsec.  
VLA observations (Tafoya et al., 2009) reveals that the ionized shell consists
 of 2 structures. An extended (1.5 arcsec) bipolar structure with P.A.$\sim$-30\deg, similar
 to the one observed by the HST. The other structure is a central compact
 structure ($\sim$0.25 arcsec) elongated in a direction perpendicular to the bipolar structure,
similar to the dark lane observed in the HST image.
IRAS 17347 appears slightly extended in the mid-infrared images
 published by Meixner et al. (1999)
Our VISIR observations clearly resolved the bipolar structure
 with an unresolved  central core. The deconvolved images
 reveal that the nebula is not bipolar, but multipolar.
This indicates that the nebula has been shaped by precessing jets.

\subsubsection{IRAS 19327$+$3024} 
BD+30\deg3639, is one of the best studied dual dust chemistry PN.
 It is a dense, young PN with a [WC9] central star. HST imaging has
 shown that this nebula has a remarkable elliptical ``squared off''
 morphology (Harrington et al. 1997). Kinematic studies of this object
 indicate that its nebula is seen nearly pole-on, with the rotational
 symmetry axis at about P.A. 30\deg-60\deg in the plane of sky and an
 inclination of about 20\deg (Bryce \& Mellema 1999).

\begin{figure*}
\begin{center}
\includegraphics[width=26cm,angle=90]{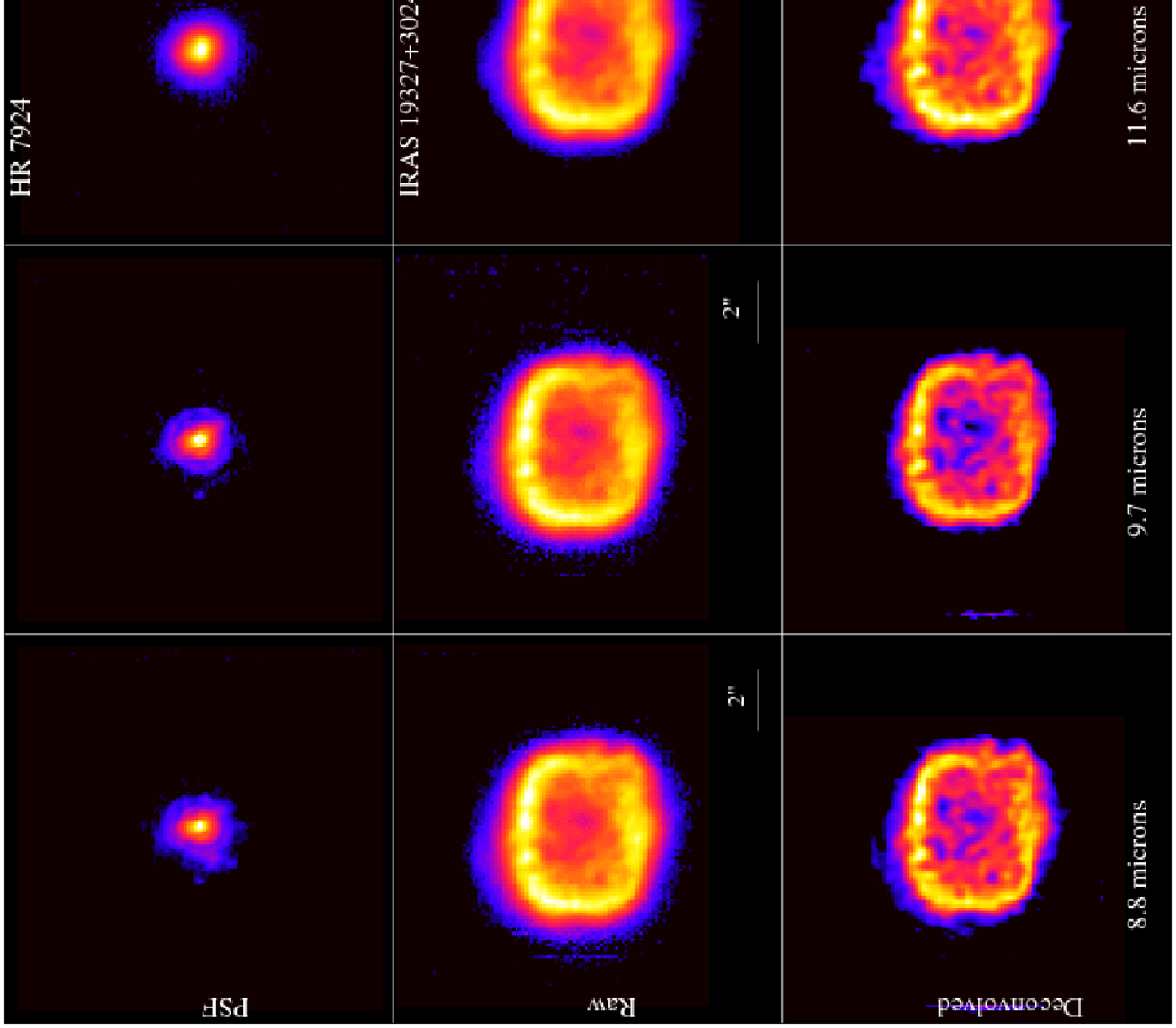}
\caption{T-Recs/Gemini South images of IRAS 19327 (BD+30\deg3639}
\label{im_19327}
\end{center}
\end{figure*}

Our Michelle images of BD+30\deg3639 at 8.8, 9.7, 11.6 and 18.5$\mu$m are 
displayed Fig.\ref{im_19327}. The morphology of these
 images is remarkably similar to the optical ones obtained by the HST, 
displaying an elliptical-rectangular shape. The images are very similar 
at the 4 observed wavelengths with dimension 6.8$\times$8.0''. The 
elliptical ring appears to be very clumpy, what is best seen on the 
deconvolved images. Note that the clumpy structures are observed at 
similar location in different images and are thus not deconvolution 
artefacts. The north east of the ring appears to be the brightest, 
while the south-east is weaker, displaying a kind of hole. This ring 
structure is surrounded by a faint elliptical halo. 
Some filamentary structure are also observed inside the ring, with a 
decreasing intensity toward the location of the central star that is 
not detected in any of our MIR images.

  \subsubsection{IRAS 21282$+$5050}   
\begin{figure*}
\begin{center}
\includegraphics[width=23cm,angle=90]{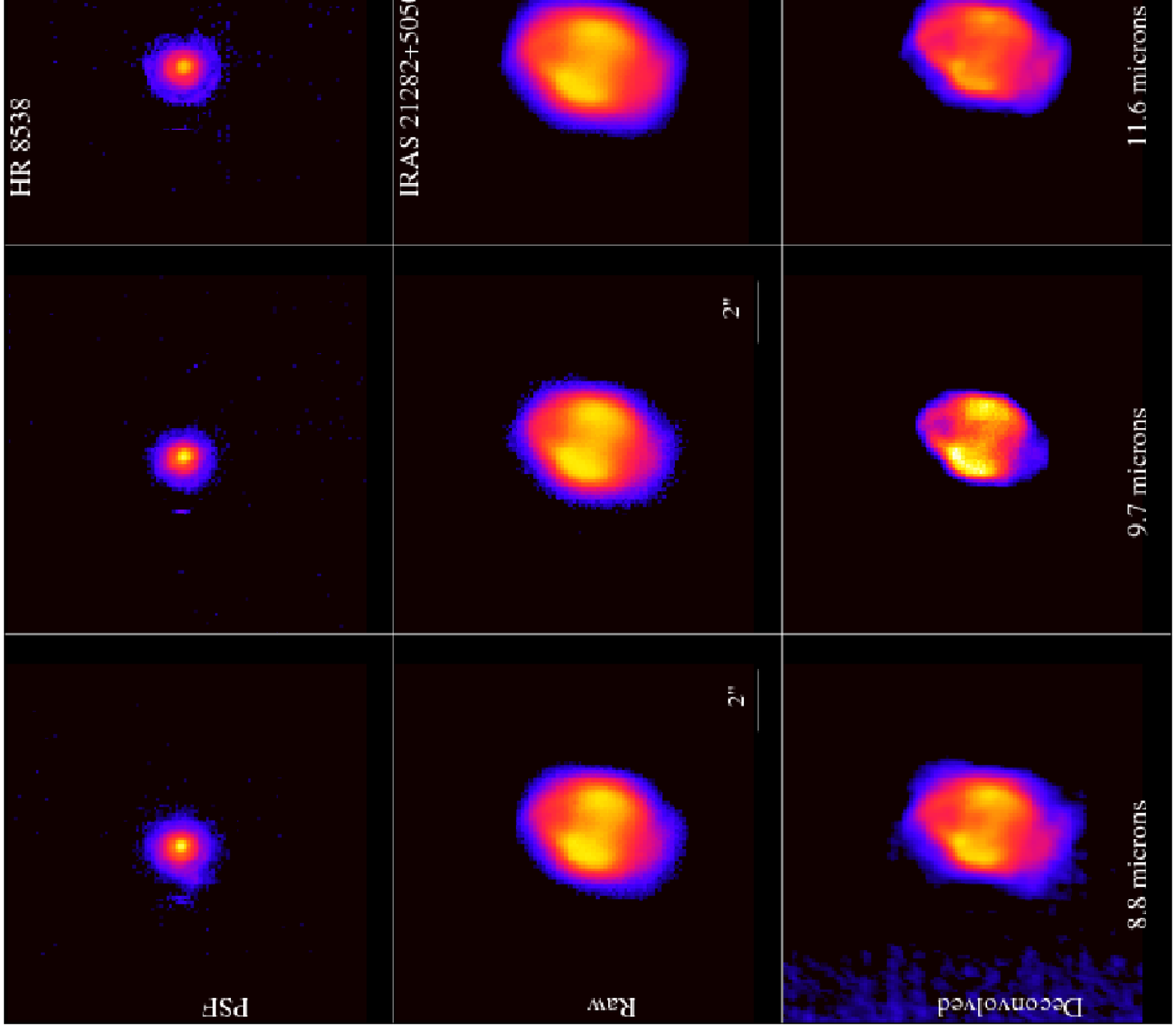}
\caption{Michelle/Gemini North burst mode images of IRAS 21282.}
\label{im_21282}
\end{center}
\end{figure*}

This is a young carbon rich PN with a [WC 11] central star and  is located at
 $\sim$ 2kpc (Shibata et al. 1989).
Michelle images of IRAS 21282+5050 at 8.8, 9.7, 11.6 and 18.1$\mu$m are 
displayed Fig.\ref{im_21282}. As already 
mentioned by Meixner et al. (1993) using MIR observations and the 
NASA 3m IRTF telescope, the structure of the nebula appears the 
same in all the images, with an elliptical nebula having a major 
axis at a P.A.$\sim$ 165\deg and two peaks lying almost East/West. 
The dimension of this extended structure is 5.4'' $\times$ 3.6''. The 
central star is not detected in all our images. The two observed 
lobes can be interpreted as the projection of a dusty torus seen roughly edge on.

\subsection{Evolved massive objects}

\subsubsection{IRAS 10215$-$5916}
AFGL 4106 is a post-red supergiant binary (Molster et al., 1999). Optical observations with the ESO/NTT telescope revealed the shape
of the ionized region in its circumstellar envelope through the H$_{\alpha}$ line (van Loon et al. 1999).
An arc, extending from roughly North-East to South-West clockwise is seen.
The TIMMI/ESO mid-infrared images of AFGL 4106 show that the dust  has an oval to box shaped distribution, with a bright unresolved 
peak centred on the central objects (Molster et al. 1999). The dust distribution shows some indications of clumpiness.
The North-West part of the nebula is fainter compared to the rest and seems slightly more extended, while the H$_{\alpha}$ 
shows a clear anti-correlation with the MIR emission.

\begin{figure*}
\begin{center}
\includegraphics[width=16cm]{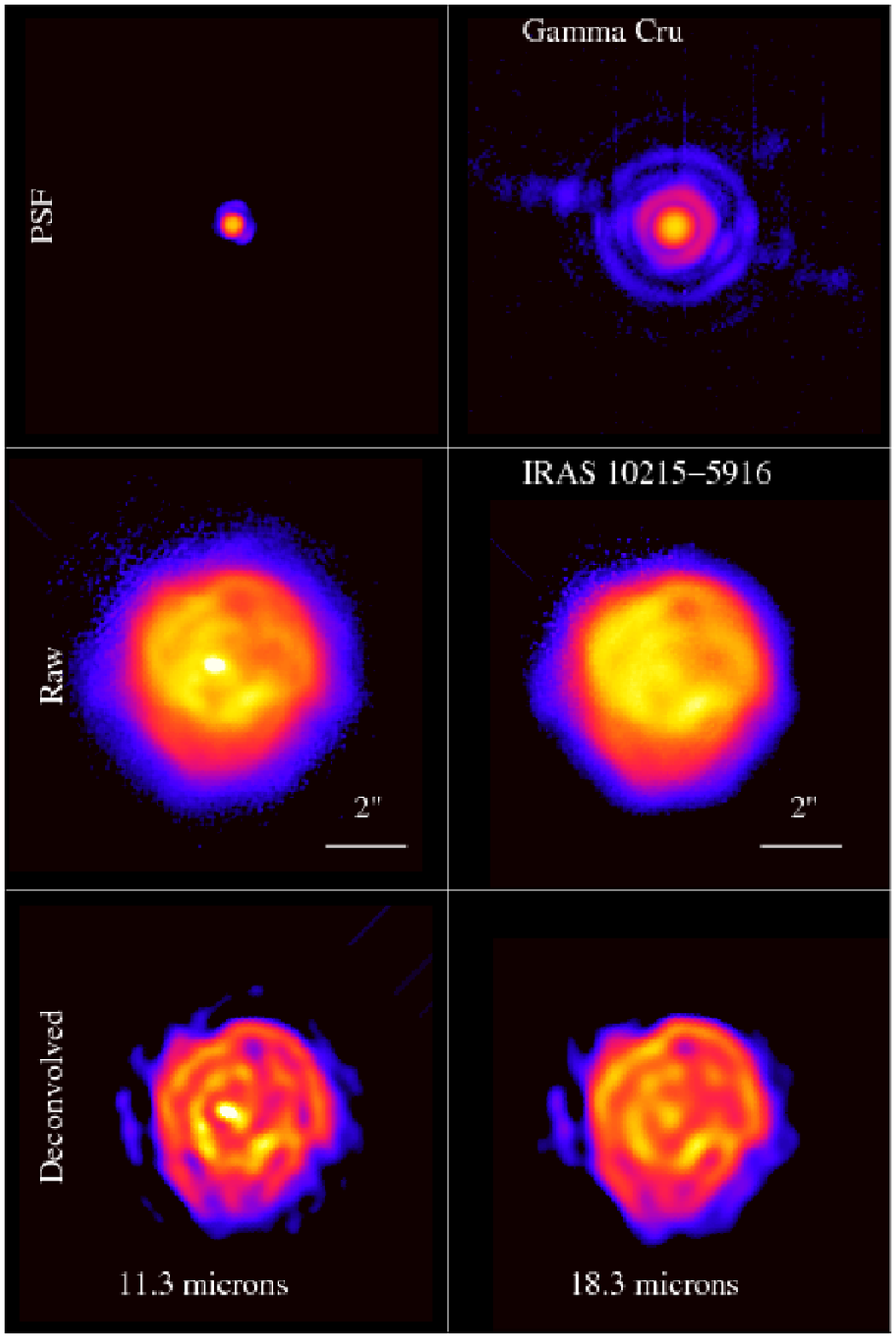}
\caption{ T-Recs mode images of IRAS 10215 (AFGL 4106).}
\label{im_10215}
\end{center}
\end{figure*}

Our T-Recs images of AFGL 4106 (Fig. \ref{im_10215}) show a similar overall morphology of the dust distribution as the one observed by Molster et al. (1999).
But the higher angular resolution of our observations allows us to resolve structure inside the envelope of AFGL 4106.
The arc seen in H$_{\alpha}$ is clearly resolved, but larger, extending from East to West clockwise. A smaller arc is clearly seen, extending from North to West anticlockwise.
Many other clumps and under densities are clearly seen. The central parts of the nebula show a bright clump, extended in a direction perpendicular to the elliptical
nebula. The arcs seen could form a spiral structure, similar to the pinwheel nebula observed around WR\,104 (Tuthill et al., 2008).
The pinwheel structure observed in WR\,104 is due to dust formation triggered by the interaction of the wind from the mass-losing star and the orbiting companion.
AFGL 4106 is known to be a binary system, and it is thus very likely that the spiral structure we observe is also due to a wind-binary companion interaction.

  \subsubsection{IRAS 17163$-$3907} 
IRAS 17163$-$3907 (Hen 1379) was discovered by Henize in 1976 (Henize et al., 1976). Despite being
 one of the brightest mid-infrared objects in the sky, it remains poorly studied.
From near-infrared imaging, Epchtein et al. (1987) classified it as a PPN
 candidate. Its IRAS spectrum indicates the presence of silicate dust and it is unresolved
in the optical by the HST (Si{\'o}dmiak et al.), while speckle observations
 in the  L band (3.6 $\mu$m) indicate an angular dimension of 1.11$\pm$0.23 arcsec (Starck et al., 1994).
We present here the first mid-infrared images of this target.  It shows the presence of 
two concentric, almost circular, detached shells with a angular diameter of $\sim$5.5 arcsec.
A point source is clearly seen in all the images, and is brighter at shorter
 wavelengths. The deconvolved images show that the overall spherical shells are made of smaller
patchy structures.

  \subsubsection{IRAS 19114$+$0002}   
 IRAS 19114$+$0002 is classified as a yellow hypergiant or a post-AGB star,
depending on the adopted distance. It was imaged in the mid-infrared
 with the Keck 1 telescope to resolve a detached shell around this object (Jura \& Werber, 1999).
This shell as an inner diameter of $\sim$3.3 arcsec and an outer diameter
of at least 5.7 arcsec. They noted that the central star is offset from the
centre of the shell by $\sim$0.35 arcsec. This central star appears
asymmetric in the East-West direction. They explained this by imperfect chop/nod
motion. They also noted that the northern part of the almost circular shell
is brighter  and present some departure from spherical symmetry.
Gledhill \& Takami (2001) modelled the dust shell around IRAS 19114 seen in their polarimetric observations  (Gledhill et al. (2001)). They found that the observation
are well reproduced with a spherically symmetric dust distribution and a r$^{-2}$ density law.
This is an indication that during the mass-loss phase, the mass-loss rate was constant.
They estimated the dust mass of the shell to be 0.08M$_{\odot}$.
In our images we observe a similar detached shell at all wavelengths, but also
an East/West extension in the central object at 8$\mu$m.  As the PSF observations
associated with all our observations are perfectly circular and we used the same settings for
 these observations, this structure can not be due to chop/nod motion and is real. A weaker
 dusty structure seems to connect this central source to the detached shell.
 Some warmer dust is thus certainly present close to the central star.

\subsubsection{IRAS 19244$+$1115}  

IRC +10420 is a F red supergiant with a dusty circumstellar envelope.
This star is certainly in the short transition phase
between RSG and Wolf$-$Rayet stars.  It might be the only object in this transition
phase (Bl{\"o}cker 1999). It is one of the brightest source in N and Q  band
and has an intrinsic luminosity L$\sim$5 $\times$10$^5$L$_{\odot}$ for a distance estimated to
5kpc. IRC +10420 is thus very close to the Humphreys$-$Davidson limit, upper limit to the luminosity of stars.
HST observations (Humphreys et al., 1997) show that the nebula around IRC +10420 is extended ($\sim$ 15 arcsec) 
with a bright core $\sim$3 arcsec in diameter. This core has a very complex morphology. It has been observed 
in N band (Humphreys et al. 1997; Meixner et al. 1999). These images in the mid $-$infrared show that it contains
 two lobes separated by 1 arcsec. This core has a dimension $\sim$2$\times$2 arcsec. HST
observations show the presence of an elongation in the envelope with P.A. $\sim$ 215\deg.
Recent radial motion studies (Tiffany et al., 2010) indicate that we are viewing IRC +10420 nearly pole-on.
The extension they observe toward the South-West is likely to be the equatorial plane, with the South-West side poiting toward us.

Our VISIR images reveals the presence of an extended envelope, with an extension toward the South-West along
a P.A.$\sim$223\deg. It presents a more or less asymmetric morphology along that direction, as revealed by HST observations.
The high dynamic range in our image also
reveals the presence of complex structures inside the nebula.

\subsection{Other objects}


  \subsubsection{IRAS 12405$-$6219}
IRAS 12405$-$6219 has been classified as a possible PN, based on its IRAS colours
 (van de Steen \& Pottasch, 1993). Su{\'a}rez et al. (2009) noted that its near-infrared colours were very
 similar to those of H~{\sc II} regions.
Our images  show a morphology very unusual for a post-AGB star or a planetary nebula, but common
 to H~{\sc II} regions. IRAS 12405$-$6219 is thus very certainly an
 H~{\sc II} region.

  \subsubsection{IRAS 18184$-$1302} 
MWC 922 (a.k.a. the Red Square nebula) is a dust enshrouded Be star (Tuthill \& Lloyd, 2007).
 Near-infrared Palomar adaptive optics images reveals a regular and symmetric structure around 
that object (Tuthill \& Lloyd, 2007).
The images show a square-like structure, as the projection of biconal lobes, crossed by a series of rungs
 and an equatorial dark band crossing the core, with a P.A. of 46\deg.
Our mid-infrared images show the same square-like structure, even if it is not fully clear due to striping
 that affected the whole right part of all our images.
Our deconvolved images clearly reveal the presence of a core with a similar orientation as the one assessed
 from near-infrared imaging.


\section{Discussion}

\subsection{Two kinds of objects: resolved cores and detached shells}
\begin{figure}
\begin{center}
\includegraphics[width=9cm]{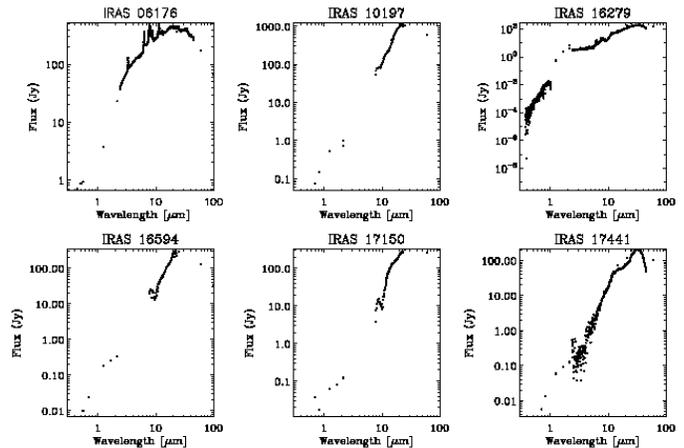}
\caption{ Spectral Energy Distribution of the resolved objects with a bright central source}
\label{sed_torus}
\end{center}
\end{figure}

\begin{figure}
\begin{center}
\includegraphics[width=9cm]{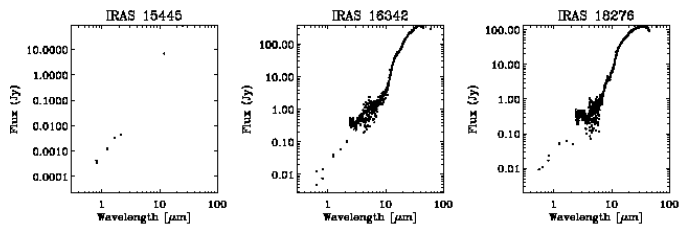}
\caption{ Spectral Energy Distribution of the resolved objects with an equatorial dark lane}
\label{sed_darklane}
\end{center}
\end{figure}

\begin{figure}
\begin{center}
\includegraphics[width=9cm]{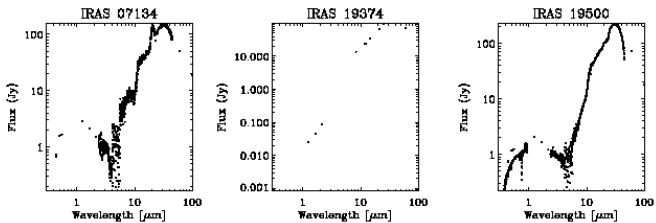}
\caption{ Spectral Energy Distribution of the resolved objects with a detashed shell}
\label{sed_detached}
\end{center}
\end{figure}

For the largest objects that are clearly resolved, we can notice that the PPNe
 observed can be divided into two categories: on one side the objects with 
a dense central core, in the form of a bright central source, resolved or not 
(IRAS 06176, IRAS 07399, IRAS 10197, IRAS 15103, IRAS 16279, IRAS 16594,
IRAS 17150, IRAS 17311, IRAS 17441), or a dark lane, resolved or not, with most
of the emission coming from the poles (IRAS 15445, IRAS 16342, IRAS 18276) indicating the presence
of a large amount of dust, making the central regions optically thick even in the mid-infrared.
On the other side, some objects do not have such a central core, and we can observe either a
detached shell or the central star (IRAS 07134, IRAS 19374 and IRAS 19500).
The objects without a central core  all have  an  elliptical morphology, while the 
objects with a central core are either bipolar or multipolar.
This can be seen in their Spectral Energy Distribution (SED), as the objects with a dense central core 
(Fig.\ref{sed_torus}) or an equatorial dark lane (Fig.\ref{sed_darklane}) have a rather
 flat SED in the near-infrared wavelength range, due to the presence
 of hot dust close to the central star. The SED of the objects with detached shells are 
characterized by the presence of a clear double-peaked distribution (Fig.\ref{sed_detached}),
 with a first peak shorter than 1 micron due to the central star, and a second peak
 due to the cool dust in the shell. The flux is much lower in the
 near-IR due to the absence of dust close to the central star.

 The shape of the SED can be affected by the orientation of the nebula too. The
orientation of the nebula affects the ratio between the  photospheric and the dust peaks
of the SED (Su et al., 2001), and only if the optic depth is significant. The presence/absence of infrared excess observed
in our sources is different and thus not related to orientation effects.
This infrared excess due to hot dust  is an indication that the dense cores play a  role in the shaping of the nebulae.
Two main classes of models have been proposed to explain the shaping of nebulae.
The first class of models is based on the Generalized Interacting Winds models described by 
Balick (1987).
In these models, a fast wind from the central star of a PPN or PN interacts with a slower wind, 
remnant of the AGB phase, assumed to be toroidal.

In the second class of models, the primary shaping agents are high speed collimated outflows 
or jets  that are created at the end of the AGB phase or at the beginning
of the PPN phase (Sahai \& Trauger, 1998). The interaction of these jets with a spherical AGB 
wind will create lobes that are in fact cavities.
If the direction of the jets changes with time, multipolar nebulae can be shaped.

Both models require the presence of a central torus/disc in the core of the nebulae. 
Our observations clearly indicate
that the bipolar and multipolar nebulae have such a central structure in their core.


\subsection{Departure from circular symmetry}
We resolved 25 PPNe in our survey. All these nebula show a clear
 departure from  circular symmetry.
Some circular shells are resolved in our survey, but only around
 massive evolved stars such as IRAS 17163 and IRAS 19114.
A dramatic change in the the distribution of the circumstellar material
 is often observed when a star evolves from the AGB phase to the PN phase
 (Frank \& Balick, 2002).
Most AGB stars have  a large scale cicularly symmetric morphology
 (Mauron \& Huggin, 2006), while PNe display a variety of morphologies from
 elliptical to bipolar or multipolar.
Parker et al. (2006), from a large optical imaging survey of PNe, found that
 80\% of the PNe show a clear sign of departure from circular symmetry, and thus that $\sim$20\%
of the PNe are spherical. The shaping of the PNe is thought to occur at the very end of the AGB phase 
or the beginning of the PPN phase. It is thus surprising that in our sample of 25 resolved PPN we do not 
find any circular ones.

The fact that we do not observe any circular PPNe could be a sample selection effect.
We selected our targets as bright IRAS 12$\mu$m  sources. To be a bright emitter at these wavelengths, 
an object needs to have some dust hot  enough ($\sim$300K), and thus not far  from the central star.
This is the case for the stars with a central core, which are aspherical, as mentioned before.
The spherical PPNe are fainter than the non spherical ones
in the mid-infrared, due to the lack of  central torus/disc emitting in this wavelength range.
At the end of the AGB phase, the envelope of the AGB progenitors of circular PPNe  are ejected and rapidly  cool down while expanding.
There are thus very few spherical PPNe that are bright in the mid-infrared. 
Furthermore those bright PPNe are compact and thus difficult to spatially resolve.
The best way to detect such spherical envelopes is thus at longer wavelengths. Such 
detached shells are actually observed in the far infrared with the Herschel
Space Observatory (Kerschbaum et al., 2010).

\subsection{Formation of S-shaped structures}
As we mentioned in Section \ref{17441}, for IRAS 17441 a tilt is observed
between the orientation of the central dusty torus we resolved and the
tips of the observed
 S-shaped structure. Such a tilt was observed by Volk et al. (2007), and
 measured to be almost 90 degrees. They suggested that a precession of the dusty torus could 
explain the observed S-shaped structure of the nebula. They estimated the 
dynamical age of the envelope, assuming a distance of 1kpc and an expansion velocity of 100km/s,
to be $\sim$100 yr. According to this, the torus should thus precess with a rate of $\sim$1\deg/yr.
As our observations were made 4 years after the observations presented by 
Volk et al. (2007), we should see a tilt of the torus
of 4\deg between the two observations. The images provided by these authors
show that the orientation of the torus they
observed is exactly the same as the one we observed. The torus in the core
of IRAS 17441 is thus not precessing at such a high
rate.
When we compare the images we obtained of IRAS 17441 and IRAS 10197, one
can see some very striking similarities. Both images reveal the presence
of an S-shaped envelope with a resolved central dusty torus. The central 
torus of IRAS 10197 is also tilted with respect to the tips of
 the S-shaped structure, as noted by Ueta et al. (2007). These authors 
 estimated the tilt to be $\sim$46\deg. Some images of IRAS 10197 were obtained
with VISIR in December 2005, two and a half years prior to our observations.
We retrieved and reduced these data to analyse them. The orientation of the torus 
appears to be exactly the same as the one we observed. Assuming a distance 
to IRAS 10197 of 2kpc, an angular extension of the S-shaped structure of
$\sim$2.5'' and an expansion of 30km/s (Sahai et al., 1999), the dynamical 
age of this structure is $\sim$800 yr. The torus would thus need to precess with a rate
of $\sim$0.06\deg/yr to explain the S-shaped structure. Such a precession 
rate is almost impossible to detect with the observations we have.

It thus appears that we can not explain the observed S-shaped structures with
a  precession of the  central tori. This could be due either to the fact that
we underestimated the dynamical age of the nebula or that another mechanism is 
responsible for the S-shaped structure.
A more plausible explanation is that the S-shaped structure is not due to the
precession of the torus itself, but to precessing outflows inside this torus.
The presence of such outflows has been observed in the PN NGC 6302
(Meaburn et al., 2008), which has a morphology very similar to those of IRAS 10197 and IRAS 17441.
These outflows are Hubble-type, which means that their velocity
is proportional to the distance from the source. 
A torus similar to the ones observed in the core of these objects 
is also seen in the core of NGC 6302 (Peretto et al., 2007).
The properties of such outflows can be theoretically described by a sudden ejection of material,
a "bullet" as described by Dennis et al. (2008).
Such bullets naturally account for mulitpolar flows, that  could  arise naturally from the
fragmentation of an explosively driven polar directed shell.
It is thus likely that the S-shaped observed in IRAS 17441 and IRAS 10197
is due to high speed outflows triggered at the end of the
AGB phase or the beginning of the PPN phase, likely during an explosive event.
 
\subsection{Chemistry and morphology} 
 Amongst the PPNe and Water Fountains clearly resolved in our survey, 18 have a known dust chemistry: oxygen, carbon-rich or a dual dust chemistry with
both carbonaceous and oxygeneous dust grains in their envelopes.
For the oxygen-rich sources, we find that 10 out of 11 are bipolar or multipolar, while the remaining one is elliptical.
For the carbon-rich sources, we find that 2 are bipolar or multipolar and 2 elliptical. The three objects with a dual dust chemistry are multipolar or bipolar.
This is in agreement with the recent work by Guzman-Ramirez et al. (2010), which shows a strong correlation between dual dust chemistry and the presence 
of an equatorial overdensity. The dual dust chemistry could be due either to the formation of PAHs in an oxygen-rich torus after CO photodissociation, or to the presence
of a long-lived O-rich disc formed before the star turned carbon-rich due to the third dredge-up. 
 
In their mid-infrared catalogue, Meixner et al. (1999) used a different morphological classification and found that most of the elliptical source they resolved are
O-rich , while the toroidal ones tend to be C-rich. Stanghellini et al. (2007) also studied the correlation between dust composition and morphologies. They 
determined, from a  study of 41 Magellanic Clouds
PNe, that all PNe with O-rich dust are bipolar or highly asymmetric. Our study agrees with this last finding, and it seems that O-rich PPNe appears to be bipolar
or multipolar. As discussed by De Marco (2009), the low C/O ratio  of these bipolar nebulae could be due to  the interaction with a binary companion
during a  common envelope phase, or in the case of single star evolution, result from conversion of carbon to nitrogen. The common phase interaction will 
lead to the ejection of the envelope earlier than in the single star evolution scenario, leading to a less efficient dredge-up of carbon, and thus a lower C/O
ratio (Izzard et al., 2006). The conversion of carbon to nitrogen occurs for massive AGB stars with the hot bottom burning process.
It is thus likely that the bipolar PPNe have progenitor with larger masses than the ellitpical ones.
 This is in agreement with the work by Corradi \& Schwartz (1995),
who showed that bipolar PNe tend to have a higher progenitor mass. Soker (1998) proposed that this could be explained in the paradigm of binary system progenitors,
as primaries that undergoes a common envelope phase, and thus become bipolar, tend to have a higher mass.

\section{Conclusions}
We imaged 93 evolved stars and nebulae in the mid-infrared using VISIR/VLT, T-Recs/Gemini South and Michelle/Gemini North.
Our observed sample contains all the post-AGB stars observable from Paranal with an IRAS 12 $\mu$m flux density larger than 10 Jy,
including PPNe, RCrB, RV Tauri stars and  Water Fountains. The sample also includes some 10 AGB stars, 4 PNe, 4 massive
evolved stars, 1 H~{\sc II} region and a Be star.

These observations allowed us to resolve 34 objects, displaying a wealth of different structures, such as resolved
central cores, dark central lanes, detached shells, S-shaped outflows.
None of the AGB and RV Tauri stars appears to be resolved, indicating
that no bright mid-infrared extended dust shells are present around these objects. Circular detached shells are
resolved around 2 massive evolved stars.

We observed two kind of PPNe:
\begin{itemize}
\item PPNe with a dense central core, in the form of a bright central source or a dark lane, resolved or not.
All these objects are bipolar or multipolar and their SEDs display a near-infrared excess due to hot dust from a dense structure in the core of the object

\item PPNe with a detached shell or a visible central star. These objects are all elliptical and have a two 
peaked SED.
\end{itemize}

None of the PPNe appears to be circular, while a significant fraction of PNe, the evolutionary phase after the PPN phase,
are  known to be spherical. This is certainly a sample bias, as we selected bright mid-infrared stars.  Spherical PPNe have
no central torus/disc emitting in this wavelength range. At the end of the AGB phase, their envelope is ejected and rapidly 
cools down while expanding, and thus starts emitting at longer wavelengths and become brighter at longer wavelengths. Such 
detached shells are actually observed in the far infrared with the Herschel Space Observatory.

Precession of the central torii has been proposed to explain the S-shaped morphology of two of the objects we observed, IRAS 10197
and IRAS 17441. Using observations from different epochs, we do not see any sign of such a precession. We propose that the multipolar 
structures observed in the envelopes of these objects are due to outflows inside the torii, 
in a scenario similar to the one proposed by Sahai \& Trauger (1998).

A large fraction of the dust in galaxies may be produced during  the late stages of the evolution of low and intermediate mass stars.
This dust is ejected to the ISM during the PPN phase. Our observations show the existence of two paths for this dust ejection,
via a detached shell or an expanding torus. To better understand the importance of PPNe for the life cycle of dust, it would be
interesting to study how the dust production by these objects is affected by these different paths.
Spatially resolved mid-infrared spectra of these sources will allow us to study the dust composition at different locations in
these PPNe and thus to better understand the dust evolution  during the  PPNe phase.

   \section*{Acknowledgements}

E.L thanks the ESO staff in Paranal, who helped making these observations successful.
M.M. acknowledges an Origin Fellowship.
R.Sz. acknowledge support from grant N203 511838 from Polish MNiSW. 
Fits and postcript images of the nebula will be made available publically via the Torun post-AGB catalogue database at http://www.ncac.torun.pl/postagb2

      \clearpage
 \appendix   
  \section{Tables}
\begin{table*}
\caption[]{Log of the observation, with the object Right Ascension, Declination, Date of the observations, 
telescope used (VLT= Very Large Telescope, GN= Gemini North, GS= Gemini South), instrument used (B = VISIR burst mode, N= VISIR normal mode, M= Michelle, T=T-Recs),
wavelength of the observations in microns, equivalent width of the filter used in microns, and properties of the object 
(PPN= Proto-Planetary Nebula, WF = Water Fountain, MES= Massive Evolved Star, RV Tau = RV Tauri star, PN=Planetary Nebula)}{\label{infostars} }
\begin{center}
\begin{tabular}{lllclllllllllllll}
\hline
 IRAS name          & RA           & Dec          & Date. Telescope, Instrument& $\lambda$ ($\Delta$$\lambda$)& properties\\
 \hline

IRAS 00245$-$0652 &00 27 06.4& $-$06 36 16.9     &01 Jul 08, VLT, B    & 8.59 (0.42)  &  AGB       \\
  &      &                                       &01 Jul 08, VLT, B    & 11.85 (2.34)   & \\
  &      &                                       &01 Jul 08, VLT, B    & 12.81 (0.21)   & \\
IRAS 00477$-$4900 &00 50 02.5& $-$48 43 47.0     &01 Jul 08, VLT, B    & 8.59 (0.42)  & AGB       \\
  &      &                                       &01 Jul 08, VLT, B    & 11.85 (2.34)   & \\
  &      &                                       &01 Jul 08, VLT, B    & 12.81 (0.21)   & \\
IRAS 01037+1219   &01 06 26.0& +12 35 53.0       &29 Jun 08, VLT, B    & 8.59 (0.42)  &  AGB     \\
(CIT 3)&      &                                  &29 Jun 08, VLT, B    & 11.85 (2.34)   & \\
  &      &                                       &29 Jun 08, VLT, B    & 12.81 (0.21)   & \\
IRAS 01246$-$3248 &01 26 58.1& $-$32 32 35.5     &29 Jun 08, VLT, B    & 8.59 (0.42)  &  AGB       \\
(R Scl)&      &                                  &29 Jun 08, VLT, B    & 11.85 (2.34)   & \\
  &      &                                       &29 Jun 08, VLT, B    & 12.81 (0.21)   & \\
IRAS 01438+1850  &01 46 35.3& +19 05 03.7        &01 Jul 08, VLT, B    & 8.59 (0.42)  &  AGB    \\
  &      &                                       &01 Jul 08, VLT, B    & 11.85 (2.34)   & \\
  &      &                                       &01 Jul 08, VLT, B    & 12.81 (0.21)   & \\
IRAS 02270$-$2619 &02 29 15.3& $-$26 05 55.7     &29 Jun 10, VLT, B    & 8.59 (0.42)  &  AGB       \\
(R For) &      &                                 &29 Jun 10, VLT, B    & 11.85 (2.34)   & \\
  &      &                                       &29 Jun 10, VLT, B    & 12.81 (0.21)   & \\

 IRAS 05113+1347   & 05 14 07.8 & +13 50 28.3    &09 Oct 07, VLT, N    &   8.59 (0.42)  & PPN        \\
 (GLMP 88)  &      &                             &09 Oct 07, VLT, N    & 11.85 (2.34)   & \\
            &      &                             &09 Oct 07, VLT, N    & 12.81 (0.21)   & \\
 IRAS 05341+0852    & 05 36 55.1 & +08 54 08.7   &17 Nov 07, VLT, N    & 8.59 (0.42)   & PPN       \\
  &      &                                       &17 Nov 07, VLT, N    & 11.85 (2.34)   & \\
  &      &                                       &17 Nov 07, VLT, N    & 12.81 (0.21)   & \\
IRAS 06176$-$1036  & 06 19 58.2 & -10 38 14.7    &27 Dec 05, GN, M     & 7.90 (0.42)  &  PPN       \\
(Red Rectangle)&      &                          &27 Dec 05, GN, M     & 8.80 (2.34)   & \\
  &      &                                       &27 Dec 05, GN, M     & 11.60 (0.21)   & \\
  &      &                                       &27 Dec 05, GN, M     & 12.50 (0.21)   & \\
  &      &                                       &27 Dec 05, GN, M     & 18.10 (0.21)   & \\
 IRAS 06530$-$0213  & 06 55 31.8 & $-$02 17 28.3 &17 Nov 07, VLT, N    &  8.59 (0.42)  & PPN        \\
  &      &                                       &17 Nov 07, VLT, N    & 11.85 (2.34)   & \\
  &      &                                       &17 Nov 07, VLT, N    & 12.81 (0.21)   & \\
 IRAS 07134+1005   & 07 16 10.3 & +09 59 48.0    &11 Dec 07, VLT, N    & 8.59 (0.42)   & PPN        \\
  &      &                                       &11 Dec 07, VLT, N    & 11.85 (2.34)   & \\
  &      &                                       &11 Dec 07, VLT, N    & 12.81 (0.21)   & \\
  &      &                                       &19 Dec 05, GN, M     &  8.80 (0.42)   & PPN         \\
  &      &                                       &19 Dec 05, GN, M     & 11.60 (2.34)   & \\
  &      &                                       &19 Dec 05, GN, M     &  12.50 (0.21)  & \\
  &      &                                       &19 Dec 05, GN, M     &  18.10 (0.21)  & \\
 IRAS 07284$-$0940 & 07 30 47.5 & $-$09 46 36.8  &14 Feb 08, VLT, N    &  8.59 (0.42)  &RV Tau        \\
   (RAFGL 1135)    &      &                      &14 Feb 08, VLT, N    & 11.85 (2.34)   & \\
                   &      &                      &14 Feb 08, VLT, N    & 12.81 (0.21)   & \\
 IRAS 07331+0021   & 07 35 41.2 & +00 14 58.0    &11 Mar 08, VLT, N    & 8.59 (0.42)   & PPN       \\
  &      &                                       &11 Mar 08, VLT, N    & 11.85 (2.34)   & \\
  &      &                                       &11 Mar 08, VLT, N    & 12.81 (0.21)   & \\
IRAS 07399$-$1435  & 07 42 16.8   & -14 42 52.1  &08 Jan 06, GN, M     & 8.80 (0.42)  &  PPN       \\
 (OH 231.8 +4.2) &      &                        &08 Jan 06, GN, M     & 9.70 (2.34)   & \\
  &      &                                       &08 Jan 06, GN, M     & 11.60 (0.21)   & \\
  &      &                                       &08 Jan 06, GN, M     & 18.10 (0.21)   & \\

 IRAS 07430+1115  & 07 45 51.4 & +11 08 19.6     &15 Mar 08, VLT, N    &  8.59 (0.42)  & PPN       \\
  &      &                                       &15 Mar 08, VLT, N    & 11.85 (2.34)   & \\
  &      &                                       &15 Mar 08, VLT, N    & 12.81 (0.21)   & \\
 IRAS 08005$-$2356  & 08 02 40.7 & $-$24 04 42.7 &23 Dec 07, VLT, N    &  8.59 (0.42)   &  PPN        \\
  &      &                                       &23 Dec 07, VLT, N    & 11.85 (2.34)   & \\
  &      &                                       &23 Dec 07, VLT, N    & 12.81 (0.21)   & \\
 IRAS 10197$-$5750 & 10 21 33.8 & $-$58 05 48.3  &21 Mar 08, VLT, N    & 8.59 (0.42)   &  PPN        \\
  (Roberts 22)&      &                           &21 Mar 08, VLT, N    & 11.85 (2.34)   & \\
              &      &                           &21 Mar 08, VLT, N    & 12.81 (0.21)   & \\
              &      &                           &07 Apr 04, GS, T     & 11.30 (0.42)   & PPN        \\
              &      &                           &07 Apr 04, GS, T     & 18.30 (2.34)   & \\

\hline \\
\end{tabular}
\end{center}
\end{table*}
\clearpage

\begin{table*}
\begin{center}
\begin{tabular}{lllclllllllllllll}
\hline
 IRAS name          & RA           & Dec          & Date. Telescope, Instrument& $\lambda$ ($\Delta$$\lambda$)& properties\\
 \hline
IRAS 10215$-$5916  & 10 23 19.5 &$-$59 32 04.8     &18 Apr 05, GS, T     & 11.30 (0.42)  &  MES      \\
(AFGL 4106)&      &                              &18 Apr 05, GS, T     & 18.30 (2.34)   & \\
 IRAS 11385$-$5517  & 11 40 58.8 & $-$55 34 25.8 &21 Mar 08, VLT, N    & 8.59 (0.42)   &  PPN       \\
  &      &                                       &21 Mar 08, VLT, N    & 11.85 (2.34)   & \\
  &      &                                       &21 Mar 08, VLT, N    & 12.81 (0.21)   & \\
 IRAS 11472$-$0800 & 11 49 48.0 & $-$08 17 20.4  &19 Feb 08, VLT, N    &  8.59 (0.42)   &  PPN      \\
  &      &                                       &19 Feb 08, VLT, N    & 11.85 (2.34)   & \\
  &      &                                       &19 Feb 08, VLT, N    & 12.81 (0.21)   & \\
 IRAS 12222$-$4652  & 12 24 53.5 &$-$47 09 07.5  &30 Jun 08, VLT, B    & 8.59 (0.42) & RV Tau\\
(CD$-$46 7908 )                       &      &   &30 Jun 08, VLT, B    & 11.85 (2.34)  & \\
                                      &      &   &30 Jun 08, VLT, B    & 12.81 (0.21)  & \\
IRAS 12405$-$6219   & 12 43 32.1 &$-$62 36 13.0  &30 Jun 08, VLT, B    & 8.59 (0.42)  &  HII       \\
  &      &                                       &30 Jun 08, VLT, B    & 11.85 (2.34)   & \\
  &      &                                       &30 Jun 08, VLT, B    & 12.81 (0.21)   & \\ 
 IRAS 12584$-$4837  & 13 01 17.8 &$-$48 53 18.7  &30 Jun 08, VLT, B    & 8.59 (0.42)& PPN  \\
 (V1028 Cen)                          &      &   &30 Jun 08, VLT, B    & 11.85 (2.34)   & \\
                                      &      &   &30 Jun 08, VLT, B    & 12.81 (0.21)   & \\
IRAS 13462$-$2807 &13 49 02.0& $-$28 22 03.5     &01 Jul 08, VLT, B    & 8.59 (0.42)  &  AGB       \\
(WHya)  &      &                                 &01 Jul 08, VLT, B    & 11.85 (2.34)   & \\
  &      &                                       &01 Jul 08, VLT, B    & 12.81 (0.21)   & \\
 IRAS 14316$-$3920  & 14 34 49.4 &$-$39 33 19.8  &30 Jun 08, VLT, B    & 8.59 (0.42)& R CrB \\
 (V854 Cen)&                                &    &30 Jun 08, VLT, B    & 11.85 (2.34)   & \\
           &                                &    &30 Jun 08, VLT, B    & 12.81 (0.21)   & \\
 IRAS 14429$-$4539  & 14 46 13.7 &$-$45 52 07.8  &30 Jun 08, VLT, B    & 8.59 (0.42)& PPN\\
  &      &                                       &30 Jun 08, VLT, B    & 11.85 (2.34)   & \\
  &      &                                       &30 Jun 08, VLT, B    & 12.81 (0.21)   & \\
IRAS 14562$-$5406  & 14 59 53.5 & $-$54 18 07.5    &09 May 04, GS, T     & 11.30 (0.42)  &  PN       \\
(Hen 2-113) &      &                             &09 May 04, GS, T     & 18.30 (2.34)   & \\
IRAS 15103$-$5754 & 15 14 18.9 &$-$58 05 20.0    &29 Jun 08, VLT, B    & 8.59 (0.42)  &  WF       \\
(GLMP 405) &      &                              &29 Jun 08, VLT, B    & 11.85 (2.34)   & \\
  &      &                                       &29 Jun 08, VLT, B    & 12.81 (0.21)   & \\
 IRAS 15373$-$5308   & 15 41 07.4 &$-$53 18 15.0 &30 Jun 08, VLT, B    & 8.59 (0.42)& PPN\\
  &      &                                       &30 Jun 08, VLT, B    & 11.85 (2.34)   & \\
IRAS 15445$-$5449   & 15 48 23.5 &$-$54 58 33.0  &01 Jul 08, VLT, B    & 11.85 (2.34)   & WF\\
  &      &                                       &01 Jul 08, VLT, B    & 12.81 (0.21)   & \\
 IRAS 15452$-$5459  & 15 49 11.5 &$-$55 08 52.0  &29 Jun 08, VLT, B    & 8.59 (0.42)& PPN \\
 IRAS 15469$-$5311  & 15 50 43.8 &$-$53 20 43.3  &01 Jul 08, VLT, B    & 8.59 (0.42)& RV Tau \\
   &      &                                      &01 Jul 08, VLT, B    & 11.85 (2.34)   & \\
   &      &                                      &01 Jul 08, VLT, B    & 12.81 (0.21)   & \\
 IRAS 15553$-$5230 & 15 59 11.4 &$-$52 38 41.0   &29 Jun 08, VLT, B    & 11.85 (2.34)&  PPN\\
 (GLMP 440) &      &                             &29 Jun 08, VLT, B    & 12.81 (0.21)   & \\
 IRAS 16239$-$1218 &16 26 43.7& $-$12 25 35.8    &01 Jul 08, VLT, B    & 8.59 (0.42)  &  AGB       \\
(VOph) &      &                                  &01 Jul 08, VLT, B    & 11.85 (2.34)   & \\
  &      &                                       &01 Jul 08, VLT, B    & 12.81 (0.21)   & \\
 IRAS 16279$-$4757 & 16 31 38.1 &$-$48 04 04.0   &29 Jun 08, VLT, B    & 8.59 (0.42)& PPN\\
  &      &                                       &29 Jun 08, VLT, B    & 12.81 (0.21)   & \\
IRAS 16333$-$4807    & 16 37 06.1 &$-$48 13 42.0 &29 Jun 08, VLT, B    & 8.59 (0.42)  &  WF      \\  
  &      &                                       &29 Jun 08, VLT, B    & 11.85 (2.34)   & \\
  &      &                                       &29 Jun 08, VLT, B    & 12.81 (0.21)   & \\
 IRAS 16342$-$3814& 16 37 40.1 &$-$38 20 17.0    &29 Jun 08, VLT, B   & 11.85 (2.34)& WF\\
 (Water foutain nebula) &    &                   &29 Jun 08, VLT, B   & 12.81 (0.21)   & \\
 IRAS 16559$-$2957  & 16 59 08.2 & $-$30 01 40.3 &15 Apr 08, VLT, N    & 8.59 (0.42)   & PPN       \\
  &      &                                       &15 Apr 08, VLT, N    &  11.85 (2.34)  & \\
  &      &                                       &15 Apr 08, VLT, N    & 12.81 (0.21)   & \\
 IRAS 16594$-$4656 & 17 03 10.0 &$-$47 00 27.0   &01 Jul 08, VLT, B   & 8.59 (0.42)& PPN\\
 (Water Lily nebula)&      &                     &01 Jul 08, VLT, B   & 11.85 (2.34)   & \\
                    &      &                     &01 Jul 08, VLT, B   & 12.81 (0.21)   & \\
 IRAS 17028$-$1004 & 17 05 37.9 & $-$10 08 34.6  &18 Apr 08, VLT, N   &  8.59 (0.42)  & PPN        \\
 (M2$-$9)   &      &                             &18 Apr 08, VLT, N   & 11.85 (2.34)   & \\
           &      &                              &18 Apr 08, VLT, N   & 12.81 (0.21)   & \\
IRAS 17047$-$5650   & 17 09 00.9 & $-$56 54 47.9   &07 Jul 05, GS, T    & 11.30 (0.42)  &  PN       \\
CPD-56\deg 8032  &      &                        &07 Jul 05, GS, T    & 18.30 (2.34)   & \\
 IRAS 17088$-$4221  & 17 12 22.6 & $-$42 25 13.0 &18 Apr 08, VLT, N   & 8.59 (0.42)    & PPN        \\
(GLMP 520) &      &                              &18 Apr 08, VLT, N   & 11.85 (2.34)   & \\
           &      &                              &18 Apr 08, VLT, N   & 12.81 (0.21)   & \\
               
\hline \\
\end{tabular}
\end{center}
\end{table*}

\begin{table*}
\begin{center}
\begin{tabular}{lllclllllllllllll}
\hline
 IRAS name          & RA           & Dec          & Date. Telescope, Instrument& $\lambda$ ($\Delta$$\lambda$)& properties\\
 \hline
 IRAS 17106$-$3046  & 17 13 51.8 & $-$30 49 40.7 &18 Apr 08, VLT, N    &  11.85 (2.34)  & PPN   \\
  &      &                                       &18 Apr 08, VLT, N    &  12.81 (0.21)  & \\
 IRAS 17150$-$3224  & 17 18 19.9 & $-$32 27 21.6 &15 Apr 08, VLT, N   & 8.59 (0.42)   & PPN       \\
 (Cotton candy nebula)  &      &                 &15 Apr 08, VLT, N   & 11.85 (2.34)   & \\
                        &      &                 &15 Apr 08, VLT, N   & 12.81 (0.21)   & \\
                        &      &                 &09 Apr 04, GS, T    & 11.30 (0.42)   &PPN       \\
                       &      &                  &09 Apr 04, GS, T    & 18.30 (2.34)   & \\
 IRAS 17163$-$3907 & 17 19 49.3 &$-$39 10 37.9   &29 Jun 08, VLT, B   & 8.59 (0.42)& MES \\
 Hen 3$-$1379 &      &                           &29 Jun 08, VLT, B   & 11.85 (2.34)   & \\
              &      &                           &29 Jun 08, VLT, B   & 12.81 (0.21)   & \\   
 IRAS 17233$-$4330   & 17 26 58.6 &$-$43 33 13.6 &30 Jun 08, VLT, B   & 11.85 (2.34)  & RV Tau \\
  &      &                                       &30 Jun 08, VLT, B   & 12.81 (0.21) & \\
 IRAS 17243$-$4348 & 17 27 53.6 &$-$43 50 46.3   &01 Jul 08, VLT, B   & 8.59 (0.42)& RV Tau\\
 (LR Sco)&      &                                &01 Jul 08, VLT, B   & 11.85 (2.34)   & \\
         &      &                                &01 Jul 08, VLT, B   & 12.81 (0.21)   & \\
 IRAS 17245$-$3951  & 17 28 04.7 & $-$39 53 44.3 &21 May 08, VLT, N   & 11.85 (2.34)   &PPN  \\
 (Walnut Nebula)    &      &                      &21 May 08, VLT, N   & 12.81 (0.21)   & \\
 IRAS 17311$-$4924  & 17 35 02.5&$-$49 26 26.3   &30 Jun 08, VLT, B   & 8.59 (0.42)& PPN \\
 (LSE 76) &      &                               &30 Jun 08, VLT, B   & 11.85 (2.34)   & \\
          &      &                               &30 Jun 08, VLT, B   &  12.81 (0.21)  & \\
 IRAS 17347$-$3139  & 17 38 01.3 & $-$31 40 58.0 &26 May 08, VLT, N    & 8.59 (0.42)   & PPN        \\
 (GLMP 591)  &      &                            &26 May 08, VLT, N    & 11.85 (2.34)   & \\
             &      &                            &26 May 08, VLT, N    & 12.81 (0.21)   & \\
 IRAS 17441$-$2411 & 17 47 08.3 &$-$24 12 59.9   &01 Jul 08, VLT, B    & 8.59 (0.42)& PPN\\
 (Silkworm nebula) &      &                      &01 Jul 08, VLT, B    & 11.85 (2.34)   & \\
                   &      &                      &01 Jul 08, VLT, B    & 12.81 (0.21)   & \\
                   &      &                      &31 May 08, VLT, N    & 8.59 (0.42)  & PPN        \\
                   &      &                      &31 May 08, VLT, N    & 11.85 (2.34)   & \\
                   &      &                      &31 May 08, VLT, N    & 12.81 (0.21)   & \\
 IRAS 17516$-$2525 & 17 54 43.5 &$-$25 26 27.0   &30 Jun 08, VLT, B    & 8.59 (0.42)& PPN\\
  &      &                                       &30 Jun 08, VLT, B    & 11.85 (2.34)   & \\
  &      &                                       &30 Jun 08, VLT, B    & 12.81 (0.21)   & \\
 IRAS 17530$-$3348  & 17 56 18.5 &$-$33 48 43.3  &30 Jun 08, VLT, B    &  8.59 (0.42)& RV Tau \\
(Al Sco)  &      &                               &30 Jun 08, VLT, B    & 11.85 (2.34)   & \\
          &      &                               &30 Jun 08, VLT, B    & 12.81 (0.21)   & \\
 IRAS 17534+2603   & 17 55 25.2 & +26 03 59.9    &21 May 08, VLT, N    & 8.59 (0.42)  & PPN        \\
  &      &                                       &21 May 08, VLT, N    & 11.85 (2.34)   & \\
  &      &                                       &21 May 08, VLT, N    & 12.81 (0.21)   & \\
IRAS 18043$-$2116   & 18 07 21.2 & $-$21 16 14.0 &29 Jun 08, VLT, B    & 12.81 (0.21)   & WF\\
 IRAS 18071$-$1727  & 18 10 06.1 & $-$17 26 34.5 &21 Jun 08, VLT, N    & 8.59 (0.42)    & PPN       \\
  &      &                                       &21 Jun 08, VLT, N    &  11.85 (2.34)  & \\
  &      &                                       &21 Jun 08, VLT, N    & 12.81 (0.21)   & \\
 IRAS 18095+2704  & 18 11 30.7 & +27 05 15.5     &22 May 08, VLT, N    & 8.59 (0.42)  & PPN       \\
  &      &                                       &22 May 08, VLT, N    & 11.85 (2.34)   & \\
  &      &                                       &22 May 08, VLT, N    & 12.81 (0.21)   & \\
 IRAS 18123+0511  & 18 14 49.4 &+05 12 56.0      &29 Jun 08, VLT, B    & 8.59 (0.42)& RV Tau\\
  &      &                                       &29 Jun 08, VLT, B    & 11.85 (2.34)   & \\
  &      &                                       &29 Jun 08, VLT, B    & 12.81 (0.21)   & \\
 IRAS 18135$-$1456 & 18 16 25.6 &$-$14 55 15.0   &29 Jun 08, VLT, B    & 8.59 (0.42)& PPN\\
  &      &                                       &29 Jun 08, VLT, B    & 11.85 (2.34)   & \\
  &      &                                       &29 Jun 08, VLT, B    & 12.81 (0.21)   & \\
  OH 12.8$-$0.9    & 18 16 49.2 & $-$18 15 01.8  &29 Jun 08, VLT, B    & 8.59 (0.42)  &  WF       \\
  &      &                                       &29 Jun 08, VLT, B    & 11.85 (2.34)   & \\
  &      &                                       &29 Jun 08, VLT, B    & 12.81 (0.21)   & \\
 IRAS 18184$-$1302 & 18 21 15.9 & $-$13 01 27.0  &21 May 08, VLT, N    & 8.59 (0.42)  & Be       \\
 (MWC 922)    &      &                           &21 May 08, VLT, N    &  11.85 (2.34)  & \\
              &      &                           &21 May 08, VLT, N    & 12.81 (0.21)   & \\
 IRAS 18184$-$1623 & 18 21 18.9 & $-$16 22 29.0  &21 May 08, VLT, N    & 8.59 (0.42)  & PPN        \\
  &      &                                       &21 May 08, VLT, N    &  11.85 (2.34)  & \\
 IRAS 18276$-$1431 & 18 30 30.6 & $-$14 28 55.8  &21 Jun 08, VLT, N    & 8.59 (0.42)  & PPN        \\
 (V* V445 Sct)&      &                           &21 Jun 08, VLT, N    & 11.85 (2.34)   & \\
              &      &                           &21 Jun 08, VLT, N    & 12.81 (0.21)   & \\
  &      &                                       &01 Jul 08, VLT, B    & 8.59 (0.42)  &  PPN       \\
  &      &                                       &01 Jul 08, VLT, B    & 11.85 (2.34)   & \\
  &      &                                       &01 Jul 08, VLT, B    & 12.81 (0.21)   & \\

\hline \\
\end{tabular}
\end{center}
\end{table*}

\begin{table*}
\begin{center}
\begin{tabular}{lllclllllllllllll}
\hline
 IRAS name          & RA           & Dec          & Date. Telescope, Instrument& $\lambda$ ($\Delta$$\lambda$)& properties\\
 \hline
IRAS 18286$-$0959  & 18 31 22.7 &$-$09 57 22.0   &30 Jun 08, VLT, B    & 8.59 (0.42)  &  WF       \\
  &      &                                       &30 Jun 08, VLT, B    & 11.85 (2.34)   & \\
  &      &                                       &30 Jun 08, VLT, B    & 12.81 (0.21)   & \\
IRAS 18450$-$0148  & 18 47 40.8 &$-$01 44 57.0   &01 Jul 08, VLT, B    &  11.85 (2.34)  &  WF       \\
(W43A) &      &                                  &01 Jul 08, VLT, B    &  12.81 (0.21) & \\

IRAS 18460$-$0151  & 18 48 42.8 &$-$01 48 40.0   &01 Jul 08, VLT, B    & 8.59 (0.42)  &  WF       \\
  &      &                                       &01 Jul 08, VLT, B   & 11.85 (2.34)   & \\
  &      &                                       &01 Jul 08, VLT, B   & 12.81 (0.21)   & \\
 IRAS 19016$-$2330  & 19 04 43.5 &$-$23 26 08.8  &30 Jun 08, VLT, B   & 8.59 (0.42)& PPN\\
  &      &                                       &30 Jun 08, VLT, B   & 11.85 (2.34)   & \\
  &      &                                       &30 Jun 08, VLT, B   & 12.81 (0.21)   & \\
 IRAS 19075+0921   & 19 09 57.1 & +09 26 52.2    &21 May 08, VLT, N   & 8.59 (0.42)  & PPN       \\
  &      &                                       &21 May 08, VLT, N   &  11.85 (2.34)  & \\
  &      &                                       &21 May 08, VLT, N   & 12.81 (0.21)   & \\
 IRAS 19114+0002  & 19 13 58.6 & +00 07 31.9     &25 Apr 08, VLT, N    & 8.59 (0.42)  & MES     \\
 (AFGL 2343) &      &                            &25 Apr 08, VLT, N    &  11.85 (2.34)  & \\
             &      &                            &25 Apr 08, VLT, N    & 12.81 (0.21)   & \\
 IRAS 19125+0343     & 19 15 01.1 &+03 48 42.7   &29 Jun 08, VLT, B    & 8.59 (0.42)& RV Tau  \\
 (BD+03 3950)       &      &                     &29 Jun 08, VLT, B    & 11.85 (2.34)   & \\
                    &      &                     &29 Jun 08, VLT, B    &  12.81 (0.21)  & \\
IRAS 19126$-$0708 &19 15 23.4& $-$07 02 49.9     &29 Jun 08, VLT, B    & 8.59 (0.42)  &  AGB       \\
(W Aql) &      &                                 &29 Jun 08, VLT, B    & 11.85 (2.34)   & \\
  &      &                                       &29 Jun 08, VLT, B    & 12.81 (0.21)   & \\
 IRAS 19132$-$3336  & 19 16 32.7 &$-$33 31 20.3  &29 Jun 08, VLT, B    &  8.59 (0.42)& R Cbr \\
(RY Sgr)  &      &                               &29 Jun 08, VLT, B    & 11.85 (2.34)   & \\
          &      &                               &29 Jun 08, VLT, B    & 12.81 (0.21)   & \\
IRAS 19134+2131    & 19 15 35.2 &+21 36 34.0     &01 Jul 08, VLT, B    & 11.85 (2.34)   & WF  \\
  &      &                                       &01 Jul 08, VLT, B    & 12.81 (0.21)   & \\
IRAS 19175$-$0807 &19 20 18.0& $-$08 02 10.6     &30 Jun 08, VLT, B    & 8.59 (0.42)  &  AGB       \\
(V1420 Aql)&      &                              &30 Jun 08, VLT, B    & 11.85 (2.34)   & \\
  &      &                                       &30 Jun 08, VLT, B    & 12.81 (0.21)   & \\
 IRAS 19192+0922  & 19 21 36.5 & +09 27 56.5     &21 May 08, VLT, N    & 8.59 (0.42)   & PPN        \\
  &      &                                       &21 May 08, VLT, N    &  11.85 (2.34)  & \\
  &      &                                       &21 May 08, VLT, N    & 12.81 (0.21)   & \\
 IRAS 19244+1115  & 19 26 48.0 & +11 21 16.7     &21 May 08, VLT, N    & 8.59 (0.42)  & MES        \\
 (IRC +10420)    &      &                        &21 May 08, VLT, N    & 11.85 (2.34)   & \\
                 &      &                        &21 May 08, VLT, N    & 12.81 (0.21)   & \\
IRAS 19327+3024  &   19 34 45.2  & +30 30 58.9   &26 Aug 05, GN, M     & 8.80 (0.42)  &  PN       \\
(BD+30\deg 3639)&      &                         &26 Aug 05, GN, M     & 9.70  (2.34)   & \\
  &      &                                       &26 Aug 05, GN, M     & 11.60 (0.21)   & \\
  &      &                                       &26 Aug 05, GN, M     & 18.10 (0.21)   & \\
 IRAS 19343+2926 & 19 36 18.9 & +29 32 50.0      &21 May 08, VLT, N    & 8.59 (0.42)  & PPN        \\
 (Min Footprint)&      &                         &21 May 08, VLT, N    &  11.85 (2.34)  & \\
                &      &                         &21 May 08, VLT, N    & 12.81 (0.21)   & \\
 IRAS 19374+2359  & 19 39 35.5 & +24 06 27.1     &24 Jul 08, VLT, N    & 8.59 (0.42)  & PPN       \\
  &      &                                       &24 Jul 08, VLT, N    & 11.85 (2.34)   & \\
  &      &                                       &24 Jul 08, VLT, N    & 12.81 (0.21)   & \\
 IRAS 19386+0155 & 19 41 08.3 & +02 02 31.3      &26 May 08, VLT, N    & 8.59 (0.42)  & PPN       \\
(V1648 Aql)  &      &                            &26 May 08, VLT, N    & 11.85 (2.34)   & \\
             &      &                            &26 May 08, VLT, N    & 12.81 (0.21)   & \\
 IRAS 19454+2920 & 19 47 24.8 & +29 28 10.8      &24 Jul 08, VLT, N    & 8.59 (0.42)  & PPN        \\
  &      &                                       &24 Jul 08, VLT, N    & 11.85 (2.34)   & \\
  &      &                                       &24 Jul 08, VLT, N    & 12.81 (0.21)   & \\
 IRAS 19477+2401 & 19 49 54.9 & +24 08 53.3      &19 Jul 08, VLT, N    & ?8.59 (0.42)  & PPN        \\
 (Cloverleaf Nebula) &      &                    &19 Jul 08, VLT, N    & 11.85 (2.34)   & \\
                     &      &                    &19 Jul 08, VLT, N    & 12.81 (0.21)   & \\
 IRAS 19480+2504  & 19 50 08.3 & +25 12 00.9     &17 Jul 08, VLT, N    & 8.59 (0.42)   & PPN        \\
  &      &                                       &17 Jul 08, VLT, N    & 11.85 (2.34)   & \\
  &      &                                       &17 Jul 08, VLT, N    & 12.81 (0.21)   & \\
 IRAS 19500$-$1709 & 19 52 52.7 & $-$17 01 50.3  &25 Apr 08, VLT, N    &  8.59 (0.42)   & PPN       \\
 (V5112 Sgr)&      &                             &25 Apr 08, VLT, N    & 11.85 (2.34)   & \\
            &      &                             &25 Apr 08, VLT, N    & 12.81 (0.21)   & \\
 IRAS 20004+2955    & 20 02 27.4 & +30 04 25.5   &21 May 08, VLT, N    & 8.59 (0.42)  & PPN       \\
  &      &                                       &21 May 08, VLT, N    & 11.85 (2.34)   & \\
  &      &                                       &21 May 08, VLT, N    & 12.81 (0.21)   & \\

\hline \\
\end{tabular}
\end{center}
\end{table*}
\begin{table*}
\begin{center}
\begin{tabular}{lllclllllllllllll}
\hline
 IRAS name          & RA           & Dec          & Date. Telescope, Instrument& $\lambda$ ($\Delta$$\lambda$)& properties\\
 \hline
 IRAS 20043+2653  & 20 06 22.7 & +27 02 10.6     &17 Jul 08, VLT, N    & 8.59 (0.42)  & PPN        \\
 (GLMP 972)  &      &                            &17 Jul 08, VLT, N    & 11.85 (2.34)   & \\
             &      &                            &17 Jul 08, VLT, N    & 12.81 (0.21)   & \\
 IRAS 20077$-$0625  & 20 10 27.9 & $-$06 16 13.6 &25 Apr 08, VLT, N    & 8.59 (0.42) &PPN&        \\
  &      &                                       &25 Apr 08, VLT, N    & 11.85 (2.34)   & \\
  &      &                                       &25 Apr 08, VLT, N    & 12.81 (0.21)   & \\
 IRAS 20547+0247   & 20 57 16.4 &+02 58 44.0     &29 Jun 08, VLT, B    & ?8.59 (0.42)& PPN\\
 (U Equ)   &      &                              &29 Jun 08, VLT, B    & ?11.85 (2.34)   & \\
           &      &                              &29 Jun 08, VLT, B    & ?12.81 (0.21)   & \\
IRAS 21032$-$0024 &21 05 51.7& $-$00 12 40.3     &29 Jun 08, VLT, B    & 8.59 (0.42)  &  AGB       \\
(RV Aqr)&      &                                 &29 Jun 08, VLT, B    & 11.85 (2.34)   & \\
  &      &                                       &29 Jun 08, VLT, B    & 12.81 (0.21)   & \\
IRAS 21282+5050   & 21 29 58.4  &+51 03 59.8     &03 Sep 05, GN, M     & 8.80 (0.42)  &  PPN       \\
  &      &                                       &03 Sep 05, GN, M     & 9.70  (2.34)   & \\
  &      &                                       &03 Sep 05, GN, M     & 11.60 (0.21)   & \\
  &      &                                       &03 Sep 05, GN, M     & 18.10 (0.21)   & \\
IRAS 22196$-$4612 &22 22 44.2& $-$45 56 52.6     &30 Jun 08, VLT, B    & 8.59 (0.42)  &  AGB       \\
(pi Gru) &      &                                &30 Jun 08, VLT, B    & 11.85 (2.34)   & \\
  &      &                                       &30 Jun 08, VLT, B    & 12.81 (0.21)   & \\
IRAS 22327$-$1731 &22 35 27.5& $-$17 15 26.9     &30 Jun 08, VLT, B    & 8.59 (0.42)  & PPN       \\
(HM Aqr)&      &                                 &30 Jun 08, VLT, B    & 11.85 (2.34)   & \\
  &      &                                       &30 Jun 08, VLT, B    & ?12.81 (0.21)   & \\
 IRAS 23166+1655 &23 19 12.4  & +17 11 35.4      &16 Jul 08, VLT, N    & 8.59 (0.42)  &  PPN       \\
 (RAFGL 3068)   &      &                         &16 Jul 08, VLT, N    & 11.85 (2.34)   & \\
                &      &                         &16 Jul 08, VLT, N    & 12.81 (0.21)   & \\

\hline \\
\end{tabular}
\end{center}
\end{table*}

\begin{table*}
\caption[Infrared photometry of the observed stars. J, H and K$_s$ are magnitude from 2MASS. 
F$_{12}$,  F$_{25}$,  F$_{60}$ and    F$_{100}$ are the IRAS fluxes (in Jy) at 12, 25, 60 and 100$\mu$m respectively. 
In the last column, the value 1 or 2 indicate respectively that an IRAS or an ISO spectrum of the object exist.]{\label{photostars}MIR spectra: 1: IRAS, 2: ISO}
\begin{center}
\begin{tabular}{lllclllllllllllll}
\hline
 IRAS name     &J &H &K$_s$ &F$_{12}$&  F$_{25}$&     F$_{60}$&     F$_{100}$ & MIR spectra&\\
 \hline


 IRAS 00245$-$0652&  1.581  & 0.621 & 0.236 &1.16e+02&5.90e+01&1.15e+01 &4.49e+00& 1\\ 
 IRAS 00477$-$4900&  3.184  & 2.228 & 1.862 &1.92e+01&1.01e+01&1.44e+00 &1.14e+00& 1\\ 
 IRAS 01037+1219  & 7.437   & 4.641 & 2.217 &1.16e+03 &9.68e+02 &2.15e+02 &7.21e+01&1,2\\
 IRAS 01246$-$3248&  1.973  & 0.695 & $-$0.117&1.62e+02&8.21e+01&5.48e+01 &2.32e+01&1,2\\
 IRAS 01438+1850  &  2.016  & 1.012 & 0.722 &7.67e+01&3.99e+01&6.60e+00 &2.80e+00 &1\\ 
 IRAS 02270$-$2619&  4.230  & 2.537 & 1.349 &2.54e+02&7.53e+01&1.60e+01 &5.06e+00&1,2\\
 IRAS 05113+1347  &  9.020  & 8.423 & 8.171 &3.78e+00&1.53e+01&5.53e+00 &1.67e+00&1\\
 IRAS 05341+0852  &  10.009 & 9.405 & 9.108 &4.51e+00&9.85e+00&3.96e+00 &8.01e+00&2\\
 IRAS 06176$-$1036&  6.577  & 5.145 &3.655  & 4.21e+02 & 4.56e+02 & 1.73e+01  & 6.62e+01& 1,2\\
 IRAS 06530$-$0213&  9.651  & 8.909 & 8.512 &6.11e+00&2.74e+01&1.51e+01 &4.10e+00&\\
 IRAS 07134+1005  &  6.868  & 6.708 & 6.606 &2.45e+01&1.17e+02&5.01e+01 &1.87e+01&1,2\\
 IRAS 07284$-$0940&  4.925  & 4.269 & 4.042 &1.24e+02&8.84e+01&2.66e+01 &9.54e+00&1\\
 IRAS 07331+0021  &  5.816  & 5.322 & 4.940 &1.53e+01&6.81e+01&1.85e+01 &3.68e+00&1\\
 IRAS 07399$-$1435&  9.863  & 8.281 & 6.546  & 1.90e+01 & 2.26e+02 & 5.48e+02  & 2.94e+01& 1,2\\
 IRAS 07430+1115  &  8.836  & 8.211 & 7.766 &7.68e+00&2.99e+01&1.07e+01 &2.53e+00&1\\
 IRAS 08005$-$2356&  7.974  & 6.923 & 5.685 &1.80e+01&5.18e+01&2.98e+01 &1.04e+01&1\\
 IRAS 10197$-$5750&  9.877  & 8.966 & 7.399 &2.00e+02&1.09e+03&5.88e+02 &2.73e+02&1,2\\
 IRAS 10215$-$5916&4.406    &3.432  & 2.970 &2.01e+02&1.76e+03& 8.51e+02& 1.81e+02& 1,2\\
 IRAS 11385$-$5517&  5.947  & 5.138 & 3.991 &9.26e+01&1.38e+02&1.93e+02 &1.04e+02&1,2\\
 IRAS 11472$-$0800&  9.657  & 9.047 & 8.630 &1.14e+01&1.42e+01&1.78e+00 &1.00e+00&1\\
 IRAS 12222$-$4652&  6.941  & 6.380 & 5.338 &3.25e+01&3.32e+01&7.99e+00 &2.41e+00&1\\
 IRAS 12405$-$6219&  16.689 & 13.928& 11.689&1.66e+01&1.09e+02&2.51e+02 &4.30e+02& 1\\ 
 IRAS 12584$-$4837&  10.177 & 9.424 & 7.805 &3.61e+01&4.88e+01&1.30e+01 &3.31e+00&1,2\\
 IRAS 13462$-$2807& $-$1.737& $-$2.689& $-$3.215&4.20e+03&1.19e+03&1.95e+02 &7.22e+01&1,2\\
 IRAS 14316$-$3920&  6.106  & 5.695 & 4.875 &2.30e+01&7.82e+00&1.51e+00 &1.03e+00&1\\
 IRAS 14429$-$4539&  10.636 & 9.818 & 9.133 &1.46e+01&3.33e+01&1.36e+01 &2.91e+00&1\\
 IRAS 14562$-$5406&   9.821 &8.934  & 7.531 &9.24e+01&3.10e+02& 1.77e+02&7.13e+01& 1,2\\
 IRAS 15103$-$5754&  15.191 & 12.718& 10.643&1.08e+01&1.02e+02&1.26e+02 &1.03e+02& 1\\ 
 IRAS 15373$-$5308&  15.037 & 11.664& 8.756 &3.80e+01&5.62e+01&4.25e+01 &1.18e+02&1\\
 IRAS 15445$-$5449&  15.315 & 13.723& 12.982&6.88e+00&8.72e+01&1.13e+03 &2.18e+03&  \\
 IRAS 15452$-$5459&  11.034 & 8.837 & 6.973 &8.71e+01&2.43e+02&2.74e+02 &4.01e+02&1,2\\
 IRAS 15469$-$5311&  7.190  & 6.235 & 4.967 &4.88e+01&4.21e+01&1.55e+01 &2.78e+02&1\\
 IRAS 15553$-$5230&  13.380 & 11.534& 9.859 &9.99e+00&7.00e+01&4.96e+01 &2.83e+02&1,2\\
 IRAS 16239$-$1218&  3.972  & 2.566 & 1.731 &2.90e+01&8.35e+00&1.99e+00 &1.66e+00&1\\
 IRAS 16279$-$4757&  8.660  & 6.605 & 5.490 &4.30e+01&2.68e+02&1.63e+02 &2.65e+02&1,2\\
 IRAS 16333$-$4807&  11.353 & 10.504& 10.184&9.33e+00&4.30e+01&8.93e+01 &1.13e+02& 1\\ 
 IRAS 16342$-$3814&  11.608 & 10.589& 9.569 &1.62e+01&2.00e+02&2.90e+02 &1.39e+02&1,2\\
 IRAS 16559$-$2957&  11.596 & 10.713& 9.347 &9.17e+00&3.24e+01&1.64e+01 &4.18e+00&1\\
 IRAS 16594$-$4656&  9.881  & 9.002 & 8.260 &4.49e+01&2.98e+02&1.31e+02 &3.44e+01&1,2\\
 IRAS 17028$-$1004&  11.198 & 9.177 & 6.996 &5.05e+01&1.10e+02&1.24e+02 &7.58e+01&1\\
 IRAS 17047$-$5650&  9.513  & 8.499 & 6.862 &1.43e+02& 2.57e+02&1.99e+02&9.20e+01& 1,2 \\
 IRAS 17088$-$4221&  13.208 & 11.891& 11.328&4.27e+01&1.28e+02&1.07e+02 &3.69e+01&1\\
 IRAS 17106$-$3046&  9.975  & 8.906 & 8.316 &4.01e+00&6.24e+01&5.12e+01 &1.73e+01&1\\
 IRAS 17150$-$3224&  11.099 & 10.219& 9.391 &5.79e+01&3.22e+02&2.68e+02 &8.24e+01&1,2\\
 IRAS 17163$-$3907&  4.635  & 3.021 & 2.407 &1.24e+03&1.15e+03&6.63e+02 &5.92e+02&1\\
 IRAS 17233$-$4330&  10.423 & 9.592 & 8.371 &1.70e+01&1.34e+01&3.67e+00 &3.21e+01&1\\
 IRAS 17243$-$4348&  8.035  & 7.358 & 6.462 &1.08e+01&8.77e+00&3.69e+00 &5.41e+00&1\\
 IRAS 17245$-$3951&  11.234 & 10.375& 9.716 &3.36e+00&4.47e+01&3.82e+01 &9.77e+01&1\\
 IRAS 17311$-$4924&  9.793  & 9.543 & 9.203 &1.83e+01&1.51e+02&5.87e+01 &1.78e+01&1,2\\
 IRAS 17347$-$3139&  15.097 & 12.932& 10.302&1.90e+01&1.00e+02&1.25e+02 &2.49e+02&1\\
 IRAS 17441$-$2411&  11.088 & 10.132& 9.380 &4.28e+01&1.91e+02&1.06e+02 &2.78e+01&1,2\\
 IRAS 17516$-$2525&  8.695  & 6.850 & 5.082 &5.16e+01&1.16e+02&1.00e+02 &2.92e+02&1,2\\
 IRAS 17530$-$3348&  6.864  & 6.179 & 5.485 &1.76e+01&1.14e+01&2.95e+00 &4.68e+01&1\\
 IRAS 17534+2603  &  4.998  & 4.239 & 3.632 &9.75e+01&5.45e+01&1.34e+01 &6.04e+00&1,2\\
 IRAS 18043$-$2116&  14.546 & 13.404& 13.042&6.60e+00&6.76e+00&1.66e+01 &2.37e+02&1 \\ 
 IRAS 18071$-$1727&  15.889 & 14.149& 12.713&2.37e+01&7.66e+01&8.35e+01 &3.05e+02&1\\
 IRAS 18095+2704  &  7.366  & 6.728 & 6.438 &4.51e+01&1.26e+02&2.78e+01 &5.64e+00&1,2\\
 IRAS 18123+0511  &  7.974  & 7.399 & 6.729 &1.07e+01&1.10e+01&4.21e+00 &1.85e+00&1\\
 IRAS 18135$-$1456&  15.417 & 13.751& 13.130&3.10e+01&1.24e+02&1.58e+02 &4.29e+02&1\\
 OH 12.8$-$0.9    &  17.041 & 15.725& 11.639&1.16e+01&1.69e+01&1.39e+01 &2.89e+02& \\ 
 IRAS 18184$-$1302&  8.960  & 7.396 & 5.704 &3.36e+02&5.98e+02&2.53e+02 &4.36e+02&1\\
 IRAS 18184$-$1623&  5.136  & 4.537 & 4.106 &7.00e+01&3.25e+02&1.17e+02 &5.84e+02&1\\
 IRAS 18276$-$1431&  11.670 & 10.810& 9.450 &2.26e+01&1.32e+02&1.20e+02 &3.86e+01&1,2\\

\hline \\
\end{tabular}
\end{center}
\end{table*}

\begin{table*}
\begin{center}
\begin{tabular}{llllllllllllllll}
\hline
 IRAS name     &J &H &K$_s$ &F$_{12}$&  F$_{25}$&     F$_{60}$&     F$_{100}$ &MIR spectra \\
 \hline
 IRAS 18286$-$0959&  15.431 & 13.562& 12.674&2.49e+01&2.45e+01&1.84e+01 &4.05e+02& 1\\ 
 IRAS 18450$-$0148&  16.115 & 14.794& 13.219&2.37e+01&1.04e+02&2.95e+02 &2.52e+03& 1 \\
 IRAS 18460$-$0151&  13.732 & 13.813& 13.435&2.09e+01&3.27e+01&2.77e+02 &2.91e+02& \\ 
 IRAS 19016$-$2330&  12.500 & 11.343& 9.966 &1.26e+01&5.75e+01&2.80e+01 &9.54e+00&1\\
 IRAS 19075+0921  &  16.920 & 15.536& 14.666&1.33e+02&1.64e+02&5.49e+01 &7.80e+01&\\
 IRAS 19114+0002  &  5.371  & 4.998 & 4.728 &3.13e+01&6.48e+02&5.16e+02 &1.68e+02&1,2\\
 IRAS 19125+0343  &  7.903  & 7.076 & 5.650 &2.89e+01&2.65e+01&7.81e+00 &2.80e+01&1\\
 IRAS 19126$-$0708&  1.534  & 0.238 & $-$0.556&1.58e+03&6.70e+02&1.12e+02 &3.60e+01&1,2\\
 IRAS 19132$-$3336&  5.577  & 5.423 & 5.139 &7.72e+01&2.62e+01&5.43e+00 &4.60e+00&1,2\\
 IRAS 19134+2131  &  16.543 & 14.926& 13.464&5.06e+00&1.56e+01&8.56e+00 &3.95e+00& 1\\ 
 IRAS 19175$-$0807&  5.996  & 3.593 & 1.828 &3.84e+02&1.93e+02&4.79e+01 &1.56e+01&1\\
 IRAS 19192+0922  &  9.449  & 6.759 & 4.821 &1.27e+02&1.55e+02&4.14e+01 &9.78e+00&1\\
 IRAS 19244+1115  &  5.466  & 4.544 & 3.612 &1.35e+03&2.31e+03&7.18e+02 &1.86e+02&1\\
 IRAS 19327+3024  &  9.306  & 9.231 & 8.108 &8.93e+01&2.34e+02&1.62e+02 &7.00e+01& 1,2\\
 IRAS 19343+2926  &  9.908  & 7.929 & 6.209 &1.75e+01&5.98e+01&1.18e+02 &6.80e+01&1,2\\
 IRAS 19374+2359  &  12.038 & 10.866& 9.735 &2.36e+01&9.82e+01&7.09e+01 &7.68e+02&1\\
 IRAS 19386+0155  &  7.951  & 7.069 & 6.011 &1.74e+01&4.74e+01&1.86e+01 &3.79e+00&1,2\\
 IRAS 19454+2920  &  11.853 & 10.749& 10.426&1.73e+01&8.96e+01&5.44e+01 &1.47e+01&1,2\\
 IRAS 19477+2401  &  12.611 & 10.752& 9.606 &1.12e+01&5.49e+01&2.71e+01 &3.80e+01&1,2\\
 IRAS 19480+2504  &  15.216 & 14.036& 13.559&2.08e+01&6.79e+01&4.32e+01 &2.67e+01&1,2\\
 IRAS 19500$-$1709&  7.228  & 6.970 & 6.858 &2.78e+01&1.65e+02&7.34e+01 &1.82e+01&1,2\\
 IRAS 20004+2955  &  4.766  & 4.305 & 3.793 &3.17e+01&3.70e+01&4.66e+00 &3.35e+01&1,2\\
 IRAS 20043+2653  &  17.431 & 14.870& 10.604&1.79e+01&4.20e+01&2.03e+01 &7.48e+00&1\\
 IRAS 20077$-$0625&  6.906  & 3.923 & 2.059 &1.26e+03&1.06e+03&2.16e+02 &6.37e+01&1\\
 IRAS 20547+0247  &  11.561 & 10.132& 8.405 &4.55e+01&3.38e+01&1.00e+01 &2.84e+00&1\\
 IRAS 21032$-$0024&  4.046  & 2.355 & 1.239 &3.08e+02&1.16e+02&2.24e+01 &8.55e+00&1,2\\
 IRAS 21282+5050  &11.504   & 10.709& 9.551 &5.10e+01&7.44e+01& 3.34e+01&1.50e+01& 1,2\\
 IRAS 22196$-$4612& $-$0.715& $-$1.882& $-$2.351&9.08e+02&4.37e+02&7.73e+01 &2.33e+01&1,2\\
 IRAS 22327$-$1731&  8.276  & 7.609 & 6.705 &5.57e+00&4.66e+00&2.11e+00 &1.01e+00&2\\
 IRAS 23166+1655 &17.165&15.402&10.379&&&&&1\\




\hline \\
\end{tabular}
\end{center}
\end{table*}


\begin{table*}
\caption[]{Observed fluxes  and sizes of objects, position angle of the resolved nebula, and name of teh associated PSF standard}

\begin{center}
\begin{tabular}{llllllllllllllll}
\hline  
               &         &   \multicolumn{3}{c}{Source} &  \multicolumn{3}{c}{PSF} &  \\ 
               &$\lambda$& F &FWHM &Size&P.A. &   FWHM&   \\
IRAS name     & ($\mu$m)& (Jy)&(arcsec)& arcsec$\times$arcsec& $\rm ^{o}$&(arcsec)& name &\\
 \hline

00245$-$0652  & 8.59  & 73.2  & 0.26 &unresolved  &   & 0.24   &HD 196321\\
              & 11.85 & 82.5  & 0.33 &unresolved  &   & 0.30   &\\
              & 12.81 & 84.2  & 0.38 &unresolved  &   & 0.32   &\\
00477$-$4900  & 8.59  & 15.9  & 0.31 &unresolved  &   & 0.30   &HD 196321\\
              & 11.85 & 13.8  & 0.31 & unresolved &   & 0.30   &\\
              & 12.81 & 12.0  & 0.32 &unresolved  &   & 0.32   &\\
01037+1219    & 8.59  &501.5 & 0.67& unresolved (saturated) &  &0.23&HD 198048\\
              &11.85  &709.2 &0.63&unresolved (saturated)& &0.30&\\
              &12.81  &789.4 & 0.47&unresolved (saturated)&&0.32&\\
01246$-$3248  & 8.59  &170.9   &0.33 &  unresolved (saturated) &   &   0.23 &HD 198048\\
              & 11.85 &132.3   &0.37  &  unresolved (saturated) &   &   0.30 &\\
              & 12.81 &78.4   &0.33  &  unresolved  &   &   0.33 &\\             
01438+1850    &  8.59 & 56.0  & 0.72 & unresolved (saturated)  &   & 0.43  &HD 196321\\
              & 11.85 & 63.8  & 0.52 &  unresolved (saturated)  &   & 0.30   &\\
              & 12.81 & 53.9  & 0.33 &  unresolved  &   & 0.32   &\\ 

02270$-$2619  & 8.59  &164.4   & 0.33& unresolved  &   &   0.30 &HD 196321\\
              & 11.85 &131.5   & 0.37 & unresolved  &   &   0.30 &\\
              & 12.81 & 71.9  & 0.33 & unresolved  &   &    0.32&\\
05113+1347    &  8.59 & 0.8  & 0.52 & unresolved  &   & 0.55   &HD 31421\\
              & 11.85 & 3.8  & 0.57 &  unresolved &   & 0.49   &\\
              & 12.81 & 4.1  & 0.51 &  unresolved &   & 0.44   &\\
05341+0852    &  8.59 & 2.4  & 0.32 & unresolved  &   & 0.26   &HD 39400\\
              & 11.85 & 5.0  & 0.34 &    unresolved  & &0.30    &\\
              & 12.81 & 4.8  & 0.36 & unresolved  &   & 0.33   &\\
06176$-$1036  &7.90   &365.7 &0.31  &3.3$\times$5.9& &0.22 &HD 59381\\
              &8.80   &313.1 &0.31  &3.3$\times$5.9& &0.22 &\\
              &11.60  &375.7 &0.36  &3.3$\times$5.9& &0.26 &\\
              &12.50  &408.9 &0.34  &3.3$\times$5.9& &0.28 &\\
              &18.10  &251.9 &0.47  &3.3$\times$5.9& &0.38 &\\
06530$-$0213  &  8.59 & 1.4  & 0.35 & unresolved  &   & 0.28   &HD 49293\\
              & 11.85 & 6.1  & 0.67 & unresolved  &   & 0.13   &\\
              & 12.81 & 1.5  & 0.05 & unresolved  &   & 0.32   &\\
07134+1005    & 8.59  & 4.9  & 0.25 & 4.8$\times$4.6 &  N/A  & 0.23   & HD 58207\\
              & 11.85 & 14.6  & 1.19 & 5.0$\times$4.7 & 27   &  0.34  &\\
              & 12.81 & 16.3  & 1.46 & 4.8$\times$4.7 & 23    & 0.37   &\\

07284$-$0940  & 8.59  & 98.2  & 0.28 & unresolved  &   & 0.30   & HD 59381\\
              & 11.85 & 126.7  & 0.33 & unresolved  &   & 0.34   &\\
              & 12.81 & 103.4  & 0.36 & unresolved  &   & 0.33   &\\
07331+0021    &  8.59 & 6.5  & 0.30 & unresolved  &   & 0.26   &HD 61935\\
              & 11.85 & 16.7  & 0.36 & unresolved  &   &    0.35&\\
              & 12.81 & 17.5  & 0.40 & unresolved  &   &    0.39&\\
07399$-$1435  &8.80   &18.3 &0.34 &4.1$\times$6.1 & &0.34 & Alpha CMa\\
              &9.70   &6.3 &0.36 &2.6$\times$4.3 & &0.36 &\\
              &11.60  &18.2 &0.54 &4.3$\times$6.7 & &0.33 &\\
              &18.10  &22.6 &0.81 &5.4$\times$6.7 & &0.41 &\\
07430+1115    &  8.59 & 3.0  &  & unresolved &   &   &HD 62721 \\
              & 11.85 & 9.6  & 0.41 & unresolved &   & 0.32   &\\
              & 12.81 & 9.7  & 0.42 &  unresolved&   & 0.33   &\\
08005$-$2356  & 8.59  & 14.0  & 0.28 & unresolved  &   & 0.27   &HD 67523\\
              & 11.85 & 16.7  & 0.34 & unresolved  &   & 0.32   &\\
              & 12.81 & 17.5  & 0.36 &  unresolved &   & 0.33   &\\
10197$-$5750  & 8.59  & 97.1  & 0.74 & 4.4$\times$3.2 & 44    & 0.25   &HD 91942\\
              & 11.85 & 181.6  & 0.75 & 4.9$\times$3.6 & 42     & 0.31   &\\
              & 12.81 & 215.4  & 0.75 & 4.7$\times$3.5 & 42     & 0.34   &\\
10215$-$5916  &11.30  & 216.1 & 0.72 &  3.4$\times$3.3 & N/A&0.42 & Gamma Gru\\ 
              &18.30  & 1395.33 & 4.01 &  3.4$\times$3.3 &N/A &0.53 &\\ 
11385$-$5517  &  8.59 & 69.3  & 0.28 & unresolved  &   & 0.25  &HD 102461\\
              & 11.85 & 86.6  & 0.31 &  unresolved &   &0.28    &\\
              & 12.81 & 87.1  & 0.34 & unresolved  &   & 0.33   &\\

\hline \\     
\end{tabular}
\end{center}
\end{table*}

   \begin{table*}
\caption[]{Observed fluxes  and sizes of objects.}

\begin{center}
\begin{tabular}{llllllllllllllll}
\hline  
11472$-$0800  &  8.59 &      & 0.30 &  unresolved  &   &  0.30  &HD 99167\\
              & 11.85 & 7.1  & 0.33 & unresolved  &   & 0.32   &\\
              & 12.81 & 5.1  & 0.34 & unresolved  &   & 0.36   &\\              
12222$-$4652  & 8.59  & 23.2  & 0.23 & unresolved &   & 0.23   &HD 111915\\
              & 11.85 & 28.8  & 0.28 & unresolved  &   & 0.29    &\\
              & 12.81 & 24.4  & 0.33 & unresolved  &   & 0.32   &\\
12405$-$6219  & 8.59  & 4.9  & 0.54 & 2.3$\times$2.1 & 128   & 0.23   &HD 111915\\
              & 11.85 & 12.7  & 0.67 & 3.1$\times$2.8 & 133   & 0.30   &\\
              & 12.81 & 16.7  & 1.04 & 3.1$\times$2.9 & 126   & 0.32   &\\

12584$-$4837  &  8.59 & 27.4  & 0.26 & unresolved  &   & 0.23   &HD 111915\\
              & 11.85 & 28.6  & 0.30 & unresolved  &   & 0.30   &\\
              & 12.81 & 22.3  & 0.32 & unresolved  &   & 0.32   &\\
 13462$-$28071 & 8.59  & 1075.4  & 0.99 & 2.3$\times$2.0 & 43   & 0.27   &HD 124294\\
              & 11.85 & 1522.8  & 1.01 & 2.2$\times$2.0 & 39   & 0.31   &\\
              & 12.81 & 1527.8  & 0.76 & 1.9$\times$1.8 & 45   & 0.33   &\\                 
14316$-$3920  &  8.59 & 19.1  & 0.24 & unresolved  &   & 0.24   &HD 111915\\
              & 11.85 & 12.5  & 0.31 &  unresolved &   & 0.30  &\\
              & 12.81 & 11.6  & 0.33 &  unresolved &   & 0.33  &\\
14429$-$4539  &  8.59 & 10.3  & 0.32 & unresolved  &   & 0.24  &HD 111915\\
              & 11.85 & 14.6  & 0.36 & unresolved  &   & 0.30  &\\
              & 12.81 & 15.2  & 0.40 & unresolved  &   & 0.33  &\\
    
14562$-$5406  &11.30  & 87.1 & 1.39&6.0$\times$4.6 &92 & 0.39& Alpha Cen\\
              &18.30  & 246.4 & 1.62&5.6$\times$4.4 &82 & 0.55& \\
15103$-$5754  & 8.59  & 1.0  & 0.34 & 1.8$\times$1.3 & 32   & 0.25   &HD 133550\\
              & 11.85 & 5.6  & 0.52 & 2.5$\times$2.3 & 32   & 0.31   &\\
              & 12.81 & 13.8  & 0.53 & 3.0$\times$2.0 & 32   & 0.33   &\\              
15373$-$5308  & 8.59  & 31.3  & 0.24 & unresolved  &   & 0.24  &HD 133774\\
              & 11.85 & 33.8  & 0.30 & unresolved  &   & 0.30  &\\
 15445$-$5449 & 11.85 & 3.2  & 0.32 & 3.6$\times$2.9 & -3  & 0.32   &HD 124294\\
              & 12.81 & 6.8  & 0.51 & 3.1$\times$2.0 & 1    & 0.33   &\\             
15452$-$5459  &  8.59 & && nodetection     &   &  &HD 133774\\

15469$-$5311  &  8.59 & 40.3  & 0.24 & unresolved  &   & 0.26  &HD 124294\\
              & 11.85 & 40.3  & 0.25 & unresolved  &   & 0.31  &\\
              & 12.81 & 39.0  & 0.32 & unresolved  &   & 0.33  &\\
  
 15553$-$5230              & 11.85 &   & 0.52 & 3.7$\times$3.6 & 91    & 0.30  &HD 133774\\
              & 12.81 &   & 0.45 & 3.4$\times$3.2 & 83    & 0.33  &\\             
16239$-$1218  & 8.59  & 30.7  & 0.25 & unresolved  &   & 0.26 &  HD 124294\\
              & 11.85 & 21.4  & 0.28 & unresolved  &   & 0.31   &\\
              & 12.81 & 21.4  & 0.34 & unresolved  &   & 0.33   &\\              
16279$-$4757  &  8.59 & 20.2  & 0.28 & 6.1$\times$4.0&10   & 0.25 &HD 163376\\ 
              & 12.81 & 30.1  & 0.42 & 7.2$\times$6.2 & 7  & 0.3  &\\
16333$-$4807  & 8.59  & 3.6  & 0.30 & problem &     & 0.25   &HD 163376\\
              & 11.85 & 7.3  & 0.39 & 5.0$\times$3.7 & -7   & 0.31   &\\
              & 12.81 & 13.8  & 0.43 & 5.1$\times$3.4 & -12    & 0.33   &\\             
 
16342$-$3814  & 11.85 & 7.7  & 0.83 & 4.3$\times$4.2  & 79  &  0.30 &HD 163376\\
              & 12.81 & 18.1  &  0.55&  3.6$\times$3.4 & 85  &  0.33 &\\
16559$-$2957  &  8.59 & 5.2  & 0.33 & unresolved  &   & 0.37   &HD 152980\\
              & 11.85 & 8.6  & 0.36 & unresolved  &   & 0.40   &\\
              & 12.81 & 10.9  & 0.41 & unresolved  &   & 0.40   &\\             
16594$-$4656  & 8.59  & 16.6?  & 1.73 & 4.9$\times$3.9 & 84   & 0.33  &HD 124294\\ 
              & 11.85 & 46.0  & 1.56 & 5.9$\times$4.5 & 83 & 0.30  &\\
              & 12.81 & 46.0  & 1.65 & 5.2$\times$4.2 & 82   & 0.32  &\\             
17028$-$1004  &  8.59 & 44.7  & 0.33 & unresolved  &   & 0.33   &HD 159187\\
              & 11.85 & 48.4  & 0.36 &   unresolved&  &   0.35 &\\
              & 12.81 & 56.9  & 0.38 &  unresolved &   & 0.37   &HD 155066\\
17047$-$5650  &11.30  & 155.2&0.62 &5.6$\times$4.6 &8 & 0.40& Eta Sgr\\      
              &18.30  & 179.3&1.10 &4.7$\times$4.1 & 10& 0.57&\\  
17088$-$4221  & 8.59  & 15.0  & 0.31 & unresolved  &   & 0.30   &\\
              & 11.85 & 27.0  & 0.41 & unresolved  &   & 0.36   &\\
              & 12.81 & 59.1  & 0.41 & unresolved  &   & 0.36   &\\
17106$-$3046                & 11.85 & 2.9  & 0.51 & 3.9$\times$3.9 & N/A & 0.33   &HD 157236\\
              & 12.81 & 3.7  & 0.56 & 4.2$\times$3.7 & N/A   & 0.35   &\\

\hline \\     
\end{tabular}
\end{center}
\end{table*}

   \begin{table*}
\caption[]{Observed fluxes  and sizes of objects.}

\begin{center}
\begin{tabular}{llllllllllllllll}
\hline  
17150$-$3224  & 8.59  & 18.3  & 0.50 & 3.7$\times$3.5 & -36   & 0.32   &HD 159433\\
              & 11.85 & 50.3  & 0.58 &4.2$\times$3.6 & -58   & 0.34   &\\
              & 12.81 & 81.0  & 0.56 &4.1$\times$3.6 & -62   & 0.38   &\\
17163$-$3907  & 8.59  &253   & 1.23 & 5.6$\times$5.6 & N/A   & 0.25   &HD 163376\\
              & 11.85 &892   & 1.53 & 5.9$\times$5.8 & N/A   & 0.30   &\\
              & 12.81 &910   & 1.52 & 6.0$\times$5.8 & N/A   & 0.33   &\\
17233$-$4330  & 11.85 & 10.4  & 0.29 & unresolved  &   & 0.30   &HD 163376\\
              & 12.81 & 10.5  & 0.33 & unresolved  &   & 0.33   &\\
17243$-$4348  &  8.59 & 8.6  & 0.25 & unresolved  &   & 0.24   &HD 124294\\
              & 11.85 & 7.8  & 0.29 & unresolved  &   & 0.30   &\\
              & 12.81 & 8.6  & 0.31 & unresolved  &   & 0.32   &\\
17245$-$3951  & 11.85 & 2.2  & 0.47 &unresolved   &   & 0.32   &HD 161892\\\
              & 12.81 & 2.9  & 0.49 & unresolved  &   & 0.35   &\\    
17311$-$4924  & 8.59  & 3.9  & 1.13 & 3.0$\times$2.8 & -90   & 0.24   &HD 163376\\
              & 11.85 & 16.8  & 1.30 & 5.5$\times$5.1 & -92   & 0.30   &\\
              & 12.81 & 19.1  & 1.33 & 3.8$\times$3.6 & -88   & 0.33   &\\
17347$-$3139  & 8.59  & 7.6  & 0.57 & 4.1$\times$3.5 & -42  &0.34    &HD 159881\\
              & 11.85 & 17.3  & 0.48 & 3.9$\times$3.1 & -41   & 0.36   &\\
              & 12.81 & 30.5  & 0.52 & 4.2$\times$3.3 & -41   & 0.38   &\\             
17441$-$2411  &  8.59 & 12.1  & 0.71 & 3.7$\times$3.5 & 16   & 0.58   &HD 196321\\ 
              & 11.85 & 39.5  & 0.76 & 4.2$\times$3.9 & 40   & 0.30   &\\
              & 12.81 & 48.1  & 0.66 & 4.0$\times$3.6 & 21   & 0.33   &\\       
17516$-$2525  &  8.59 & 46.6  & 0.25 & unresolved  &   & 0.24   &HD 163376\\ 
              & 11.85 & 42.0  & 0.29 & unresolved  &   & 0.30   &\\
              & 12.81 & 42.0  & 0.33 & unresolved  &   & 0.33   &\\
17530$-$3348  & 8.59  & 17.2  & 0.20 & unresolved  &   & 0.22   &HD 163376\\ 
              & 11.85 & 17.0  & 0.30 & unresolved  &   & 0.31   &\\
              & 12.81 & 14.3  & 0.33 & unresolved  &   & 0.32   &\\             
17534+2603    &  8.59 & 85.4  & 0.34 & unresolved  &   & 0.35   &HD 163993\\
              & 11.85 & 87.2  & 0.37 & unresolved  &   & 0.42   &\\
              & 12.81 & 77.3  & 0.38 & unresolved  &   & 0.39   &\\
18043$-$2116  & 12.81 & 1.5      & 0.36 & unresolved &   & 0.32   &HD 174387\\          
18071$-$1727  & 8.59  & 11.7  & 0.34 & unresolved  &   & 0.33   &HD 167036\\
              & 11.85 & 22.2  & 0.38 & unresolved  &   & 0.35   &\\
              & 12.81 & 41.5  & 0.38 & unresolved  &   & 0.38   &\\
18095+2704    & 8.59  & 18.8  & 0.39 & unresolved &   & 033  &HD 169414\\
              & 11.85 & 38.7  & 0.41 & unresolved &   & 0.35   &\\
              & 12.81 & 29.0  & 0.39 & unresolved &   & 0.41   &\\
18123+0511    &  8.59 & 7.3  & 0.32 & unresolved &   & 0.24  &HD 163376\\ 
              & 11.85 & 8.9  & 0.30 & unresolved &   & 0.30   &\\
              & 12.81 & 8.4  & 0.31 & unresolved &   & 0.33   &\\
18135$-$1456  & 8.59  & 7.2  & 0.34 & unresolved &   & 0.23   &HD 163376\\
              & 11.85 & 17.9  & 0.37 & unresolved &  & 0.30   &\\
              & 12.81 & 37.0  & 0.42 & unresolved &    & 0.32   &\\  
OH12.8$-$0.9  & 8.59  &9.3& 0.24  & unresolved &   & 0.24      &HD 163376\\
              & 11.85 &9.9&   0.31 &unresolved&    & 0.31    &\\
              & 12.81 &19.7&   0.34 &unresolved&     & 0.33   &\\ 
 18184$-$1302  & 8.59  &228.8  & 0.31 & 4.0$\times$3.5 & N/A   & 0.28   &HD 161892\\
              & 11.85 & 292.1  & 0.09 & 0.4$\times$0.2 & N/A   & 0.32   &\\
              & 12.81 & 345.0  & 0.07 & 1.2$\times$0.8 & N/A   & 0.33   &\\              
18184$-$1623  & 8.59  & 1.7       & 0.41 &  unresolved&   & 0.28   &HD 168415\\\
              & 11.85 & 0.5       & 0.52 &  unresolved&   & 0.31   &\\
18276$-$1431  & 8.59  & 4.4  & 0.51 & 2.5$\times$2.4  &  N/A & 0.48   &HD 181410\\
              & 11.85 & 18.9  & 0.51 & 2.7$\times$2.6 &   N/A &  0.50&\\
              & 12.81 & 22.3  & 0.51 & 2.5$\times$2.4 & N/A   & 0.52   &\\
              & 8.59  & 4.5  & 0.37 & 1.6$\times$1.5 & 8   & 0.24   &HD 196321\\
              & 11.85 & 11.9  & 0.48 & 3.5$\times$2.6 & 9   & 0.30   &\\
              & 12.81 & 17.4  & 0.45 & 2.3$\times$2.0 & 9   & 0.32   &\\              
18286$-$0959  & 8.59  & 27.0  & 0.30 & unresolved  &   & 0.23   &HD 163376\\
              & 11.85 & 37.0  & 0.31 & unresolved  &   & 0.30   &\\
              & 12.81 & 62.3  & 0.35 & unresolved  &   & 0.32   &\\
18450$-$0148  & 11.85 & 21.4  &  0.44& 4.3$\times$4.1 & 44  &  0.31  &HD 161096\\
              & 12.81 & 31.1  &  0.49 & 3.7$\times$3.3 & 32  &  0.32  &\\
18460$-$0151  &  8.59 & 9.2  & 0.24 & unresolved &   & 0.24   &HD 161096\\
              & 11.85 & 16.7  & 0.31 & unresolved &   & 0.31   &\\
              & 12.81 & 24.6  & 0.28 & unresolved &   & 0.33   &\\

\hline \\     
\end{tabular}
\end{center}
\end{table*}

\begin{table*}
\caption[]{Observed fluxes  and sizes of objects.}
\begin{center}
\begin{tabular}{llllllllllllllll}
\hline 
 19016$-$2330  &  8.59 & 8.9  & 0.39 & 2.0$\times$1.9 & N/A   & 0.23   &HD 198048\\
              & 11.85 & 11.8  & 0.39 & 2.8$\times$2.2 &N/A & 0.30   &\\
              & 12.81 & 12.8  & 0.39 & 2.2$\times$1.6 & N/A   & 0.32   &\\              
19075+0921    & 8.59  & 92.5  & 0.34 &  unresolved&   &   0.32 &HD 178690\\
              & 11.85 & 120.0  & 0.35 &unresolved   &   & 0.34   &\\
              & 12.81 & 278.9  & 0.35 & unresolved  &   & 0.34   &\\
19125+0343    & 8.59  & 25.2  & 0.26 & unresolved &    & 0.23   &HD 174387\\
              & 11.85 & 27.3  & 0.29 &  unresolved &   & 0.29   &\\ 
              & 12.81 & 22.8  & 0.33 &  unresolved &   & 0.32   &\\
19126$-$0708  & 8.59  & 644.4  & 0.79 &unresolved (saturated)  &   & 0.23   &HD 174387\\
              & 11.85 & 744.3  & 0.64 &unresolved (saturated)  &   & 0.30   &\\
              & 12.81 & 810.6  & 0.51 &unresolved  (saturated) &   & 0.32   &\\  
19114+0002    & 8.59  & 2.7  & 0.27 & 4.0$\times$3.7 &  N/A   & 0.26   &HD 178131\\
              & 11.85 & 17.1  & 0.25 & 5.2$\times$4.9 & N/A   & 0.32   &\\
              & 12.81 & 1.7  & 0.40 & 5.2$\times$5.1 & N/A  & 0.33   &\\
19132$-$3336  &  8.59 & 47.8  & 0.24 &  unresolved &   & 0.23   &HD 163376\\
              & 11.85 & 34.4  & 0.33 &  unresolved &   & 0.30   &\\
              & 12.81 & 28.8  & 0.32 &  unresolved &   & 0.32   &\\           
 19134+2131   & 11.85 & 3.9  & 0.34 &  unresolved&   & 0.30   &HD 196321\\\
              & 12.81 & 6.2  & 0.37 &  unresolved&   & 0.33   &\\               
19175$-$0807  & 8.59  & 309.3  & 0.54 & unresolved (saturated) &   & 0.23   &HD 163376\\
              & 11.85 & 295.9  & 0.42 & unresolved (saturated) &   & 0.30   &\\
              & 12.81 & 298.0  & 0.34 & unresolved &   & 0.34   &\\              

19192+0922    &  8.59 & 150.4  & 0.38 & unresolved &     & 0.31   &HD 185622\\
              & 11.85 & 181.9  & 0.37 & unresolved &     & 0.34   &\\
              & 12.81 & 159.9  & 0.38 &unresolved  &   & 0.37   &\\
19244+1115    & 8.59  & 611.6  & 0.51 & 3.3$\times$3.2&N/A   & 0.31   &HD 183439\\
              & 11.85 & 1655.1  & 0.52 & 3.8$\times$ 3.6&  N/A &  0.33  &\\
              & 12.81 & 1099.0  & 0.54 & 4.0$\times$3.7&N/A   & 0.36   &\\
19327+3024    &8.80   & 53.8&1.85 &7.5$\times$6.9 &84 & 0.23 &HR 7924 \\
              &9.70   & 58.0&2.05 &7.1$\times$6.4 &80 & 0.23 & \\ 
              &11.60  & 104.0&1.03 &7.5$\times$6.2 &80 & 0.25 & \\ 
              &18.10  & 308.0&1.97 &7.6$\times$6.7 &84 & 0.38 & \\              
19343+2926    & 8.59  & 15.2  & 0.31 & unresolved  &   & 0.30   &HD 186860\\
              & 11.85 & 14.3  & 0.34 & unresolved  &   & 0.34   &\\
              & 12.81 & 16.1  & 0.36 & unresolved  &   & 0.35   &\\
19374+2359    &  8.59 & 18.7  & 0.86 & 3.1$\times$2.9 & 9  &  0.36  &HD 185622\\
              & 11.85 & 21.6  & 0.88 & 3.2$\times$3.0 & 10  &  0.38  &\\
              & 12.81 & 24.5  & 0.91 & 3.0$\times$3.0 & N/A   & 0.40   &\\
19386+0155    &  8.59 & 14.7  & 0.39 & 2.7$\times$2.5 & 116   & 0.48   &HD 185622\\
              & 11.85 & 18.4  & 0.46 & 2.6$\times$2.6 & N/A  & 0.62   &\\
              & 12.81 & 22.5  & 0.41 & 2.2$\times$2.1 & 113   & 0.45   &\\
19454+2920    &  8.59 & 4.9  & 0.46 & 2.1$\times$1.7 & 105  &    0.33   &HD 186860\\
              & 11.85 & 16.0  & 0.52 & 2.6$\times$2.4 & 95  &    0.40   &\\
              & 12.81 & 25.5  & 0.49 & 2.4$\times$2.1 & 96 &    0.40   &\\
19477+2401    &  8.59 & 3.3  & 0.47  & unresolved &   &  0.41  &HD 186860\\
              & 11.85 & 11.4  & 0.49 & unresolved  &   & 0.43   &\\
              & 12.81 & 15.1  & 0.52 & unresolved  &   & 0.45   &\\
19480+2504    & 8.59  & 5.8  & 0.50 & unresolved &   & 0.62   &HD 186860\\
              & 11.85 & 15.5  & 0.51 & unresolved &   & 0.54   &\\
              & 12.81 & 22.8  & 0.51 &unresolved  &   & 0.54   &\\
19500$-$1709  &  8.59 &  &  & problem &   &    &\\
              & 11.85 & 29.3  & 1.12 & 3.2$\times$2.9 & 85  & 0.31  &HD 188603\\
              & 12.81 & 32.9  & 1.00 & 3.0$\times$2.8 & 87   & 0.34   &\\
20004+2955    & 8.59  & 12.4  & 0.35 & unresolved  &   & 0.28   &HD 189577\\
              & 11.85 & 31.5  & 0.38 & unresolved  &   & 0.33   &\\
              & 12.81 & 20.9  & 0.40 & unresolved  &   & 0.34   &\\
20043+2653    & 8.59  & 15.3  & 0.57 & unresolved  &   & 0.48   &HD 189577\\
              & 11.85 & 12.9  & 0.53 & unresolved  &   &  0.48  &\\
              & 12.81 & 19.7  & 0.51 & unresolved  &   & 0.48   &\\
20077$-$0625  & 8.59  & 314.8  & 0.33 & unresolved  &   & 0.38   &HD 192947\\
              & 11.85 & 410.3  & 0.36 & unresolved  &   & 0.37   &\\
              & 12.81 & 405.4  & 0.36 & unresolved  &   & 0.36   &\\
\hline \\     
\end{tabular}
\end{center}
\end{table*}
\begin{table*}
\caption[]{Observed fluxes  and sizes of objects.}
\begin{center}
\begin{tabular}{llllllllllllllll}
\hline 
20547+0247    & 8.59  & 21.3  & 0.25 & unresolved  &   & 0.23   &HD 174387\\
              & 11.85 & 35.7  & 0.34 & unresolved  &   & 0.30   &\\
              & 12.81 & 41.7  & 0.33 & unresolved  &   & 0.32   &\\            
21032$-$0024  & 8.59  & 230.9  & 0.42 & unresolved (saturated) &   & 0.24   &HD 196321\\
              & 11.85 & 194.9  & 0.33 & unresolved   &   & 0.30   &\\
              & 12.81 & 142.8  & 0.34 & unresolved  &   & 0.33   &\\  
21282$+$5050  &8.80  &30.8 & 1.84 &7.4$\times$7.1 &-16 &0.24 &HR 8538\\ 
              &9.70  &31.0 & 1.77 &6.3$\times$4.8 &-13 &0.25 &\\ 
              &11.60 &92.3 & 1.80 &7.3$\times$5.6 &-18 &0.27 &\\ 
              &18.10 &165.2 & 1.87 &6.3$\times$5.1 &-20 &0.39 &\\                
22196$-$4612  & 8.59  &  316.7 & 0.74 &unresolved (saturated)  &   & 0.24   &HD 196321\\
22327$-$1731  & 8.59  & 9.4  & 0.25 & unresolved  &   & 0.23   &HD 198048\\
              & 11.85 & 7.5  & 0.30 & unresolved  &   & 0.30   &\\
23166+1655    & 8.59  & 433.0  &0.57  & unresolved (saturated)&    &   0.37 &HD 220009\\
              & 11.85 & 856.7  & 0.62 & unresolved (saturated) &   & 0.39   &\\
              & 12.81 & 942.4  & 0.56 &unresolved (saturated) &   & 0.40   &\\


\hline \\     
\end{tabular}
\end{center}
\end{table*}


\begin{table*}
\caption[]{Morphologies of the resolved targets, dust properties (C=carbon-rich, O=oxygen-rich, C/O= dual dust chemistry)}{\label{resolved} }
\begin{center}
\begin{tabular}{llllllllllllllll}
\hline
 IRAS name     & Other names                       &MIR morpho         &C/O    & type\\
 \hline
 IRAS 06176$-$1036 & Red Rectangle                 &Core/Bipolar     &C/O& PPN\\
 IRAS 07134+1005 &                                 &Elliptical   & C& PPN\\
 IRAS 07399$-$1435 &  OH 231.8 +4.2                    &Core/Bipolar  &O & PPN\\
 IRAS 10197$-$5750 & Roberts 22                        &Core/Multipolar   &C/O&  PPN\\
 IRAS 10215$-$5916 & AFGL 4106                        &Detached shell & C/O & PRSG\\
 IRAS 12405$-$6219 &                                   &Asymmetrical  & & HII \\ 
 IRAS 14562$-$5406 &Hen 2-113                         &Tore/elliptical   &C/O& PN\\
 IRAS 15103$-$5754 & GLMP 405                         & Core/bipolar &O & WF\\
 IRAS 15445$-$5449 &                                  &Dark Lane/Bipolar   &O& WF \\
 IRAS 15553$-$5230& GLMP 440                          &Marginally resolved & & PPN\\
 IRAS 16279$-$4757&                                    &Core/Multipolar  &C/O & PAGB\\
 IRAS 16333$-$4807 &                                &Core/Multipolar &O & WF \\ 
 IRAS 16342$-$3814& Water foutain nebula              &Dark Lane/Bipolar  & O& WF\\ 
 IRAS 16594$-$4656& Water Lily nebula                  &Core/Bipolar  & C& PPN\\
 IRAS 17047$-$5650 &CPD-56\deg 8032                  &Core/Bipolar  &C/O & PN\\
 IRAS 17106$-$3046 &                                  &Marginally resolved  & O& PPN\\
 IRAS 17150$-$3224 & Cotton candy nebula              &Core/Bipolar  & O& PPN\\
 IRAS 17163$-$3907& Hen 3$-$1379                      &Detached shell/Spherical  &O & PRSG \\
 IRAS 17311$-$4924& LSE 76                            &Core/Bipolar   &C& PPN\\
 IRAS 17347$-$3139 & GLMP 591                         &Core/Multipolar  &O & PN\\
 IRAS 17441$-$2411& Silkworm nebula                   &Core/Multipolar   &O& PPN\\
 IRAS 18184$-$1302 &    MWC 922                      &Square  & & Be\\
 IRAS 18276$-$1431 & V* V445 Sct                      &Marginally resolved   &O& PPN\\

 IRAS 18450$-$0148  &W43A                             &Core/Bipolar   &O& WF \\
 IRAS 19016$-$2330&                                   & Marginally resolved  & & PPN\\
 IRAS 19114+0002 &AFGL2343                            & Detached shell & O&MES\\ 
 IRAS 19244+1115 &     IRC +10420                     & Core/extended & O& MES\\
 IRAS 19327+3024   & BD+30\deg 3639                    & Elliptical & & PN\\
 IRAS 19374+2359 &                                    &Detached shell  & O& PPN\\
 IRAS 19386+0155 &V1648 Aql                           &Core/extended  & & PPN\\
 IRAS 19454+2920 &                                    &Core/Extended  & C& PPN\\
 IRAS 19500$-$1709 & V5112 Sgr                       &Detashed Shell, no central star  & C& PPN\\
 IRAS 20043+2653 & GLMP 972                         &Core/Extended  & &     PPN\\                 
 IRAS 21282+5050   & &Toroidal  & C& PN\\

\hline \\
\end{tabular}
\end{center}
\end{table*}

\section{Images}
\clearpage




\clearpage


\clearpage


\clearpage


\begin{figure*}
\begin{center}
\includegraphics[width=18cm]{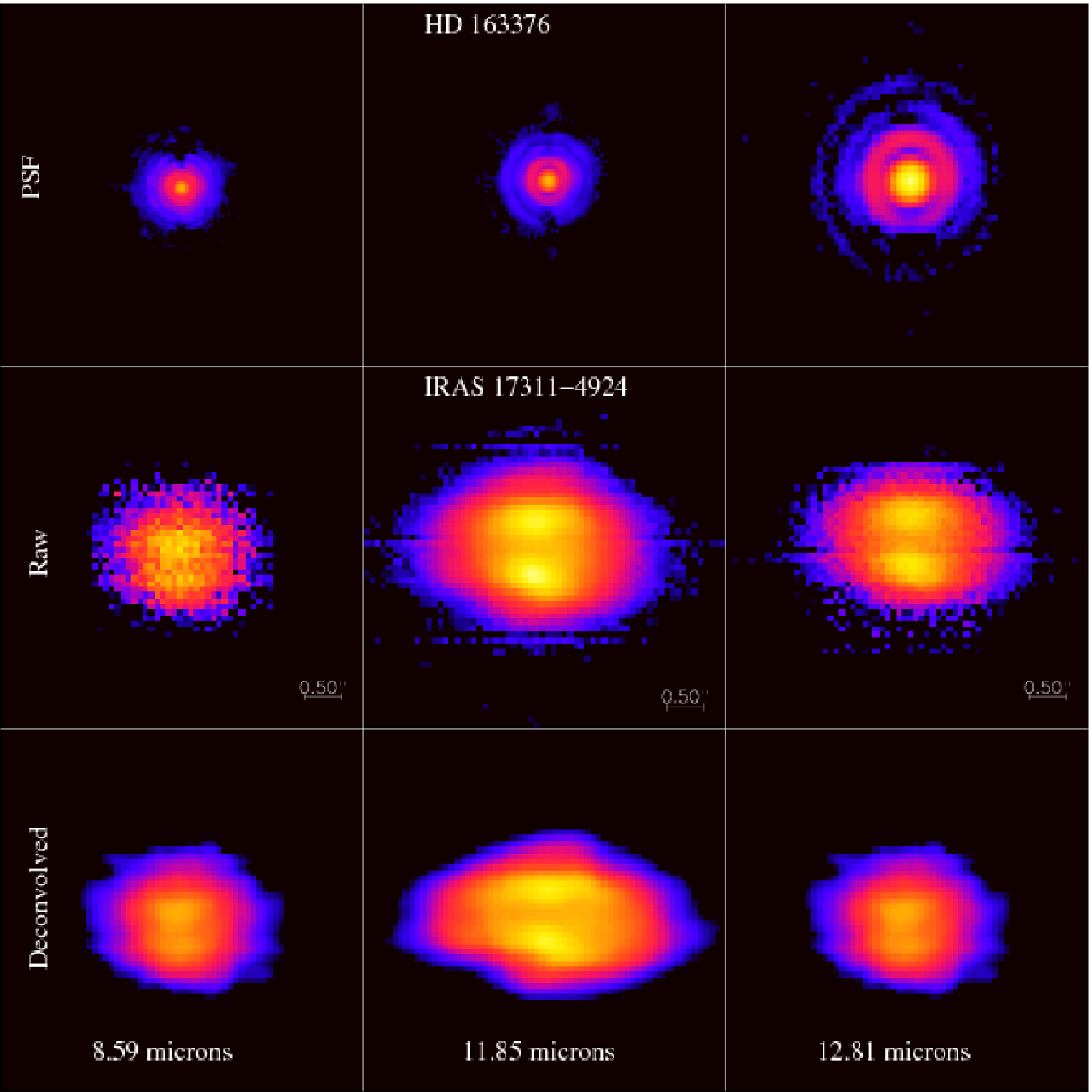}
\caption{Visir burst mode images of IRAS 17311}
\label{im_17311}
\end{center}
\end{figure*}


\begin{figure*}
\begin{center}
\includegraphics[width=18cm]{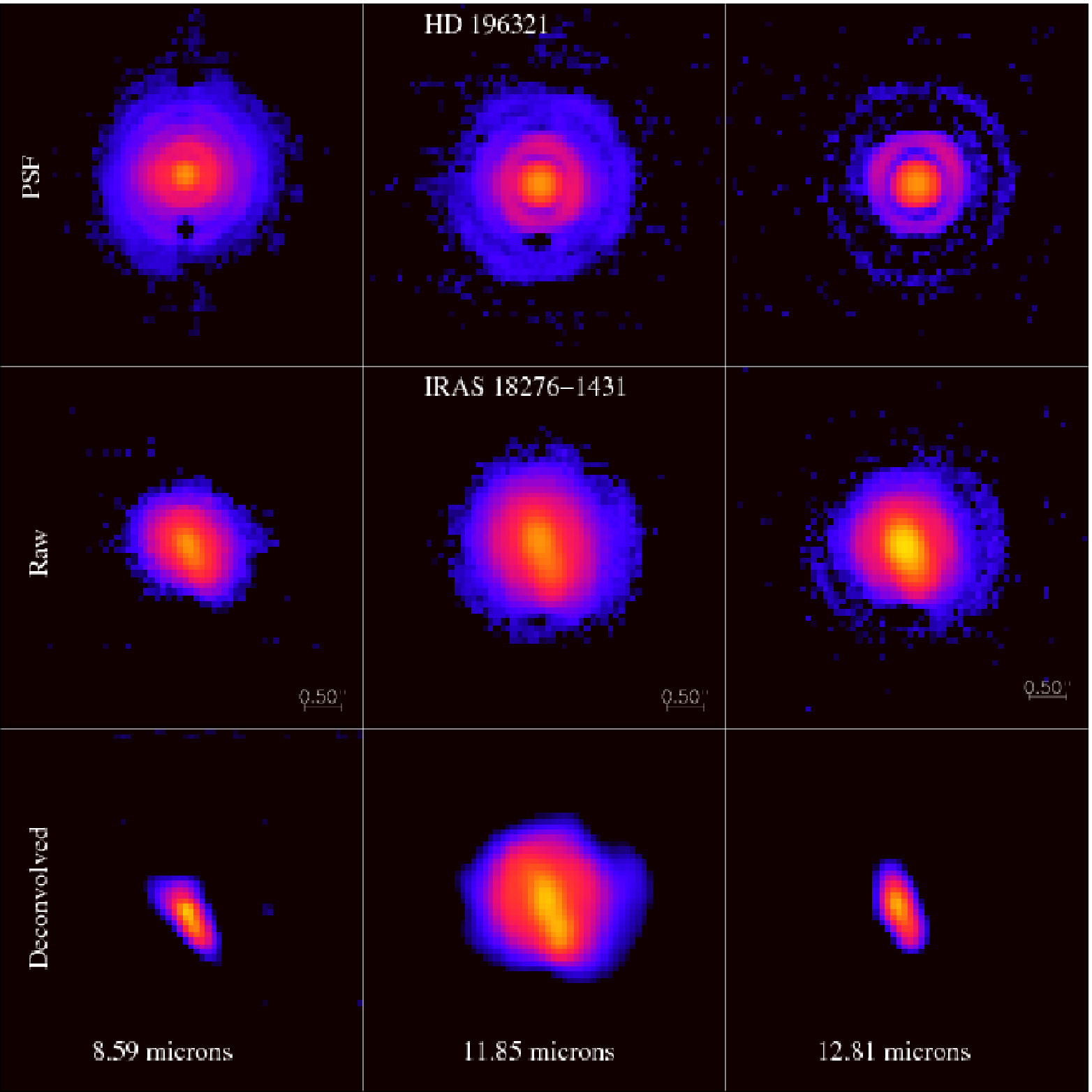}
\caption{Visir burst mode images of IRAS 18276. }
\label{im_18276}
\end{center}
\end{figure*}


\begin{figure*}
\begin{center}
\includegraphics[width=16cm]{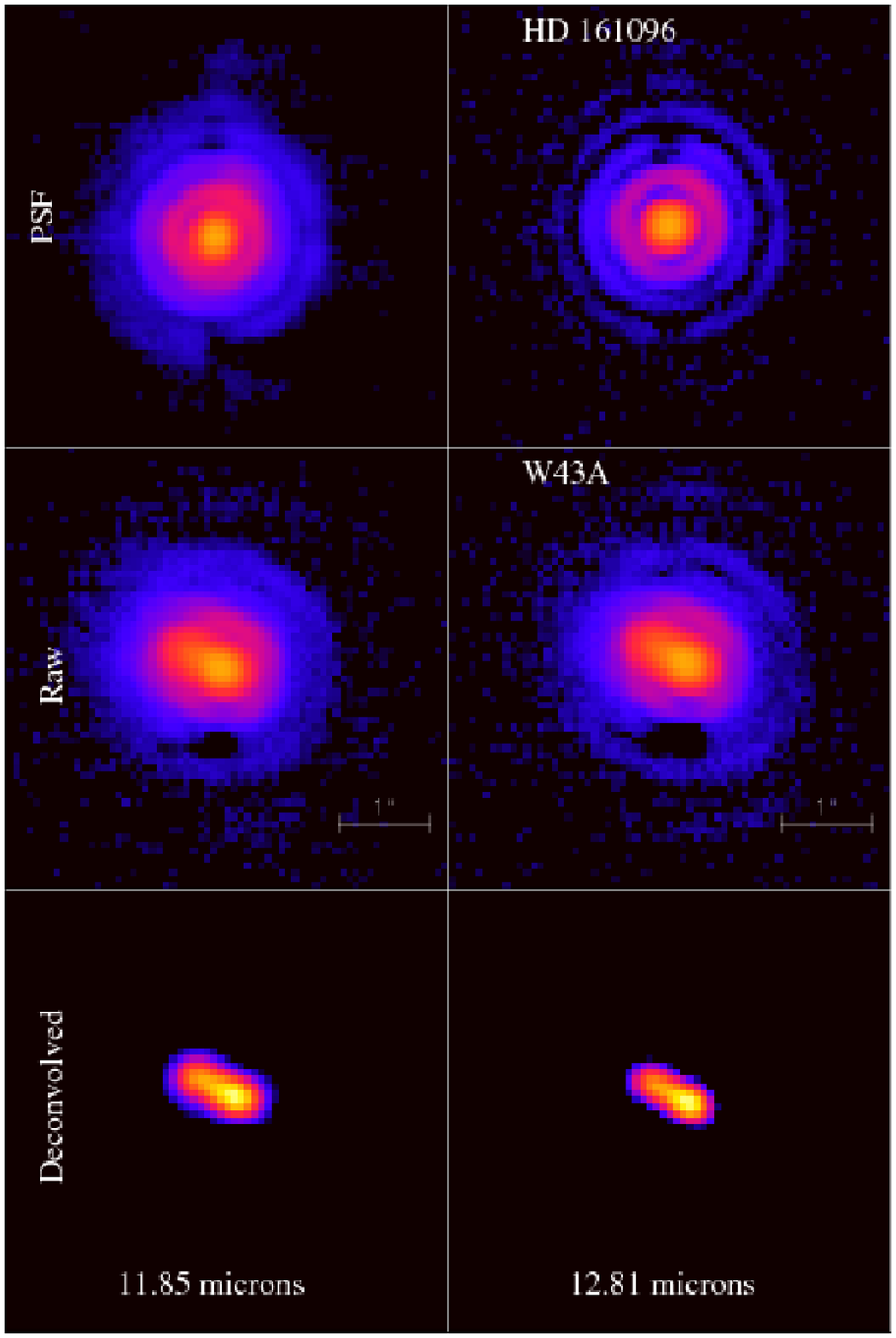}
\caption{Visir burst mode images of W43A.}
\label{im_w43a}
\end{center}
\end{figure*}

\clearpage
\begin{figure*}
\begin{center}
\includegraphics[width=18cm]{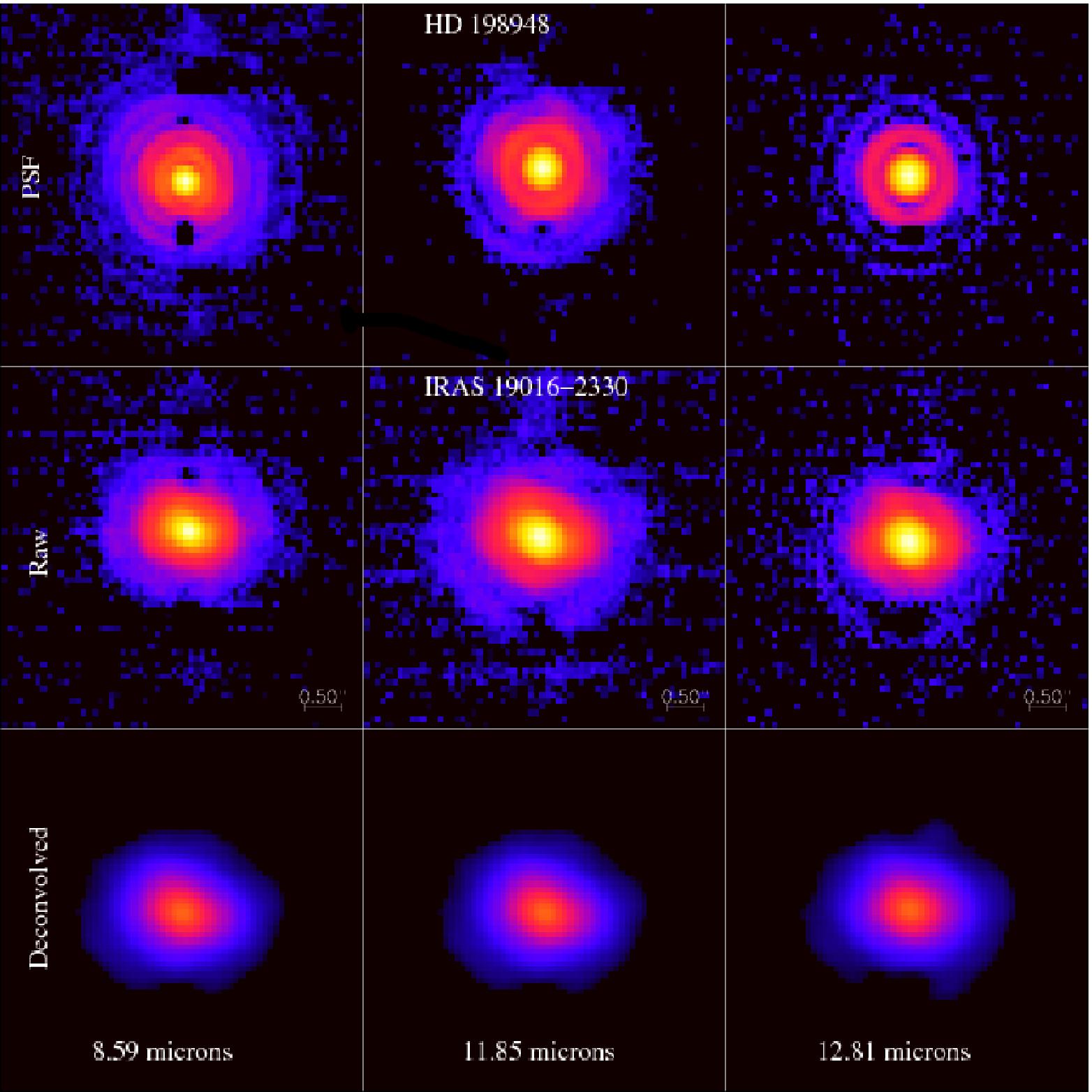}
\caption{Visir burst mode images of IRAS 19016.}
\label{im_19016}
\end{center}
\end{figure*}


\begin{figure*}
\begin{center}
\includegraphics[width=18cm]{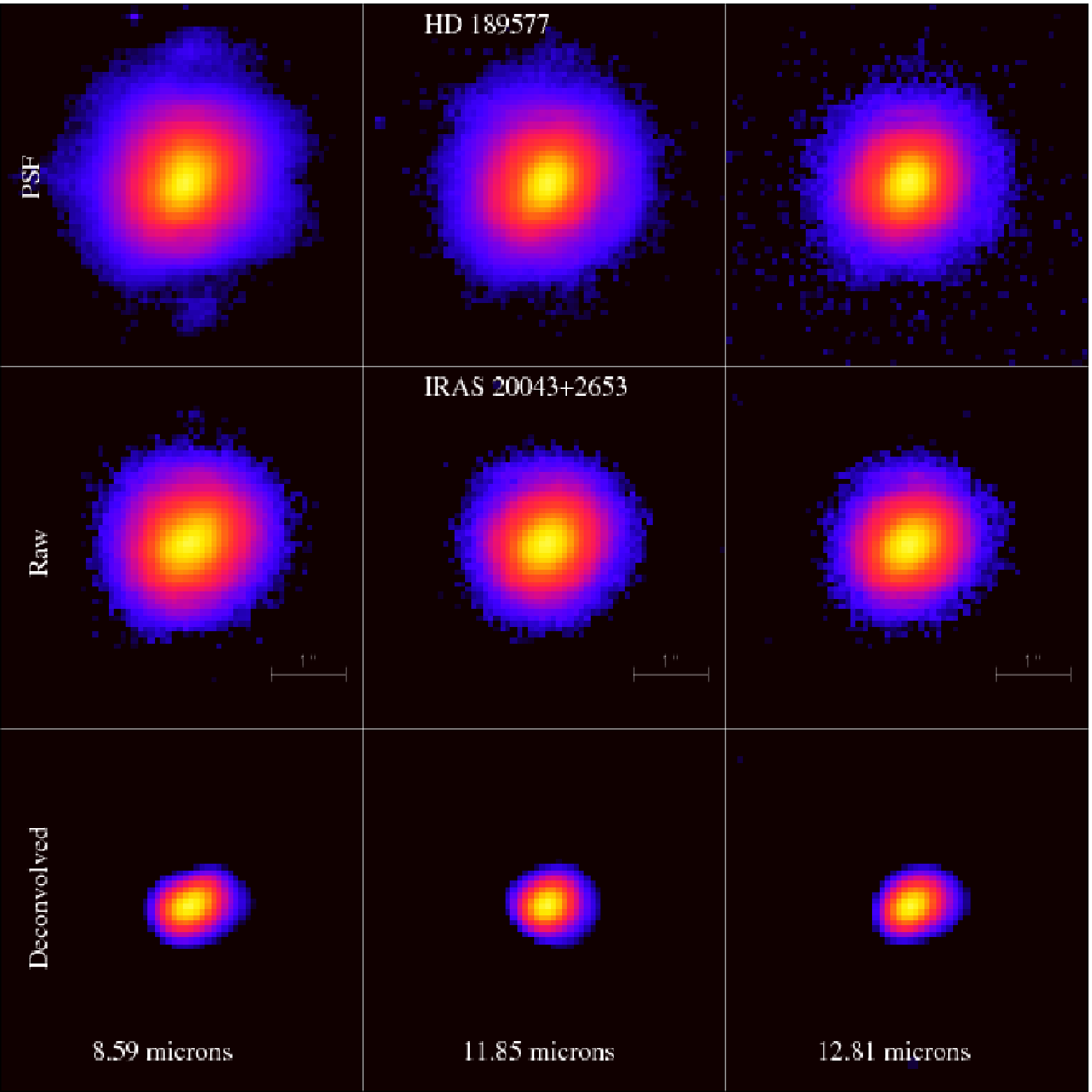}
\caption{Visir burst mode images of IRAS 20043.}
\label{im_20043}
\end{center}
\end{figure*}


\clearpage
\begin{figure*}
\begin{center}
\includegraphics[width=18cm]{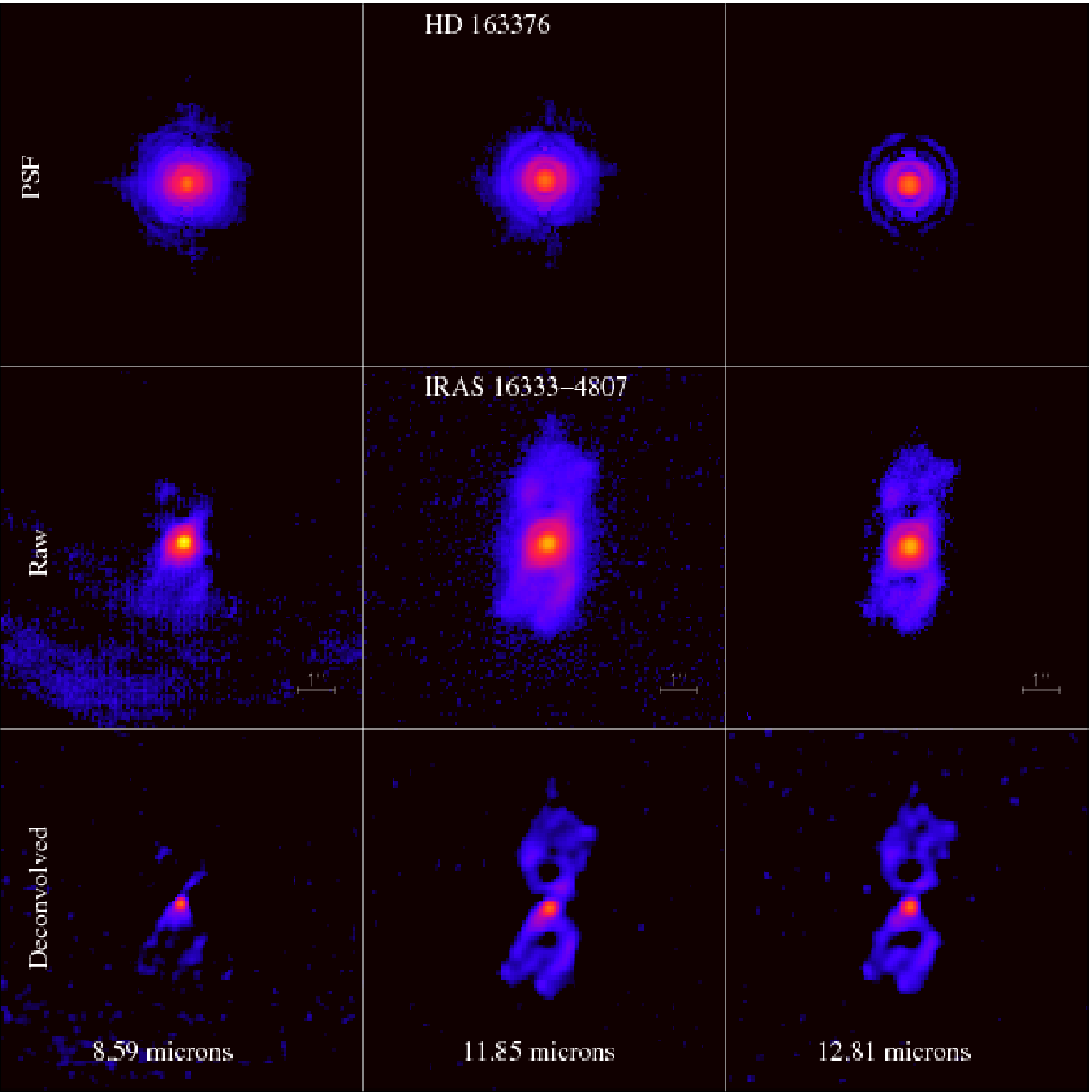}
\caption{Visir burst mode images of IRAS 16333}
\label{im_16333}
\end{center}
\end{figure*}


\begin{figure*}
\begin{center}
\includegraphics[width=18cm]{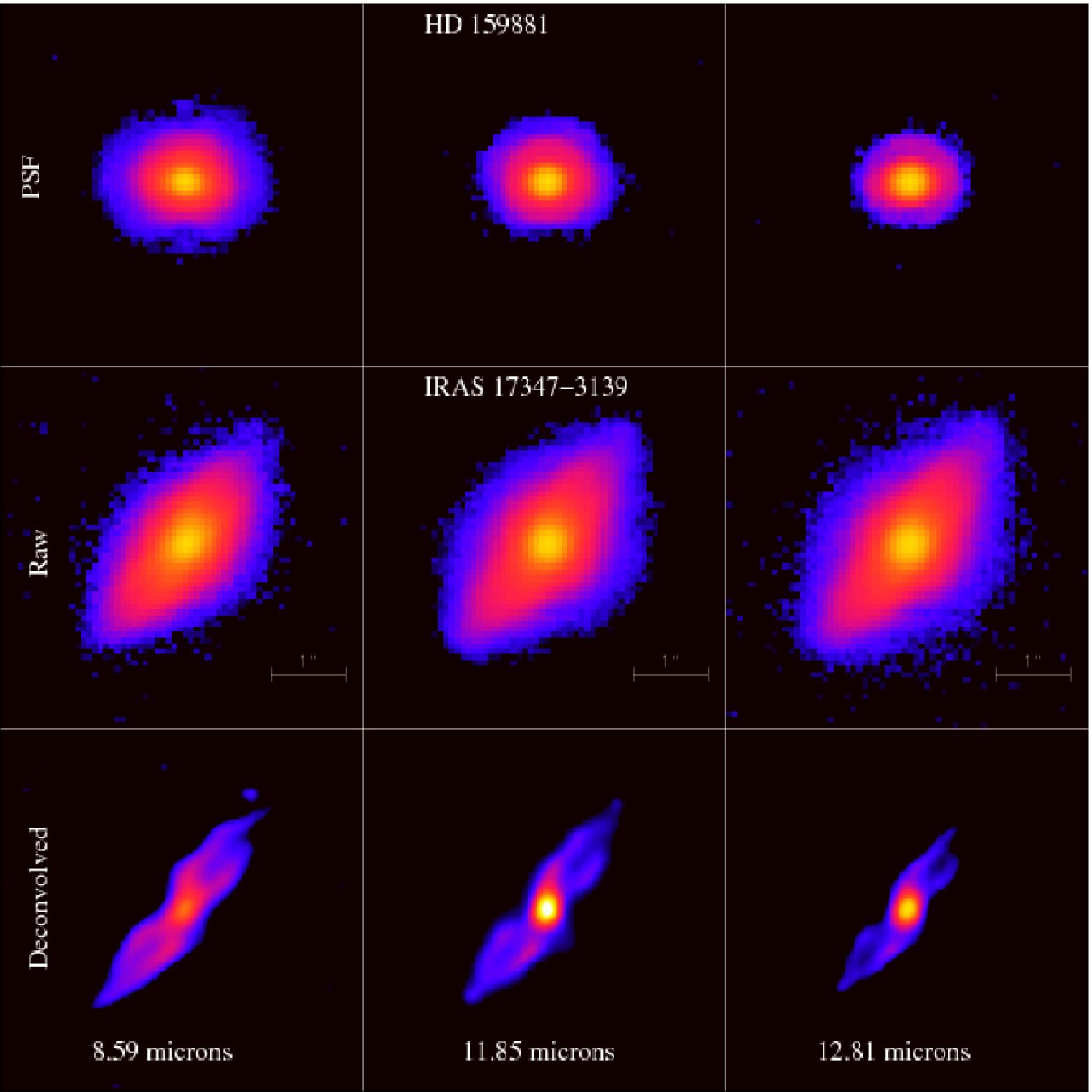}
\caption{Visir images of IRAS 17347 }
\label{im_17347}
\end{center}
\end{figure*}


\clearpage

\begin{figure*}
\begin{center}
\includegraphics[width=18cm]{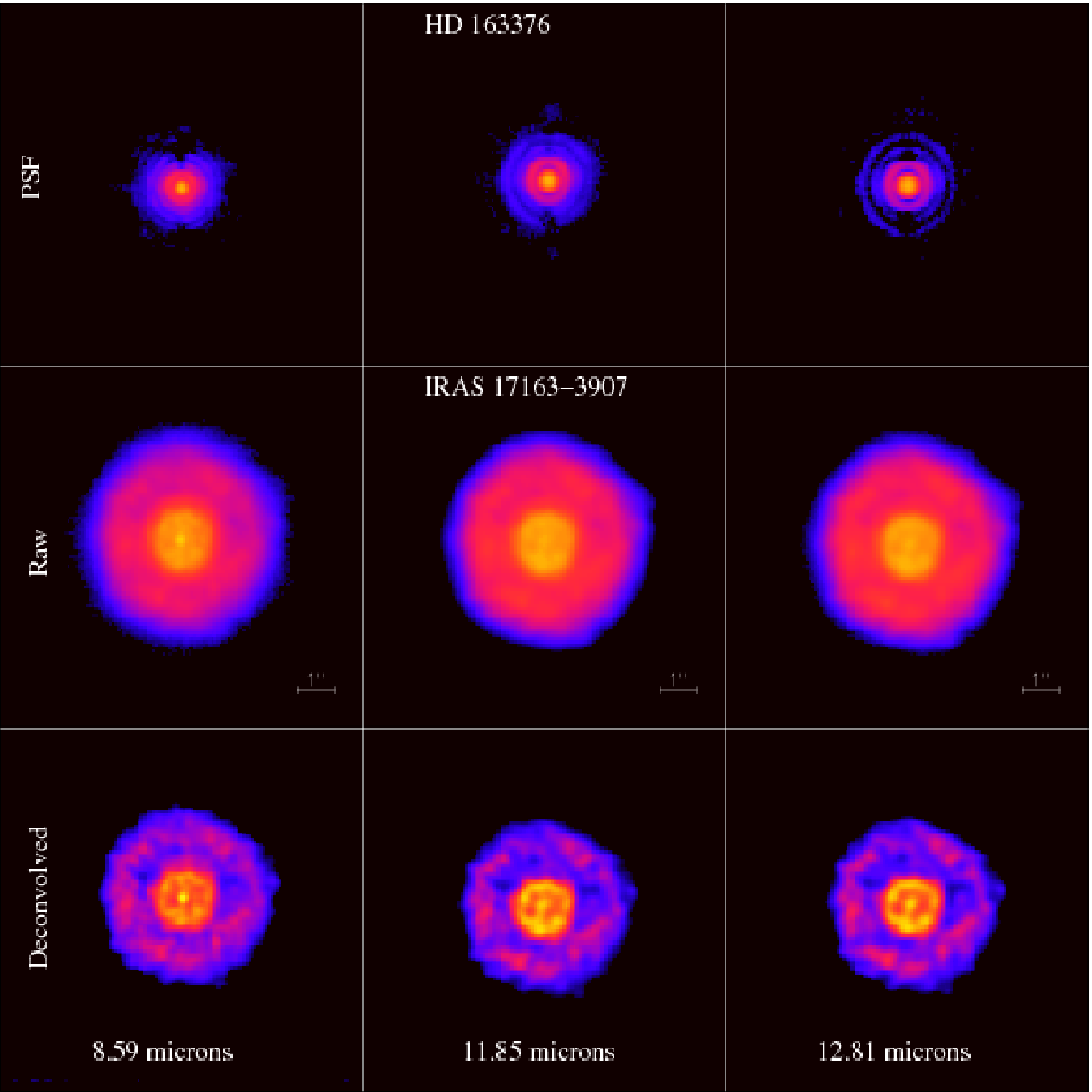}
\caption{Visir burst mode images of IRAS 17163.}
\label{im_17163}
\end{center}
\end{figure*}


\begin{figure*}
\begin{center}
\includegraphics[width=18cm]{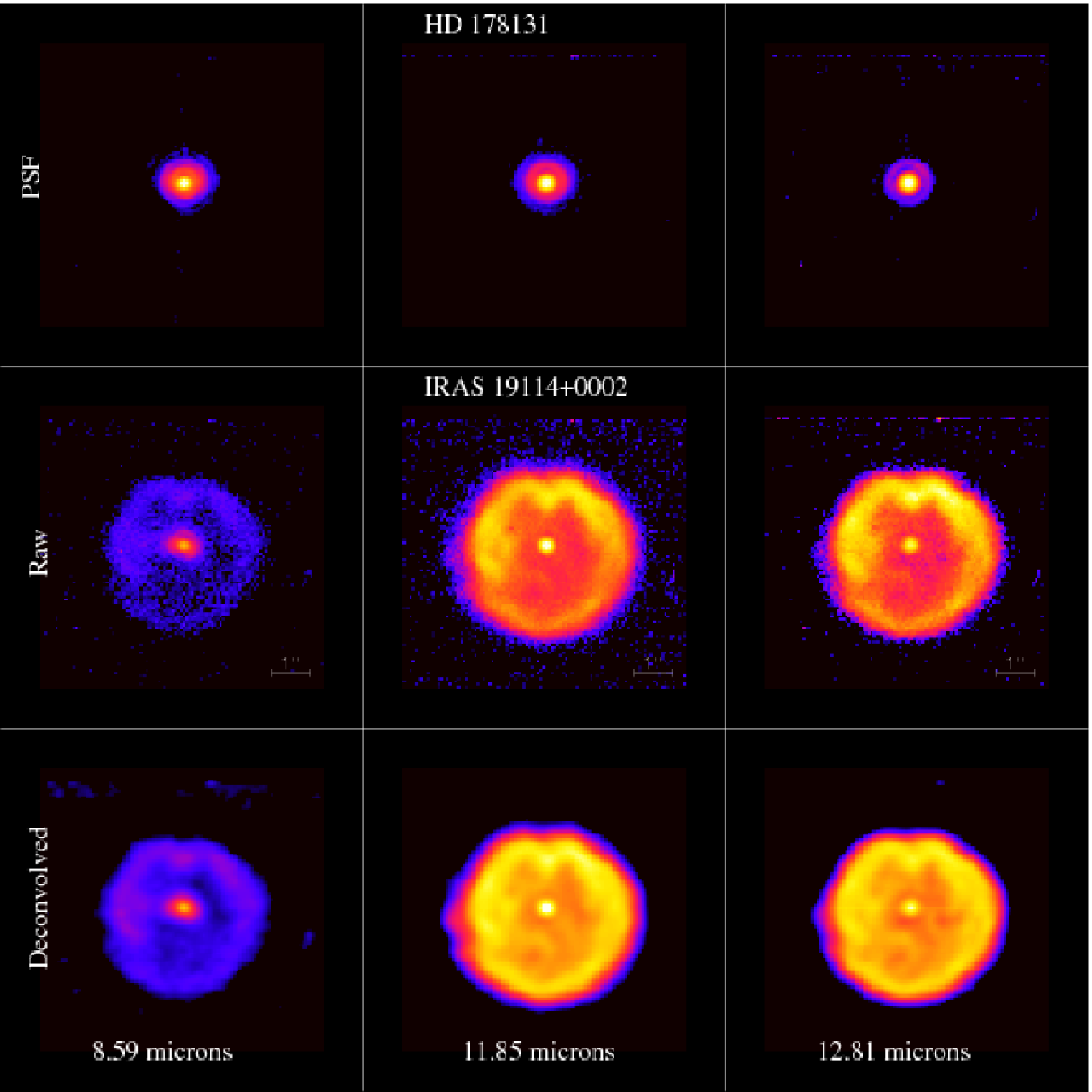}
\caption{Visir images of IRAS 19114.}
\label{im_19114}
\end{center}
\end{figure*}
\clearpage
\begin{figure*}
\begin{center}
\includegraphics[width=18cm]{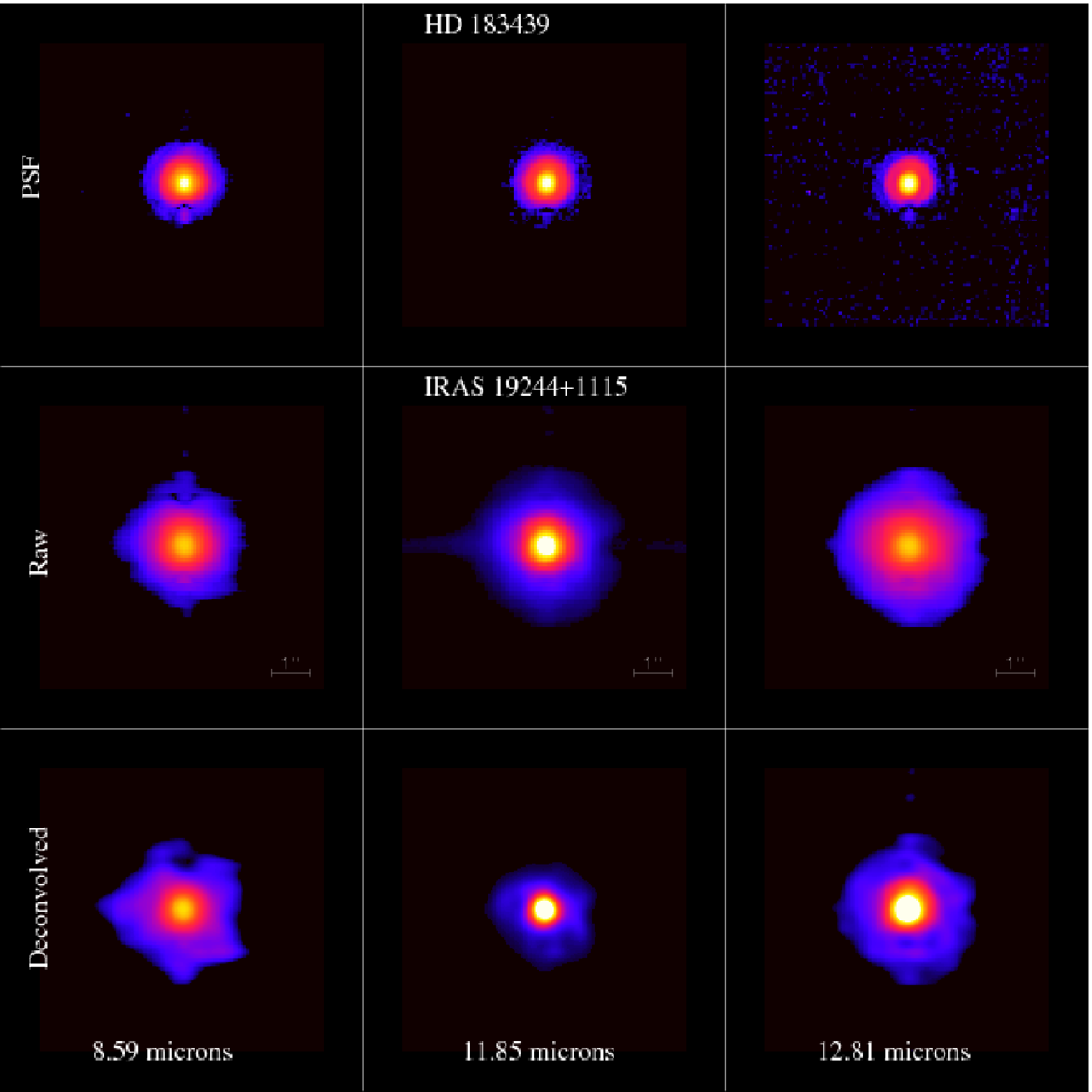}
\caption{Visir burst mode images of IRAS 19244 (IRC+10420).}
\label{im_19244}
\end{center}
\end{figure*}


\begin{figure*}
\begin{center}
\includegraphics[width=18cm]{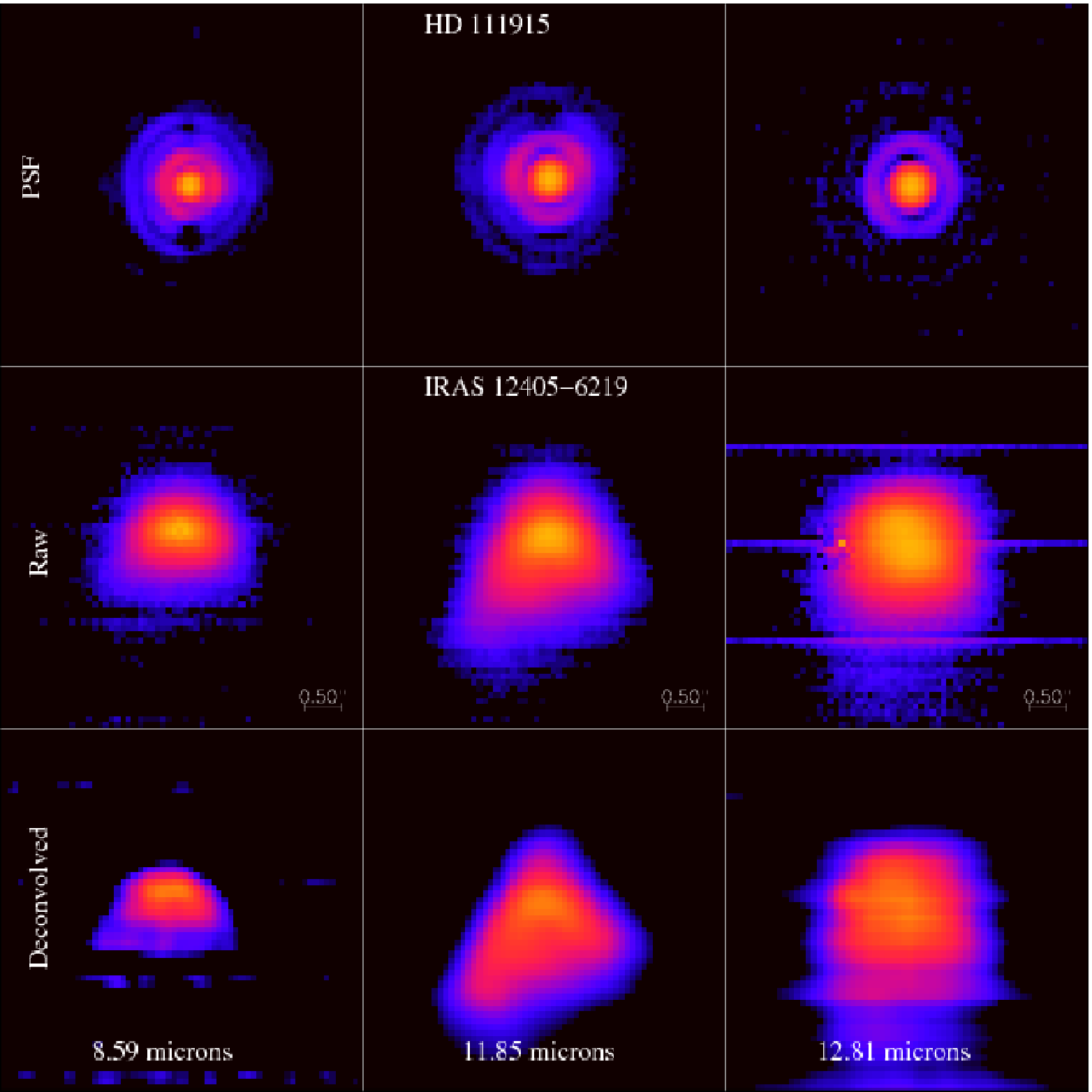}
\caption{Visir burst mode images of IRAS 12405}
\label{im_12405}
\end{center}
\end{figure*}
\clearpage
\begin{figure*}
\begin{center}
\includegraphics[width=18cm]{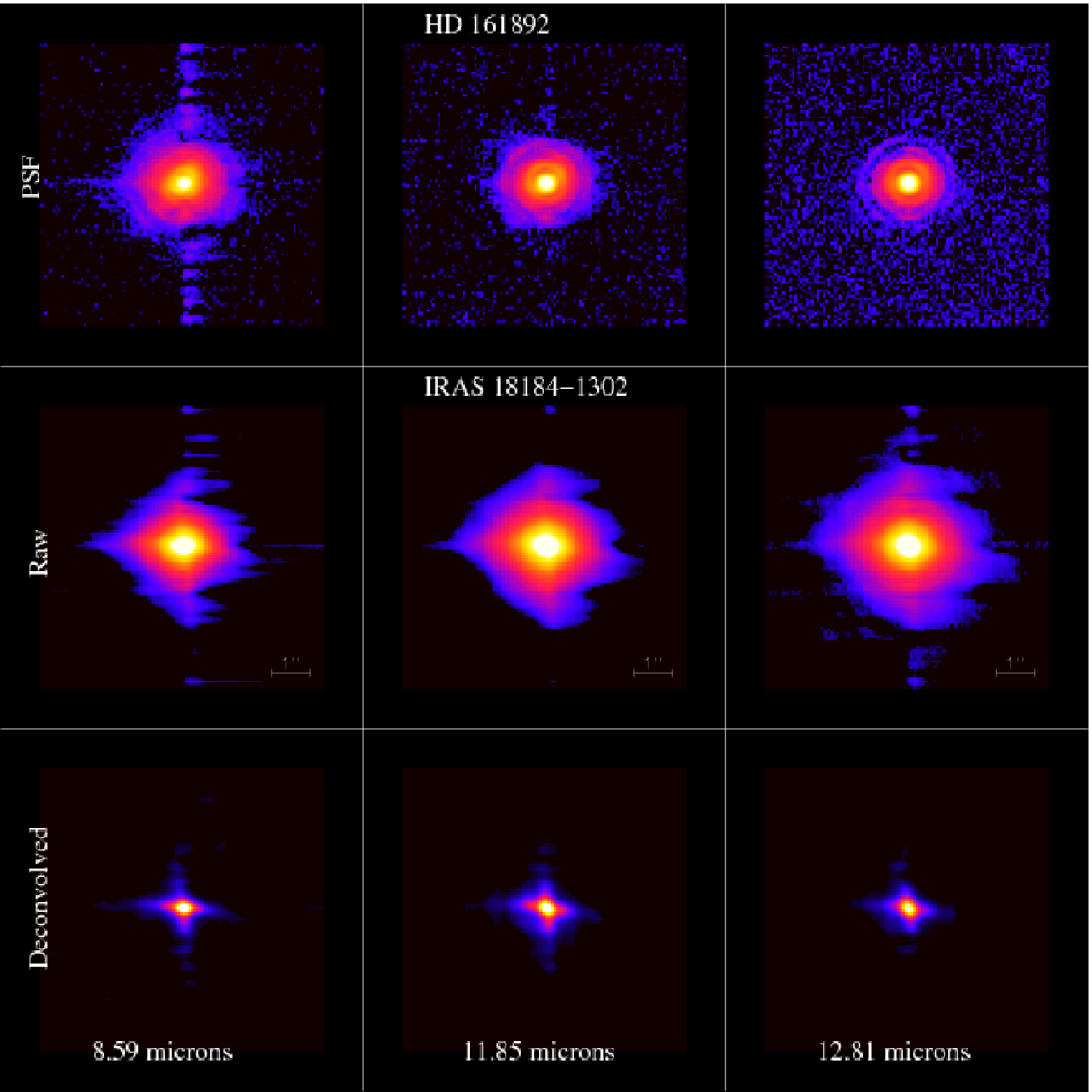}
\caption{ Visir images of IRAS 18184 (The red Square nebula).}
\label{im_18184}
\end{center}
\end{figure*}
\clearpage

\end{document}